\documentclass[twocolumn]{aastex631}

\usepackage{tabularx,rotating}
\usepackage{xcolor}
\usepackage{tablefootnote}
\usepackage{ulem} 

\newcommand{\sw}{{Swift}}
\newcommand{\cha}{{Chandra}}
\newcommand{\xmm}{XMM-{Newton}}
\shorttitle{A \sw~X-ray view of the SMS4 sample}
\shortauthors{Maselli et al.}
\graphicspath{{./}{figures/}}

\begin{document}






\title{A \sw~X-Ray View of the SMS4 Sample - X-Ray Properties of 31 Quasars and Radio Galaxies}

\correspondingauthor{Alessandro Maselli}
\email{alessandro.maselli@inaf.it}

\author[0000-0003-3760-1910]{Alessandro Maselli} 
\affiliation{INAF-Osservatorio Astronomico di Roma \\
via Frascati 33, I-00078, Monte Porzio Catone (Roma), Italy}
\affiliation{ASI-Space Science Data Center \\
via del Politecnico snc, I-00133, Roma, Italy}

\author[0000-0002-9478-1682]{William R. Forman}
\affiliation{Harvard-Smithsonian Center for Astrophysics \\
60 Garden Street, Cambridge, MA~02138, USA}

\author[0000-0003-2206-4243]{Christine Jones}
\affiliation{Harvard-Smithsonian Center for Astrophysics \\
60 Garden Street, Cambridge, MA~02138, USA}

\author[0000-0002-0765-0511]{Ralph P. Kraft}
\affiliation{Harvard-Smithsonian Center for Astrophysics \\
60 Garden Street, Cambridge, MA~02138, USA}

\author[0000-0003-3613-4409]{Matteo Perri}
\affiliation{INAF-Osservatorio Astronomico di Roma \\
via Frascati 33, I-00078, Monte Porzio Catone (Roma), Italy}
\affiliation{ASI-Space Science Data Center \\
via del Politecnico snc, I-00133, Roma, Italy}


\begin{abstract}

We present Neil Gehrels \sw~Observatory (hereafter \sw) observations of 31 sources from the SMS4 catalog, a sample of 137 bright radio sources originally designed to extend the well-studied 3CRR radio sample to the southern hemisphere.
All these sources had no \cha~or \xmm~observations: 24 of these were observed with \sw~through a dedicated proposal in 2015, and data for the remaining seven were retrieved from the \sw~archive.
The reduction and analysis of data collected by the \sw~X-ray Telescope (XRT) led to 20 detections in the 0.3--10~keV band.
We provide details of the X-ray emission in this band for these 20 detections, as well as upper limits for the remaining 11 SMS4 sources.
When statistics allowed, we investigated the extent of the X-ray emission and the hardness ratio, and we carried out a spectral analysis.
We matched the 20 X-ray-detected sources with infrared (AllWISE, CatWISE2020) and optical (GSC~2.3.2, DES~DR2) catalogs to establish associations with infrared and optical sources, and compared our results with previously published counterparts in these bands.
Requiring a detection in both the infrared and the optical bands to establish a candidate counterpart for our X-ray detections, we obtain reliable counterparts for 18 sources, while the remaining two sources need further investigation to establish firm identifications.
In the infrared, we confirm 12 previously established counterparts and provide six new candidates. 
In the optical, we find agreement with 13 previously established counterparts, while we provide an alternative candidate for five SMS4 sources.
We find that $\sim$35\% of all the SMS4 sources lie below the lower limit of 10.9~Jy for the flux density at 178~MHz established for the 3CRR sample, at variance with the values extrapolated using measurements at higher frequencies.
Therefore, for future studies where flux-density-limited samples are needed, we encourage the use of the more recent G4Jy sample.
We present the list of 56 SMS4 sources that in 2022 March remain to be observed in the X-rays with narrow-field instruments, to highlight interesting sources and organize further observational campaigns to achieve complete X-ray coverage for the whole SMS4 in the forthcoming years.

\end{abstract}



\section{Introduction} 
\label{sec:1}

Over the past few decades, observational evidence of the major role played by central supermassive black holes (SMBHs) in the evolution of their host galaxies has progressively increased.
The processes by which such an influence occurs are referred to as active galactic nucleus (AGN) feedback (for a recent review, see \citealp{2019SSRv..215....5W}).

In a simplified picture that has gained large consensus, two major feedback modes, controlled by the accretion rate onto the SMBH, have been identified (\citealp{2005MNRAS.363L..91C}).
In the so-called radio - or mechanical - mode, operating at low accretion rates, relativistic outflows from the SMBH would be able to transfer energy to the surrounding medium, over a wide range of distances (\citealp{2007ARA&A..45..117M,2012NJPh...14e5023M}).

For brightest cluster galaxies (BCGs) at the center of cooling flows (see \citealp{1994ARA&A..32..277F} for a review) these plasma jets would have the remarkable role of balancing the radiative losses of the X-ray-emitting gas in the intracluster medium (\citealp{2002MNRAS.332..729C,2015SSRv..188..141B}), giving a solution to the {\it cooling flow problem}.
More generally, the outflows would also heat the coronae surrounding massive elliptical galaxies (\citealp{1985ApJ...293..102F}), reducing the star formation rate up to its substantial quenching.
Shock fronts and cavities within these hot X-ray atmospheres, associated with the expansion of jet radio lobes, have been revealed through the comparison between detailed radio and X-ray images and have been interpreted in the framework of such interaction (see, {\it e.g.}, \citealp{2001ApJ...547L.107F,2002ApJ...567L.115J,2003ApJ...592..129K,2005ApJ...635..894F,2007ApJ...665.1057F,2011ApJ...726...86R,2015ApJ...805..112R}).

The need to properly take AGN feedback into account has been assessed also by cosmological simulations, based on a semianalytic approach (see, {\it e.g.}, \citealp{2006MNRAS.365...11C,2016ApJS..222...22C}), as well as by hydrodynamical simulations like those from the recent EAGLE (\citealp{2015MNRAS.446..521S,2015MNRAS.450.1937C}) and Illustris (\citealp{2013MNRAS.436.3031V,2018MNRAS.473.4077P}) projects.

The best way to obtain essential information about the X-ray emission from radio-loud AGNs ({\it e.g.}, radio galaxies and radio bright quasars) and their environments is to select complete, flux-limited samples from low-frequency radio surveys, which span a wide range in radio power and redshift and also are unbiased with respect to the angle between the jet and our line of sight.
So far, despite a rich collection of galaxies investigated in great detail, a consistent number of luminous radio sources remains poorly explored in X-rays using pointed observations with narrow-field instruments.

For the Northern Hemisphere, the Third  Cambridge (3C) Catalog \citep{1959MmRAS..68...37E}, along its revisions, the 3CR \citep{1962MmRAS..68..163B}  and the 3CRR \citep{1983MNRAS.204..151L}, is a premier sample for understanding the nature and evolution of powerful radio galaxies, as well as their relationship to their host galaxies and environments, on scales from parsecs to megaparsecs.
The 3CRR sample was built from a survey at 178~MHz carried out with the Cambridge interferometer and includes 173 sources at $\delta \geq 10^{\circ}$, at $|b| \geq 10^{\circ}$, and with a flux density $S_{178}$ higher than $S_{178}^{\star} = 10.9$~Jy.

In 2006, with the goal of expanding the 3CRR catalog into the Southern Hemisphere, a sample of extremely bright radio sources \citep{2006AJ....131..100B} was extracted from the Molonglo Reference Catalogue (MRC; \citealt{1981MNRAS.194..693L,1991Obs...111...72L}), based on a low-frequency survey at 408~MHz.
The first step was the compilation of the Molonglo Southern 4~Jy (MS4) sample, which included 228 sources at $-30^{\circ} \leq \delta \leq -85^{\circ}$, at $|b| > 10^{\circ}$, and with $S_{408} > 4.0$ Jy.
These 228 sources were all imaged at 843~MHz with the Molonglo Observatory Synthesis Telescope (MOST) to establish accurate positions, flux densities, and angular sizes. 
Then, the 133 MS4 sources with largest angular size (LAS) $<$ 35\arcsec~were selected for high-resolution imaging at 5~GHz with the Australia Telescope Compact Array (ATCA). 
Optical identification for all MS4 sources was pursued complementing the analysis on plates from the UK Schmidt Southern Sky Survey with {\it R}-band CCD images made with the Anglo-Australian Telescope (AAT) \citep{2006AJ....131..114B}.
Furthermore, spectral information at other frequencies was collected from the literature to derive $S_{178}$ for all MS4 sources. 
Hence, by using the same flux density threshold $S_{178}^{\star}$ as in the 3CRR catalog, a strong-source subset of 137 sources with $S_{178} \geq S_{178}^{\star}$, called SMS4, was created.

In 2020, a new sample of the brightest radio sources in the southern sky \citep{2020PASA...37...18W,2020PASA...37...17W} was generated from observations taken with the Murchison Widefield Array (MWA, \citealp{2013PASA...30....7T}) in the 72--231 MHz range.
MWA observations obtained during the first year of operations produced the Galactic and Extragalactic All-sky MWA (GLEAM; \citealp{2015PASA...32...25W}) Survey, from which the Extragalactic Catalogue (EGC; \citealp{2017MNRAS.464.1146H}) was built. 
Hence, the GLEAM 4 Jy (G4Jy) sample of 1863 sources at $\delta \leq -30^{\circ}$, at $|b| > 10^{\circ}$, and with $S_{151} > 4.0$~Jy, was established.

In recent years, significant efforts have been devoted to achieving full X-ray coverage of 3CRR sources, including a \cha~snapshot program started in 2009, beginning with sources at $z \leq 0.3$ \citep{2010ApJ...714..589M,2012ApJS..203...31M,2015ApJS..220....5M} and progressing toward higher redshifts \citep{2013ApJS..206....7M, 2018ApJS..234....7M,2018ApJS..235...32S,2020ApJS..250....7J}, coupled with follow-up observations of peculiar sources at all frequencies (see, {\it e.g.}, \citealp{2009ApJ...692L.123M,2010MNRAS.401.2697H,2016MNRAS.458..681D,2018ApJS..238...31M,2018A&A...619A..75M}).

To facilitate observing all 3CRR sources with \cha, we performed an observational campaign with \sw~\citep{2004ApJ...611.1005G} for 21 3CRR sources that, according to \cite{1983MNRAS.204..151L}, were still unidentified, to increase their multifrequency information and derive accurate X-ray fluxes.
With these observations, we detected X-ray emission with the \sw~X-ray Telescope (XRT; \citealp{2005SSRv..120..165B}) for 9 of the 21 objects, and we also associated an infrared counterpart with these nine and an additional four objects \citep{2016MNRAS.460.3829M}.
The X-ray emission for seven of these nine detected sources was then investigated in greater detail with \cha~\citep{2021ApJS..255...18M}.

In this paper, we report on a \sw~observing program of the SMS4 sample, including 31 sources with no X-ray observations in the \cha~or \xmm~archives.
A total of 24 sources were observed for the first time as a result of a dedicated proposal, while six were already present in the \sw~archive when our dedicated campaign started, in 2015 November.
The remaining source, PKS~B2148$-$555, was observed in 2019 February.

We performed a reduction and analysis of the \sw-XRT data and crossmatched the X-ray detections that we obtained with catalogs in the infrared and optical bands.
After collecting and analyzing this multifrequency information, we compared our results with those in the literature to verify the consistency of our X-ray detections with previously suggested counterparts.
The two most comprehensive and exhaustive studies that we use are \cite{2020PASA...37...18W,2020PASA...37...17W} for infrared counterparts and \cite{2006AJ....131..114B} for optical counterparts.
In the following, for the sake of simplicity, we refer to these papers as W20 and BH06, respectively.

A preliminary comparison of SMS4 with the recent G4Jy is described in Section~\ref{sec:2}; the list of observed SMS4 sources is presented in Section~\ref{sec:3}, while the reduction and analysis of \sw~X-ray data are discussed in Section~\ref{sec:4}; the multifrequency analysis is described in Section~\ref{sec:5}, and our results are summarized in Section~\ref{sec:6}.
Throughout this paper we use CGS units, unless otherwise stated.
We also assume a flat cosmology with $H_0 = 72$ km s$^{-1}$ Mpc$^{-1}$, $\Omega_M = 0.26$, and $\Omega_{\Lambda} = 0.74$ \citep{2009ApJS..180..306D}.


\begin{table*}
\footnotesize
\begin{center}
\caption{\scriptsize{List of Correspondences Established between the 31 SMS4 Sources in Our Sample and G4Jy (Top) or TGSS-ADR1 (Bottom) Sources}}
\label{tab:01}
\begin{tabular}{cccc|cccccc}
\hline
   (1)          &   (2)    & (3)      & (4)        &   (5)              &   (6)    & (7)         &  (8)                        &       (9)      &    (10)    \\   
SMS4 Name       & S$_{178}$ & LAS      & Redshift        & IAU Name           & G4Jy ID  & R.A.        & Decl.                       & Flux Density   & Morphology \\ 
                &   (Jy)   & (arcsec) &            &                    &          & ($^{h~m~s}$) & ($^{\circ}$~\arcmin~\arcsec) &      (Jy)      &            \\  
\hline          
B0007$-$446~~   & 15.0     &  33      &   (1.00)   &  J001030$-$442259  &   20     & 00 10 30.55 &        $-$44 22 57.0        & 11.88$\pm$0.02 &      s     \\ 
B0013$-$634~~   & 15.0     &  31      & ($>$ 0.56)   &  J001602$-$631005  &   27     & 00 16 02.68 &        $-$63 10 07.2        & 12.15$\pm$0.02 &      s     \\
B0049$-$433~~   & 15.0     &  18      &   (0.39)   &  J005214$-$430628  &   93     & 00 52 14.88 &        $-$43 06 29.2        & 12.45$\pm$0.01 &      s     \\
B0157$-$311~~   & 19.0     &  14      &    0.677   &  J020012$-$305324  &  213     & 02 00 12.15 &        $-$30 53 26.5        & 15.51$\pm$0.02 &      s     \\ 
B0219$-$706~~   & 16.0     &  10      &   (0.40)   &  J022008$-$702231  &  249     & 02 20 08.18 &        $-$70 22 27.8        & 11.19$\pm$0.02 &      s     \\
B0223$-$712~~   & 14.0     &  23      &   (1.27)   &  J022357$-$705949  &  257     & 02 23 57.55 &        $-$70 59 46.6        & 10.24$\pm$0.02 &      s     \\
B0245$-$558~~   & 21.0     &  38      &   (0.82)   &  J024656$-$554116  &  293     & 02 46 56.26 &        $-$55 41 22.7        & 15.33$\pm$0.02 &      d     \\
B0315$-$685~~   & 11.3     &  10      &   (1.30)   &  J031610$-$682104  &  339     & 03 16 10.10 &        $-$68 21 07.2        &  7.15$\pm$0.02 &      s     \\
B0407$-$658~~   & 59.0     &  10      &   (0.77)   &  J040820$-$654458  &  416     & 04 08 20.25 &        $-$65 45 10.7        & 50.04$\pm$0.03 &      s     \\          
B0411$-$561~~   & 11.3     &  27      &   (0.42)   &  J041247$-$560043  &  427     & 04 12 48.02 &        $-$56 00 48.8        & 10.76$\pm$0.02 &      s     \\
B0420$-$625~~   & 24.0     &  10      &   (0.81)   &  J042056$-$622337  &  446     & 04 20 56.43 &        $-$62 23 36.9        & 16.34$\pm$0.02 &      s     \\
B0453$-$301~~   & 18.0     &  50      &   (0.22)   &  J045514$-$300646  &  506     & 04 55 14.31 &        $-$30 06 47.2        & 14.81$\pm$0.02 &      s     \\
B0534$-$497~~   & 11.9     &  30      &    0.184   &  J053613$-$494420  &  563     & 05 36 13.67 &        $-$49 44 23.1        & 10.44$\pm$0.02 &      s     \\
B0546$-$445~~   & 11.4     &  14      &   (1.15)   &  J054738$-$443114  &  576     & 05 47 38.28 &        $-$44 31 14.9        &  7.60$\pm$0.02 &      s     \\
B0547$-$408~~   & 16.0     &  37      &   (0.64)   &  J054924$-$405110  &  580     & 05 49 23.49 &        $-$40 51 13.0        & 12.94$\pm$0.02 &      d     \\
B0743$-$673~~   & 13.0     &  14      &    1.512   &  J074332$-$672628  &  672     & 07 43 32.63 &        $-$67 26 28.6        & 10.20$\pm$0.03 &      s     \\ 
B0842$-$835~~   & 11.0     &  12      &   (0.82)   &  J083716$-$834440  &  718     & 08 37 14.08 &        $-$83 44 40.5        &  8.52$\pm$0.03 &      s     \\
B0842$-$754~~   & 30.0     &  12      &    0.524   &  J084125$-$754033  &  723     & 08 41 26.19 &        $-$75 40 31.1        & 20.84$\pm$0.03 &      s     \\
B0906$-$682~~   & 12.9     &  10      & ($>$ 0.56)   &  J090652$-$682940  &  752     & 09 06 52.53 &        $-$68 29 38.5        &  7.92$\pm$0.02 &      s     \\
B1017$-$421~~   & 11.6     &  88      &   (0.70)   &  J101944$-$422451  &  836     & 10 19 43.84 &        $-$42 24 50.3        &  9.69$\pm$0.02 &      d     \\
B1030$-$340~~   & 12.7     &  10      &   (0.50)   &  J103312$-$341842  &  854     & 10 33 13.07 &        $-$34 18 44.0        & 10.14$\pm$0.02 &      s     \\
B1036$-$697~~   & 11.1     &  10      &   (0.87)   &  J103828$-$700310  &  862     & 10 38 28.27 &        $-$70 03 09.0        &  6.20$\pm$0.02 &      s     \\
B1143$-$483~~   & 13.5     &  10      &   (0.33)   &  J114530$-$483606  &  950     & 11 45 30.94 &        $-$48 36 10.4        & 12.14$\pm$0.03 &      s     \\
B1413$-$364~~   & 11.3     & 174      &    0.07470 &  J141633$-$364050  & 1135     & 14 16 33.65 &        $-$36 40 47.0        &  9.69$\pm$0.06 &      d     \\ 
B1445$-$468~~   & 16.0     &  30      & ($>$ 0.56)   &  J144828$-$470136  & 1192     & 14 48 29.05 &        $-$47 01 39.9        & 10.00$\pm$0.04 &      s     \\
B1451$-$364~~   & 20.0     & 123      &   (0.43)   &  J145428$-$364006  & 1203     & 14 54 28.53 &        $-$36 39 57.3        & 16.52$\pm$0.04 &      t     \\
B1737$-$609~~   & 16.0     &  78      &   (0.41)   &  J174202$-$605519  & 1432     & 17 42 01.59 &        $-$60 55 22.4        & 13.74$\pm$0.04 &      t     \\
B2148$-$555~~   & 11.6     & 780      &    0.03880 &  J215122$-$552139  & 1732     & 21 51 28.11 &        $-$55 19 45.9        &  6.29$\pm$0.06 &      t     \\
\hline                                                                                                                                                     
B1247$-$401~~   & 12.6     &  10      &   (1.20)   & J125005.7$-$402629 &  ...     & 12 50 05.71 &        $-$40 26 29.8        & 13.17$\pm$1.32 &     ...    \\
B1346$-$391~~   & 13.8     &  19      &   (1.06)   & J134951.0$-$392251 &  ...     & 13 49 51.09 &        $-$39 22 51.1        & 12.53$\pm$1.25 &     ...    \\
B1358$-$493$^a$ & 13.0     &  51      & ($>$ 0.56)   & J140131.5$-$493235 &  ...     & 14 01 31.45 &        $-$49 32 34.8        & 14.31$\pm$1.02 &     ...    \\
\hline
\end{tabular}
\end{center}
\tablecomments{Column (1): the name in SMS4, according to the MRC or PKS (for B2148$-$555 only) designation. Column (2): the extrapolated flux density $S_{178}$. Column (3): the largest angular size of the radio source at 843~MHz. Column (4): the redshift, with lower limits and photometric estimates in parentheses. Column (5): the International Astronomical Union (IAU) name in G4Jy, according to the GLEAM designation, or in TGSS-ADR1. Column (6): the G4Jy identifier. Columns (7) and (8): R.A. and decl. of the G4Jy/TGSS-ADR1 source. Column (9): the actual flux density ($\overline{S}_{181}$ for G4Jy, or $S_{150}$ for TGSS-ADR1). Column (10): the radio source morphology, following W20 (s=single; d=double; t=triple).\\
$^a$ - For this source two TGSS components, matching the lobes of the radio galaxy in the ATCA map published in BH06, are found. The coordinates reported in this table, matching the core, come from SUMSS.}
\end{table*}

\section{Crossmatch between the SMS4 and G4Jy Samples} 
\label{sec:2}

Despite overall similarities with the 3CRR reported by \cite{2006AJ....131..100B,2006AJ....131..114B}, the SMS4 is based on the extrapolation to 178~MHz of flux density values measured at 408~MHz and similarly higher frequencies, rather than on values directly measured at 178~MHz.
Therefore, a fraction of sources with flux densities lower than $S_{178}^{\star}$ might be included in the SMS4; conversely, sources with $S_{178} > S_{178}^{\star}$ might have been excluded.

We crossmatched the SMS4 with the G4Jy sample to establish a correspondence between sources therein, with the primary goal of determining how their actual flux density values rank with respect to the $S_{178}^{\star}$ threshold established in the 3CRR.
Considering the extended nature of the radio emission, although with different angular sizes for each SMS4 source, we adopted a conservative approach in searching for G4Jy sources within a circle centered on the SMS4 coordinates and with a radius given by the corresponding LAS, as reported in \cite{2006AJ....131..100B}.

Once a match between an SMS4 source and a G4Jy source is established, we compared $S_{178}$ from SMS4 with the total, integrated flux density, measured in the 178--185 MHz range (from here on, in this paper we indicate this quantity with $\overline{S}_{181}$) from G4Jy.
For 78 G4Jy sources, multiple GLEAM components were associated by W20 with the same G4Jy source: in such cases, we compared $S_{178}$ with the sum of all the $\overline{S}_{181}$ values, each corresponding to a different GLEAM component. 
As a result, we find a reliable G4Jy counterpart for 127 SMS4 objects, and 47 of these have $\overline{S}_{181} < 10.9$~Jy.

While we were able to match 127 of the 137 SMS4 sources to G4Jy sources, 10 SMS4 sources remain unmatched.
Among these 10 unmatched sources, there are some of the very brightest sources at decl. $< 30^{\circ}$ and $\mid b \mid\,> 10^{\circ}$, which belong to a group of radio sources colloquially referred to as the {\it A-team}.
As also reported by W20, these have been masked for the GLEAM~EGC and so do not appear in the G4Jy sample.
Among these 10 SMS4 sources we find Fornax~A, Pictor~A, Centaurus~A, and a few radio galaxies in their proximity ({\it i.e.}, in the Centaurus Cluster), lying in the regions that were masked.

For each of these 10 sources, we searched in the literature for measurements at frequencies lower than 178~MHz in a) the observations performed with the Culgoora Circular Array (CCA) at 80 and 160 MHz \citep{1995AuJPh..48..143S}, in b) the survey at 145~MHz carried out with the Precision Array for Probing the Epoch of Reionization (PAPER; \citealp{2011ApJ...734L..34J}), and in c) the $1^{st}$ Alternative Data Release (ADR1) based on the survey at 150~MHz, which was carried out with the Giant Metrewave Radio Telescope (GMRT) as part of the TIFR GMRT Sky Survey (TGSS) project \citep{2017A&A...598A..78I}.
For only one source, PKS~B1318$-$434 (also known as NGC~5090), we did not find information in these three catalogs.
For all the remaining sources, we used these data to interpolate the flux densities at 178~MHz and found that $S_{178} > 10.9$~Jy for all of them, as expected.
As a result, excluding PKS~B1318$-$434, we establish that $\overline{S}_{181} < 10.9$~Jy for 47 of 136 SMS4 sources, corresponding to a fraction of $\sim$35\% of the whole SMS4 sample.

\section{Description of Our Sample of SMS4 Sources} 
\label{sec:3}

In 2015, we compiled a list of 45 SMS4 sources, classified as radio galaxies according to BH06, that were not yet observed by \sw~or \cha~or \xmm. 
From this sample, we obtained \sw~observations for 24 sources.  
We later included in our sample seven additional sources that, in 2021 October, among the just mentioned space missions, were observed in the X-rays only by \sw; all of these were classified as galaxies or quasars in BH06.

As described in Section~\ref{sec:2}, we were able to establish an SMS4-G4Jy correspondence for 28 out of 31 SMS4 sources.
Using data from the Sydney University Molonglo Sky Survey (SUMSS; \citealp{2003MNRAS.342.1117M}) and the NRAO VLA Sky Survey (NVSS; \citealp{1998AJ....115.1693C}), at higher frequency than the GLEAM survey, a morphological classification was provided by W20 for all the G4Jy sources, distinguishing single (s), double (d), triple (t), and complex (c) morphology.
The main difference between the double morphology and the triple morphology is the capability of detecting the core of the radio galaxy, in addition to the lobes; see W20 for further details on their morphology classification criteria.  
Of the 28 SMS4 sources in our sample with a G4Jy counterpart, the radio morphology is triple for three sources and is double for four sources, with all the remaining sources having a single (compact, substantially symmetric) morphology.

We notice the presence, among these 28 sources, of PKS~B2148$-$555, which is one of the 78~G4Jy sources with which, due to their extension, more than one GLEAM component was associated by~W20. 
High-resolution ATCA maps in \cite{2002MNRAS.331..717L} show the details of the emission of this relevant radio galaxy.

For the reasons described in Section~\ref{sec:2}, it was not possible to establish an SMS4-G4Jy correspondence for three (MRC~B1247$-$401, MRC~B1346$-$391, MRC~B1358$-$493) of these 31 sources.
However, for all of these, the role of G4Jy could be replaced by the TGSS, since we found matching sources in the ADR1.
The correspondence that we finally establish is one-to-one for MRC~B1247$-$401 and MRC~B1346$-$391, while for MRC~B1358$-$493 two nearby TGSS sources are found.
Comparing the position of these two sources with the high-resolution ATCA map at 5~GHz shown in \cite{2006AJ....131..114B}, one finds that they match the lobes of the radio galaxy, implying a total flux density exceeding 14~Jy.
The coordinates of the single SUMSS component, lying in the middle of the two TGSS sources, match instead the core of the radio galaxy.

As a result, we finally establish a reliable correspondence for all 31 SMS4 sources, originally detected at 408~MHz, with objects in other radio catalogs at lower frequencies, in the range 150-200~MHz.
All these correspondences are listed in Table~\ref{tab:01}, with those for the three SMS4 sources with no G4JY counterpart shown at the bottom.
Contrary to expectations from the original extrapolation, we find $\overline{S}_{181} < 10.9$~Jy for 14 of the 31 SMS4 objects.

We used SUMSS maps, available for all the 31 sources, to build radio flux density contours at 843~MHz.
To allow the comparison between the morphology of the radio source and the underlying X-ray emission, we overlay these contours on the X-ray maps (see Figure~\ref{fig:01}) that we built from \sw-XRT observations, as described in Section~\ref{sec:3}.
For G4Jy sources with double or triple morphology, we highlight the distinct SUMSS or NVSS components.
For MRC~B1358$-$493, one of the three SMS4 sources with no G4Jy counterpart, we mark the two TGSS components, matching the lobes of the radio galaxy, rather than the single SUMSS component, matching the core.  
For each source with multiple components, we assign a letter to each component, sorting them by decreasing flux density (A corresponding to the brighter component).

We describe here, as an example, the case of MRC B1737$-$609, with SUMSS contours stretched along one main direction and three resolved SUMSS components marked with A ($S_{843~MHz} \sim 2.34$~Jy), B ($S_{843~MHz} \sim 1.94$~Jy), and C ($S_{843~MHz} \sim 0.64$~Jy), in agreement with the ``triple" morphology classification by W20.
This structure is consistent with an FRII radio galaxy, with radio jets emanating from the core (C) at some intermediate angle with respect to the observer's line of sight and culminating in two lobes corresponding to A and B.
The spatial distribution of these components is not symmetric, with A and B lying at an angular distance from C of 22\arcsec~and 55\arcsec, respectively.
This asymmetry is plausibly responsible for the offset of the G4Jy centroid toward B (see Fig.~\ref{fig:02}t).

For all 31 SMS4 sources making up our sample, an optical counterpart was given in BH06 using either the plates of the UK Schmidt Southern Sky Survey or {\it R}-band CCD images made with the AAT.
However, as we discuss in Section~\ref{sec:5.3}, we give an alternative optical source for five of these BH06 candidates.
The redshift of six sources was spectroscopically measured; for the remaining 25 sources, photometric redshifts were estimated from the corresponding optical counterparts.
All these redshift values are also reported in Table~\ref{tab:01}.

As a general note, considering for the G4Jy a density of one source every 13 deg$^2$ (\citealp{2020PASA...37...18W}) and the LAS values of SMS4 sources in our sample, we estimate the probability of by-chance alignment between G4Jy and SMS4 sources to be negligible.     
In fact, LAS values are typically lower than 50\arcsec, with few exceptions (see Table~\ref{tab:01}).
Excluding the already-described case of PKS~2148$-$555, the highest LAS value (174\arcsec) in our sample is given by MRC~B1413$-$364: assuming Poisson statistics, even in this less favorable case, the probability of by-chance alignment is $\sim2\times10^{-4}$.

TGSS-ADR1 includes 0.62~million sources, down to $\sim$11~mJy.
However, for flux density values of the order of 1~Jy, the density of sources is $\sim$1 source deg$^{-2}$ (see \citealp{2017A&A...598A..78I}, their Figure 9). 
For MRC~B1358$-$493 (LAS = 51\arcsec), this implies a probability of by-chance alignment equal to $\sim6\times10^{-4}$.


\begin{figure*}
\gridline{
\fig{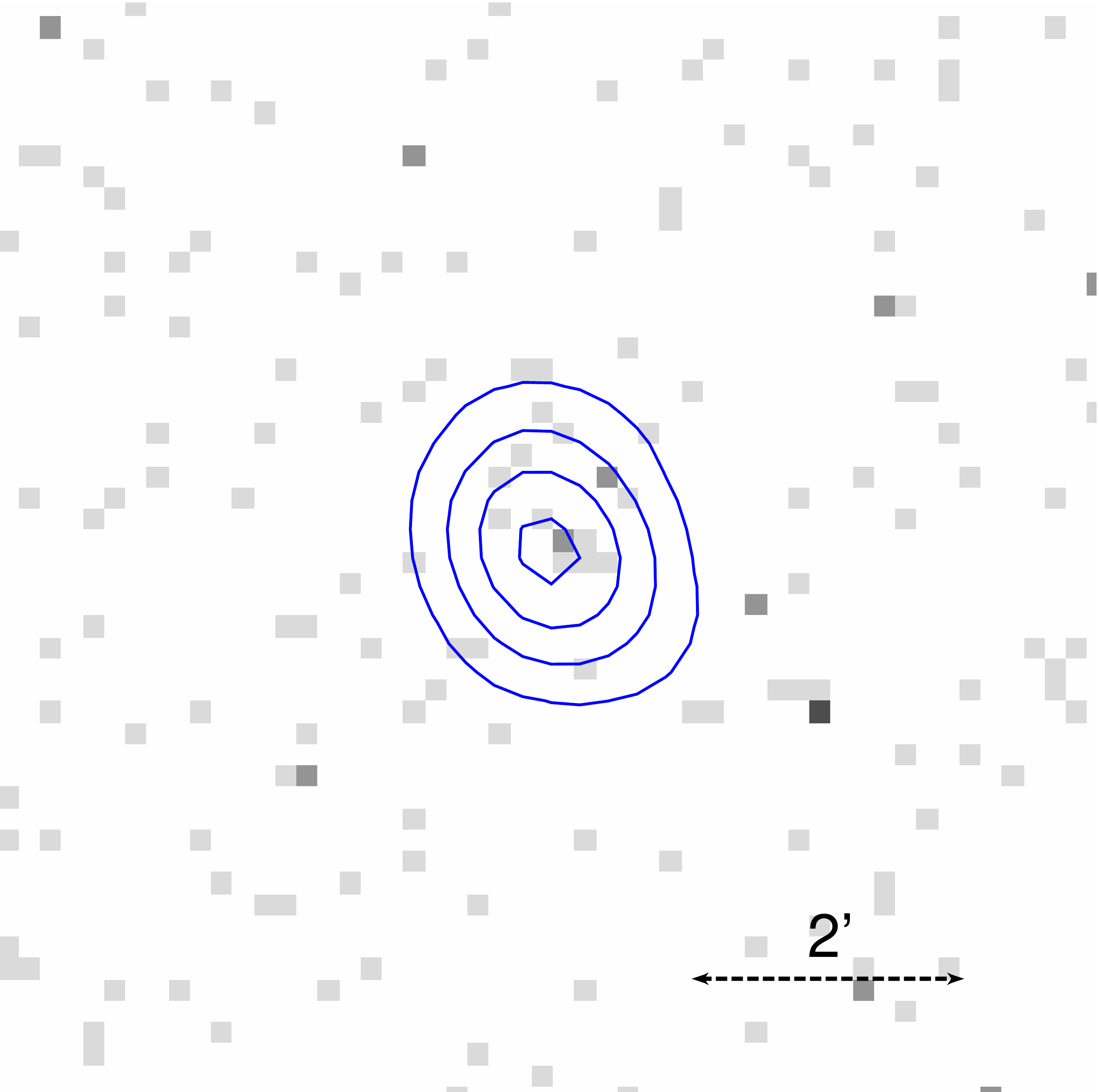}{0.30\textwidth}{(MRC~B0007$-$446)}
\fig{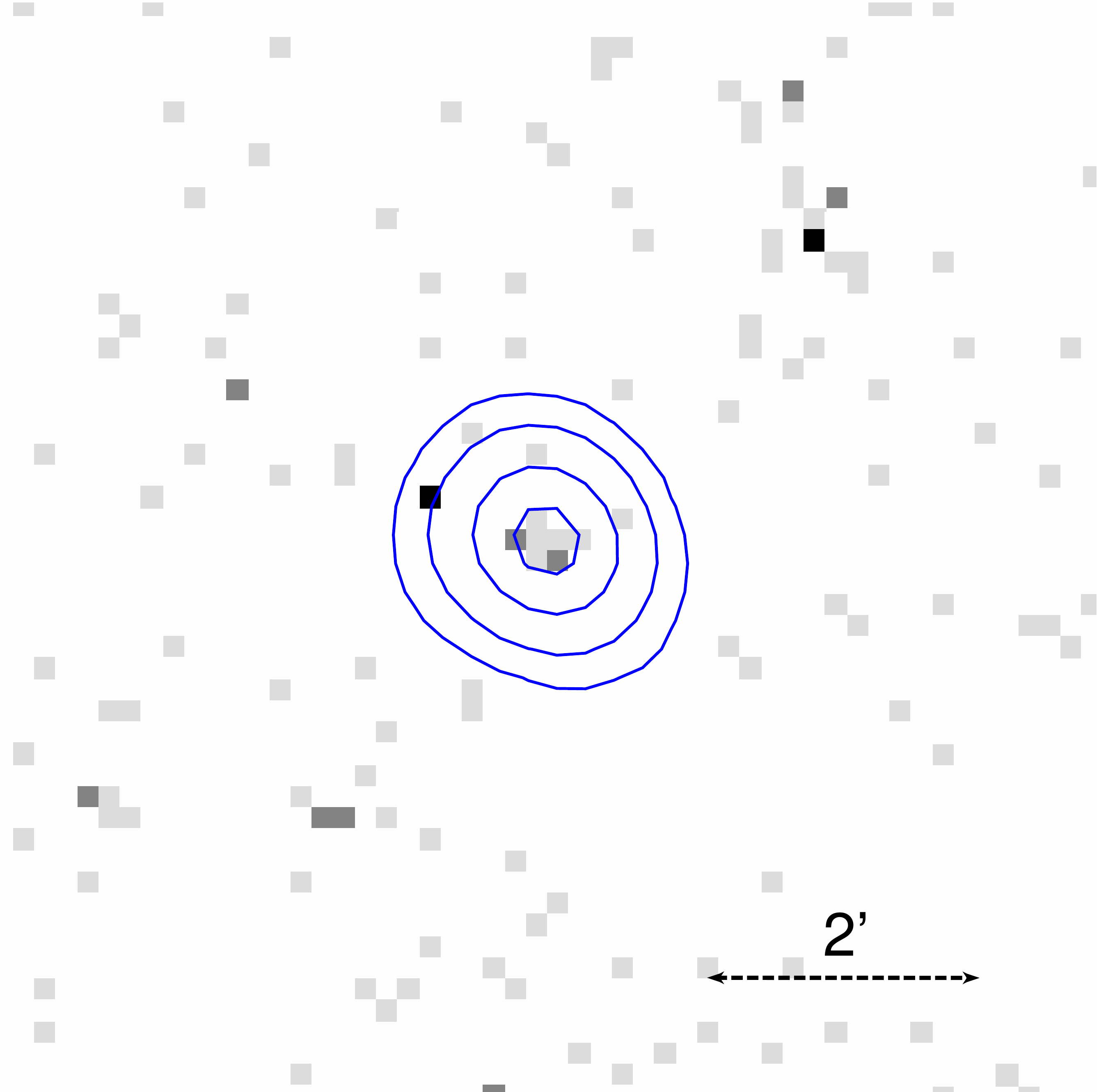}{0.30\textwidth}{(MRC~B0013$-$634)}
\fig{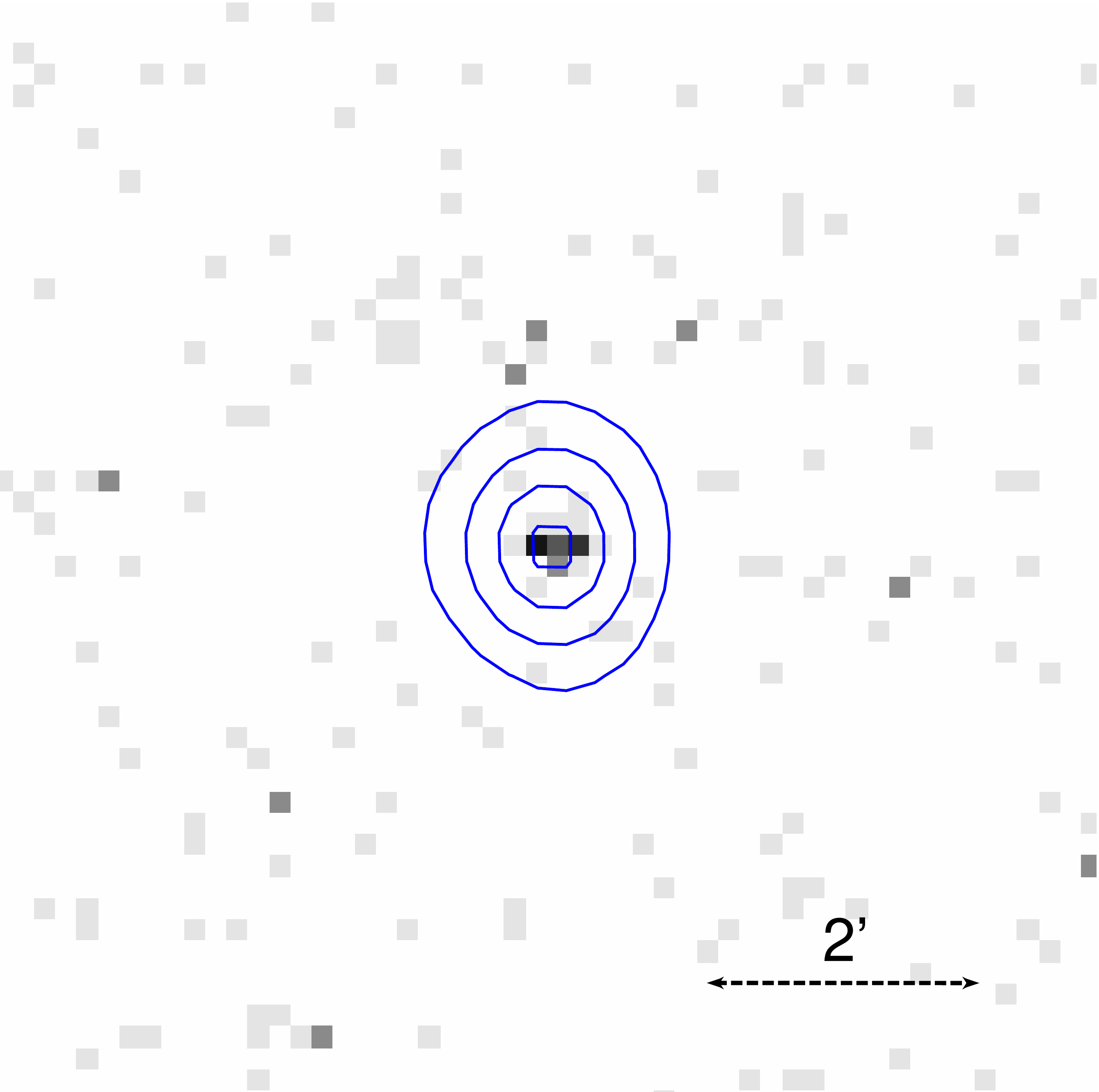}{0.30\textwidth}{(MRC~B0049$-$433)}
}
\gridline{
\fig{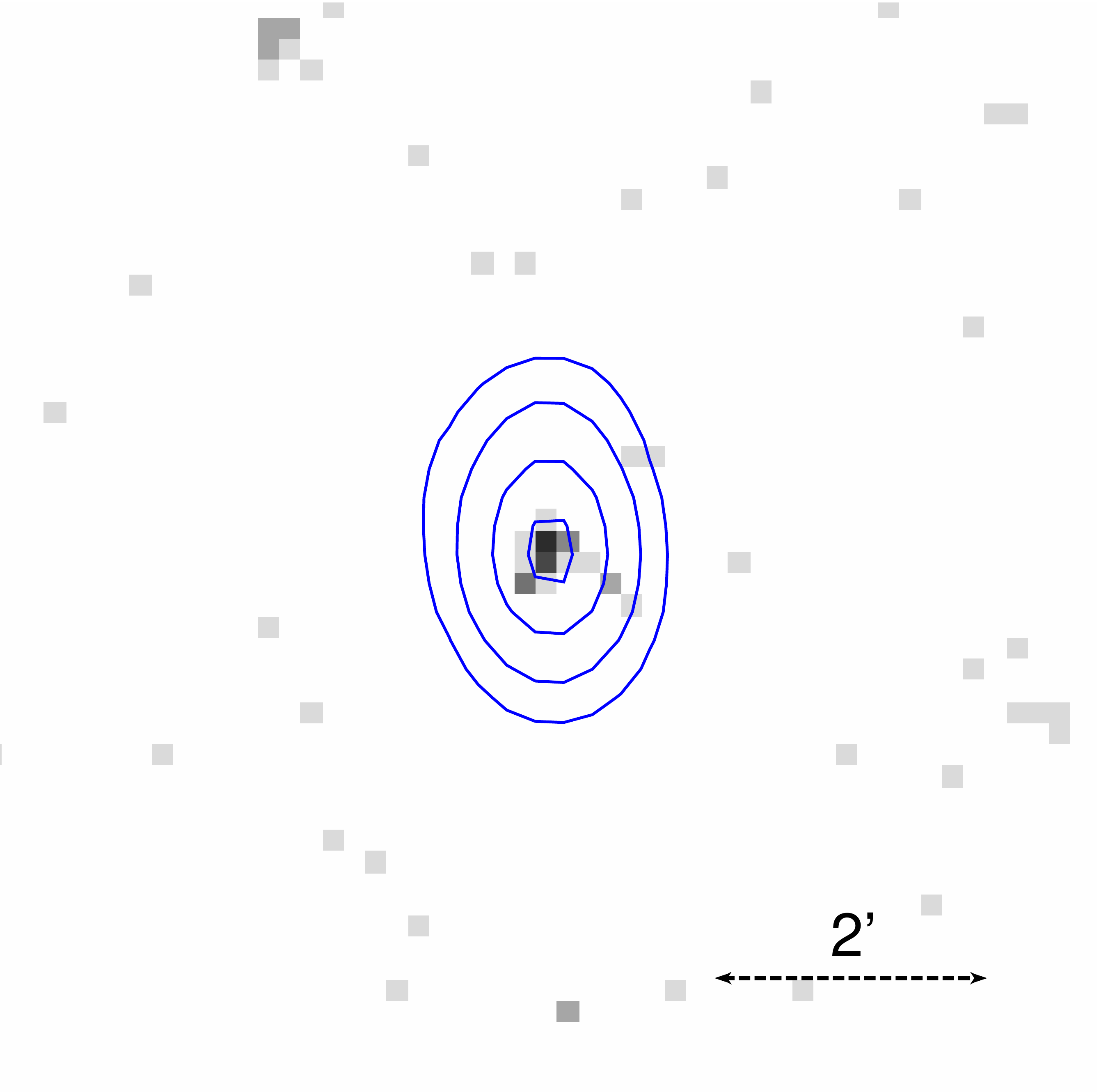}{0.30\textwidth}{(MRC~B0157$-$311)}
\fig{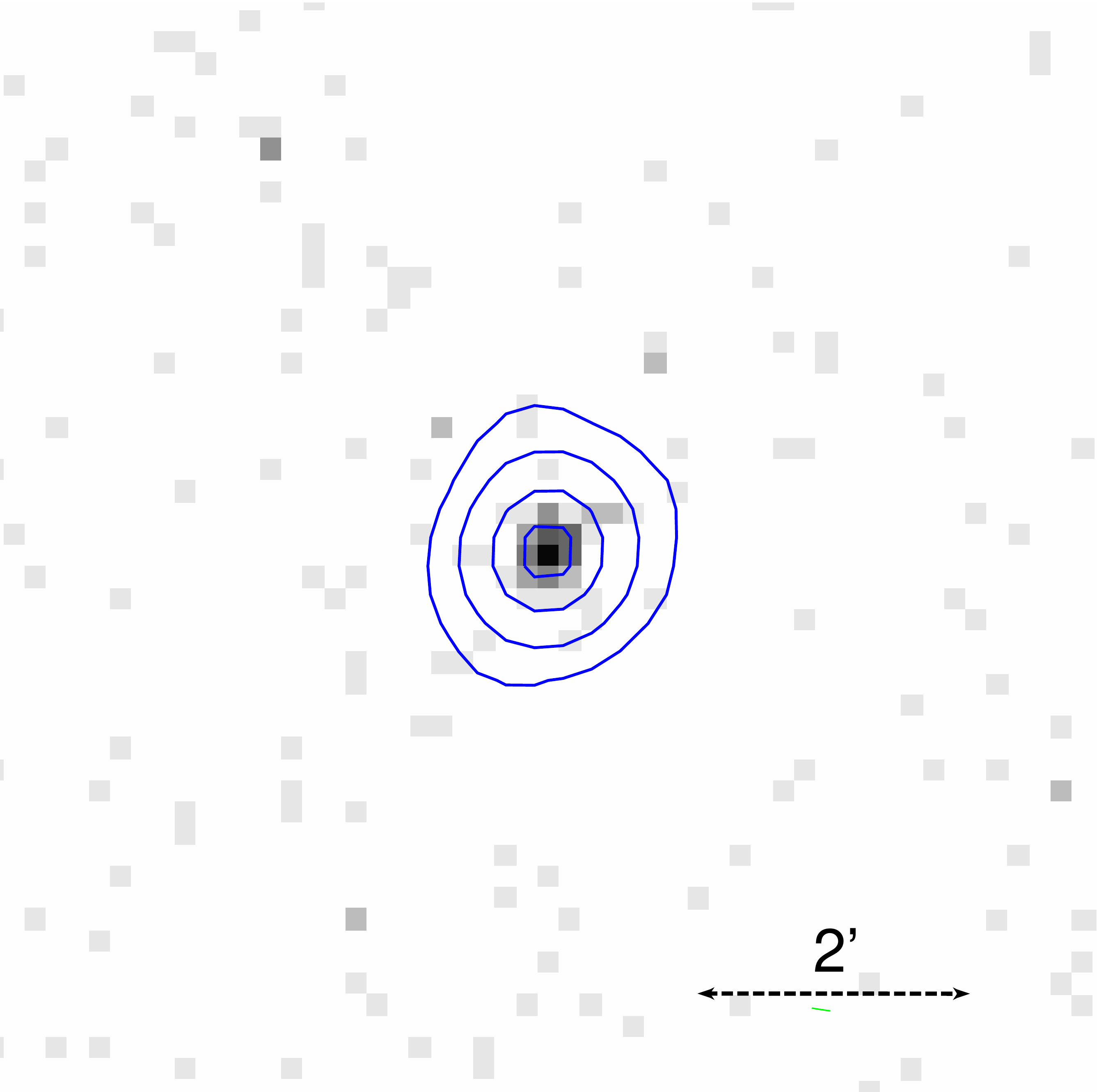}{0.30\textwidth}{(MRC~B0219$-$706)}
\fig{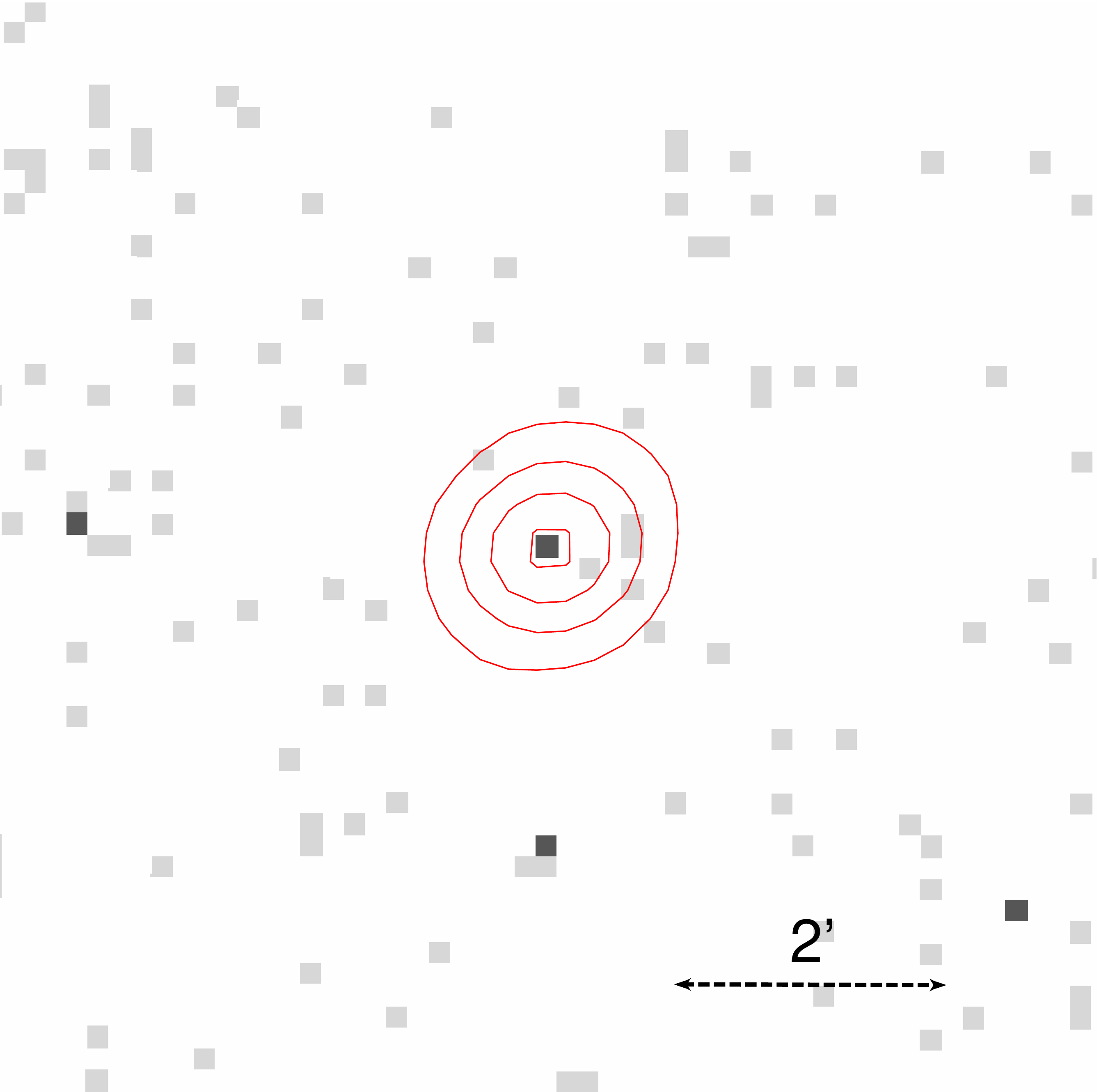}{0.30\textwidth}{(MRC~B0223$-$712)}
}
\gridline{
\fig{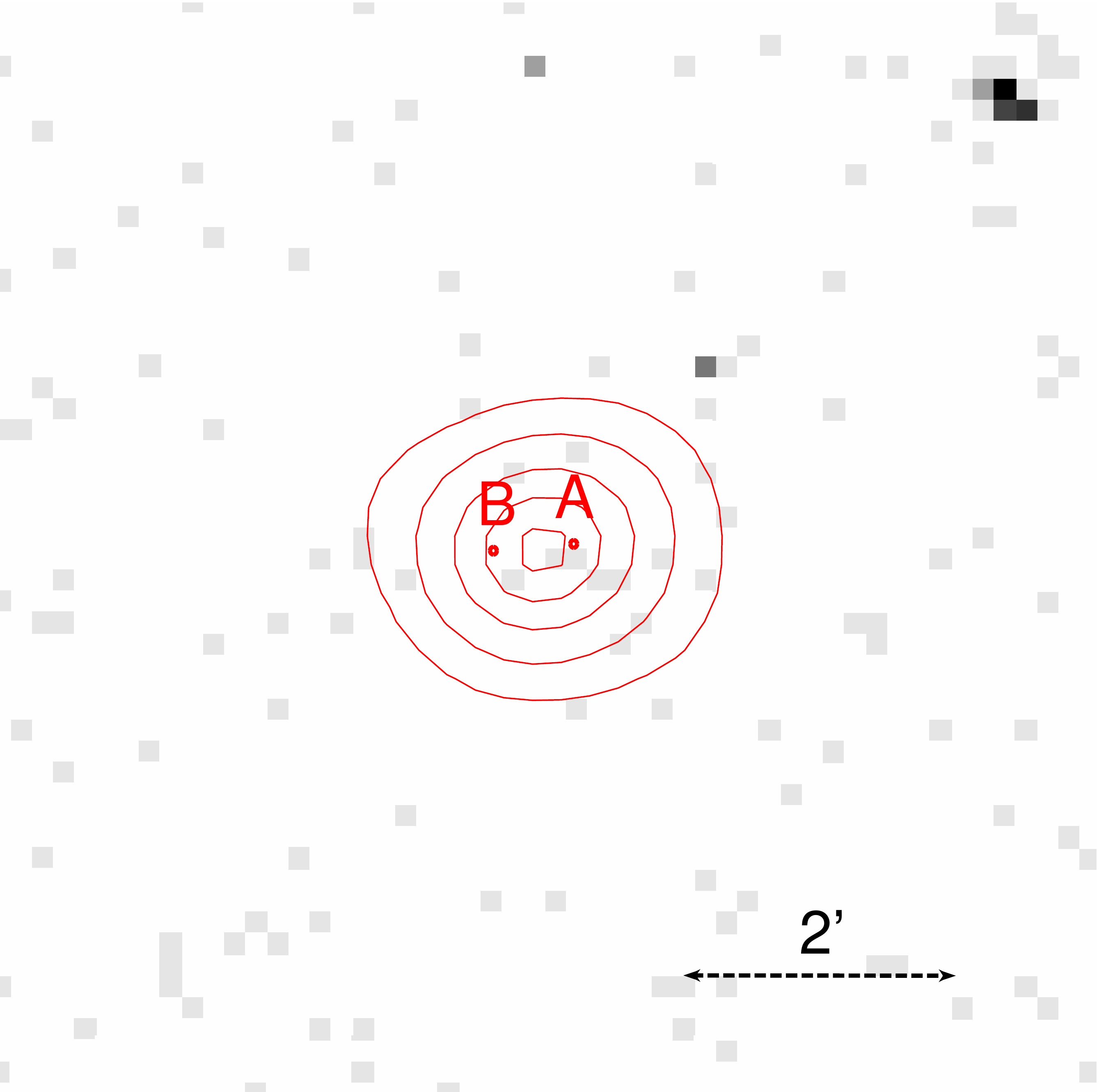}{0.30\textwidth}{(MRC~B0245$-$558)}
\fig{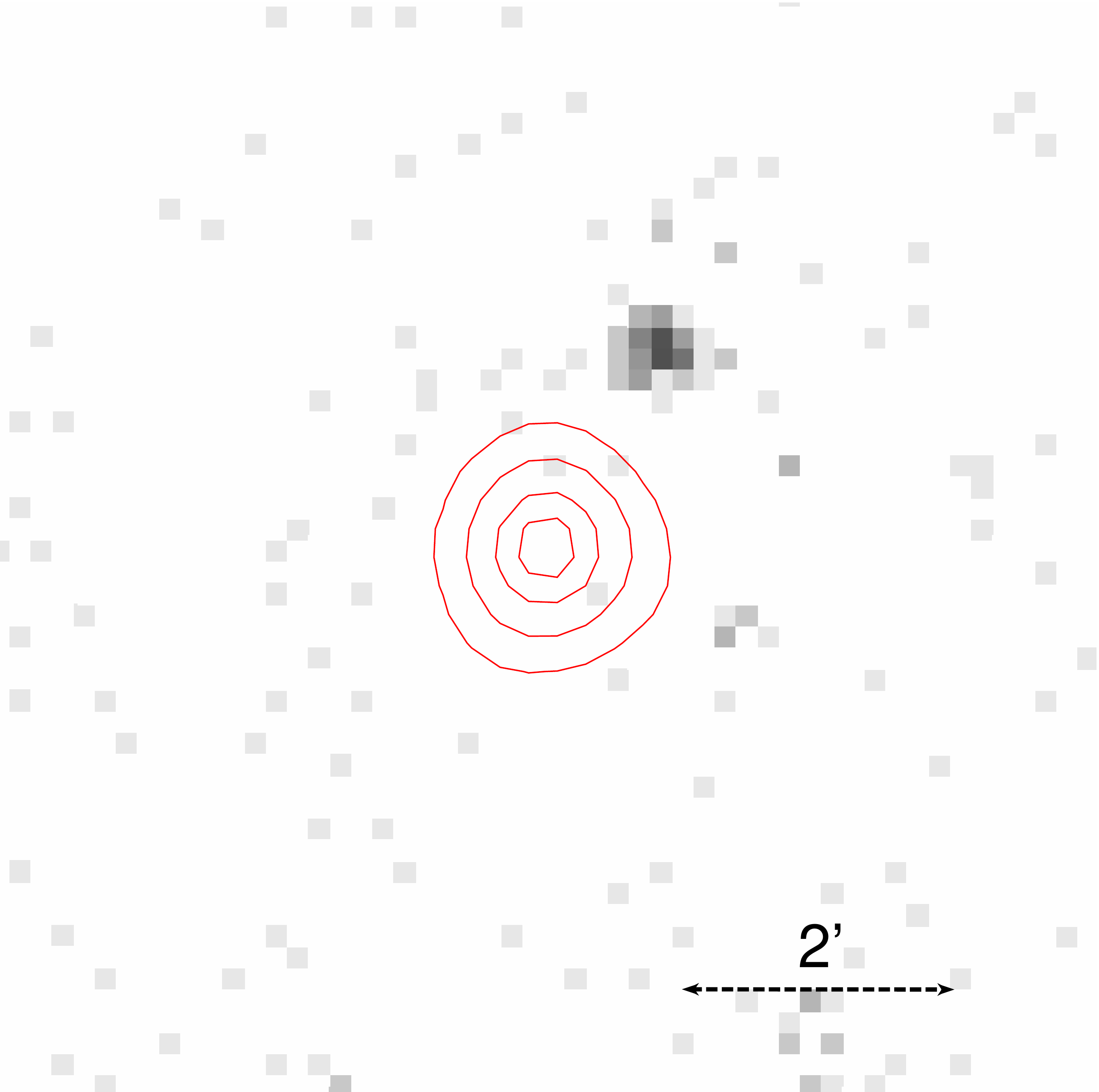}{0.30\textwidth}{(MRC~B0315$-$685)}
\fig{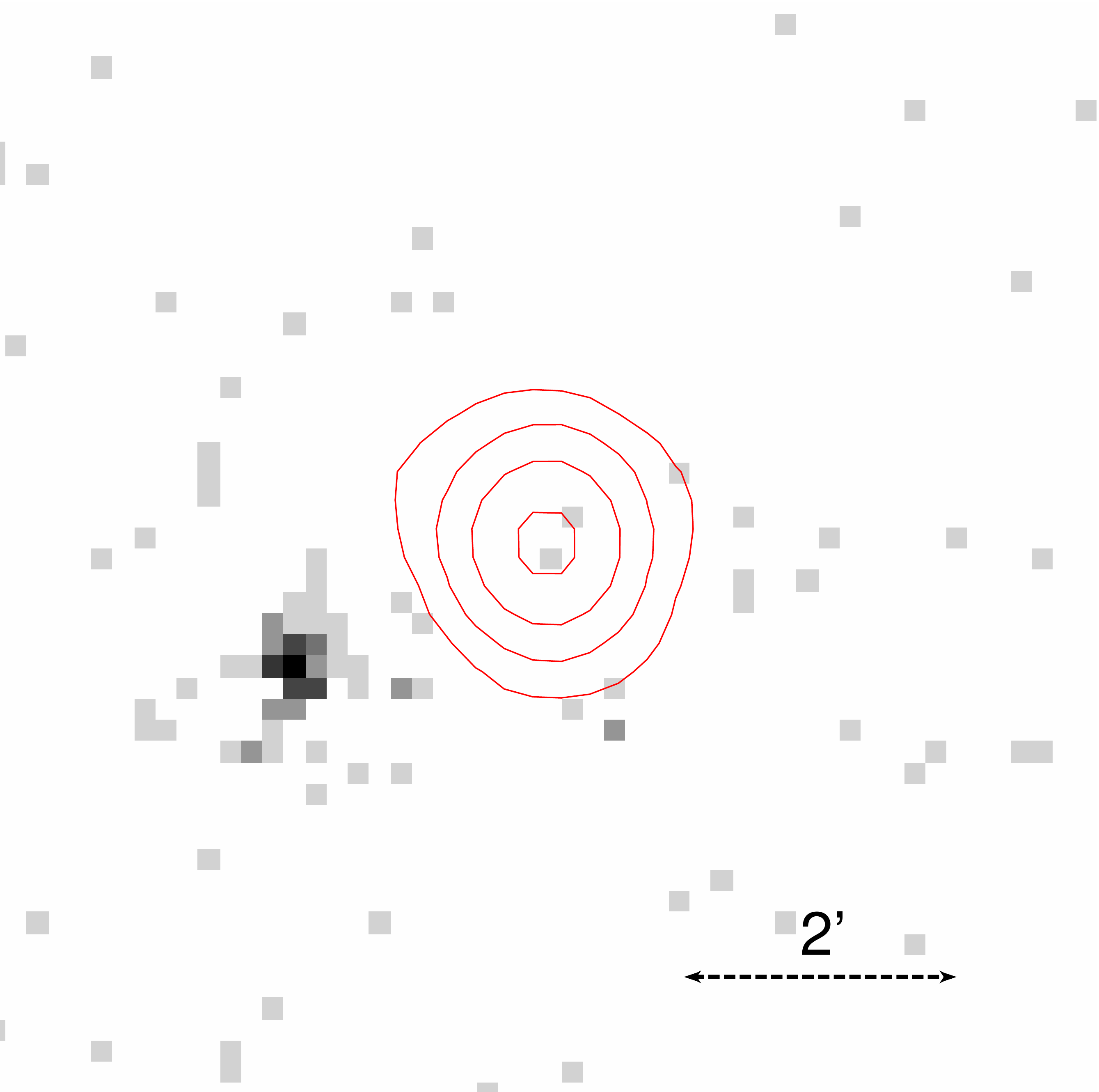}{0.30\textwidth}{(MRC~B0407$-$658)}
}
\caption{\sw~X-ray maps in the 0.3--10.0 keV band of the 31 sources included in our sample. All maps are centered at the positions of the SMS4 radio sources. Radio flux density contours at 843~MHz from SUMSS overlay the X-ray maps, and they are shown in blue for the 20 sources detected in the X-rays by \sw and in red for the 10 sources that were not detected. Contours have been selected, for each source, to best display the shape of the radio emission. For sources with double or triple morphology, following the classification in W20, distinct SUMSS components, sorted by decreasing flux density, are shown in blue; for MRC~B1358$-$493, distinct TGSS sources are shown in orange; for MRC~B1451$-$364, the core, not resolved by SUMSS but by NVSS, is shown in magenta. Although PKS~2148$-$555 was not detected as an X-ray point source, blue contours are used also in this case, since we find significant X-ray emission at the source location on arcminute scales. As discussed in the text, PKS~2148$-$555 lies within the poor cluster A3816 and is associated with the cluster BCG \citep{2002MNRAS.331..717L}. The map of PKS 2148$-$555 (20\arcmin $\times$ 20\arcmin) is binned by 16$\times$16 pixels (1 pixel = 2\farcs36). The remaining maps are binned by 4$\times$4 pixels and are 8\arcmin $\times$ 8\arcmin~except for MRC~B1737$-$609, shown at smaller scale (4\arcmin $\times$ 4\arcmin).} 
\label{fig:01}
\end{figure*}

\setcounter{figure}{0}
\begin{figure*}
\gridline{
\fig{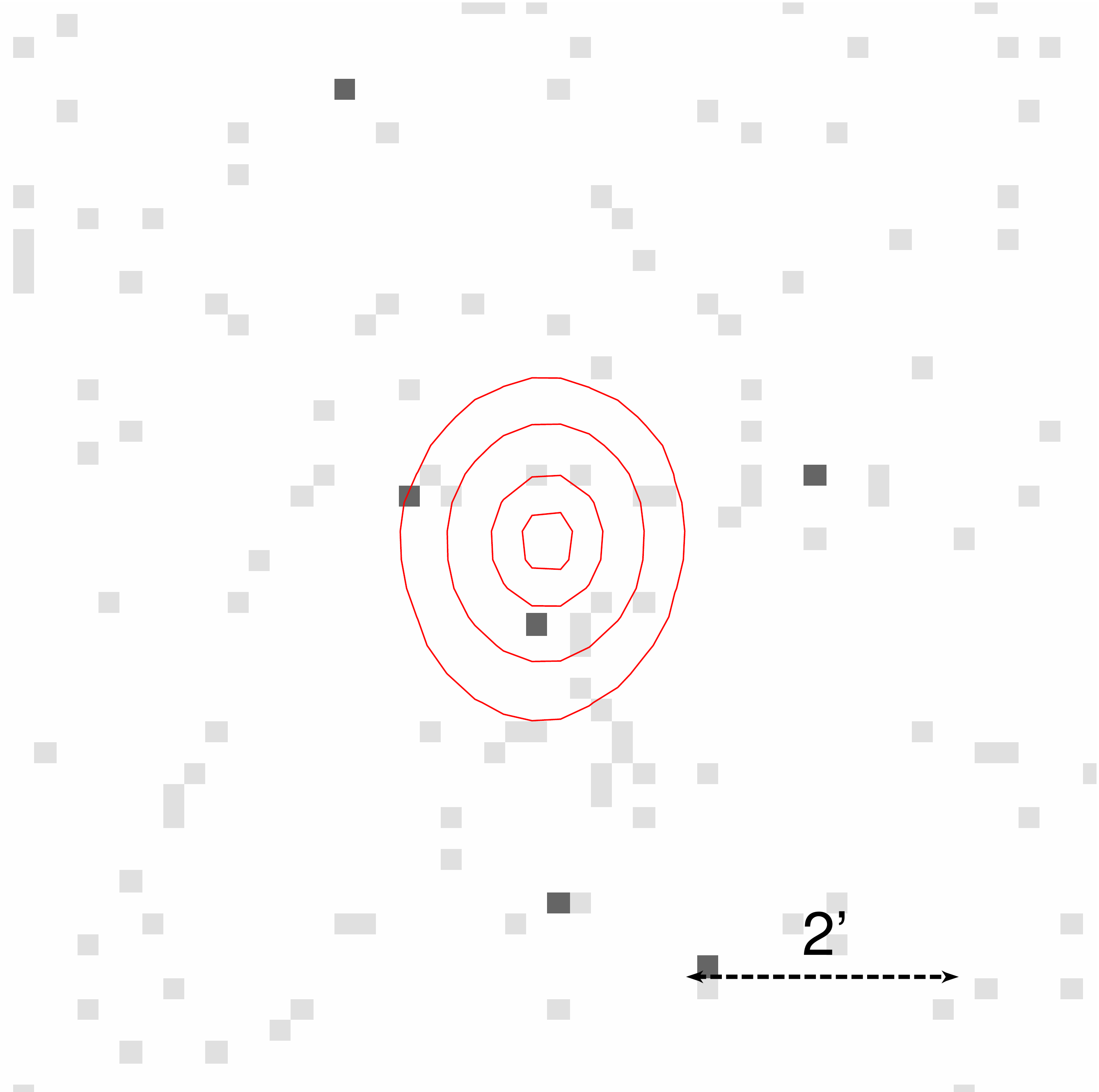}{0.30\textwidth}{(MRC~B0411$-$561)}
\fig{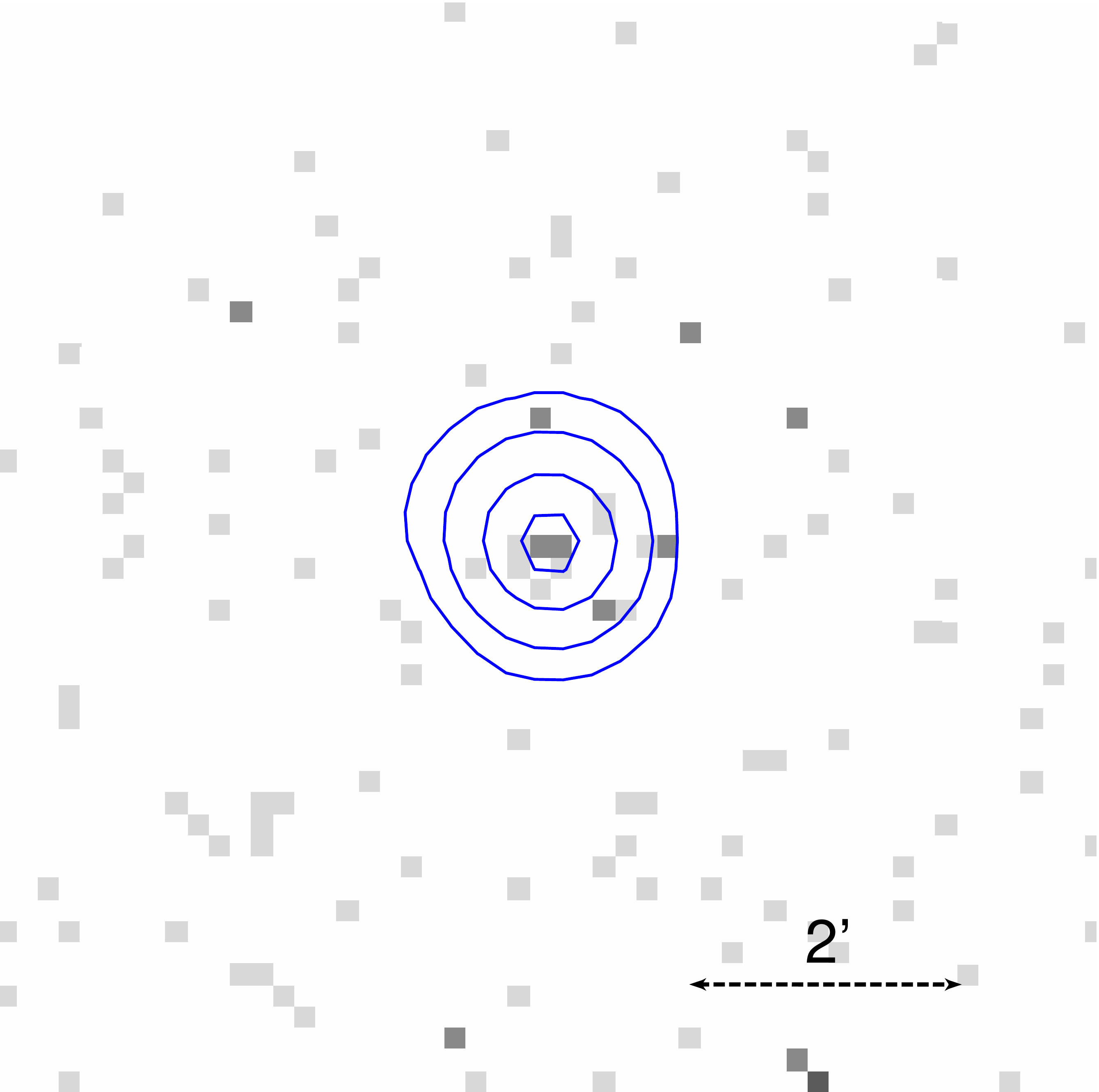}{0.30\textwidth}{(MRC~B0420$-$625)}
\fig{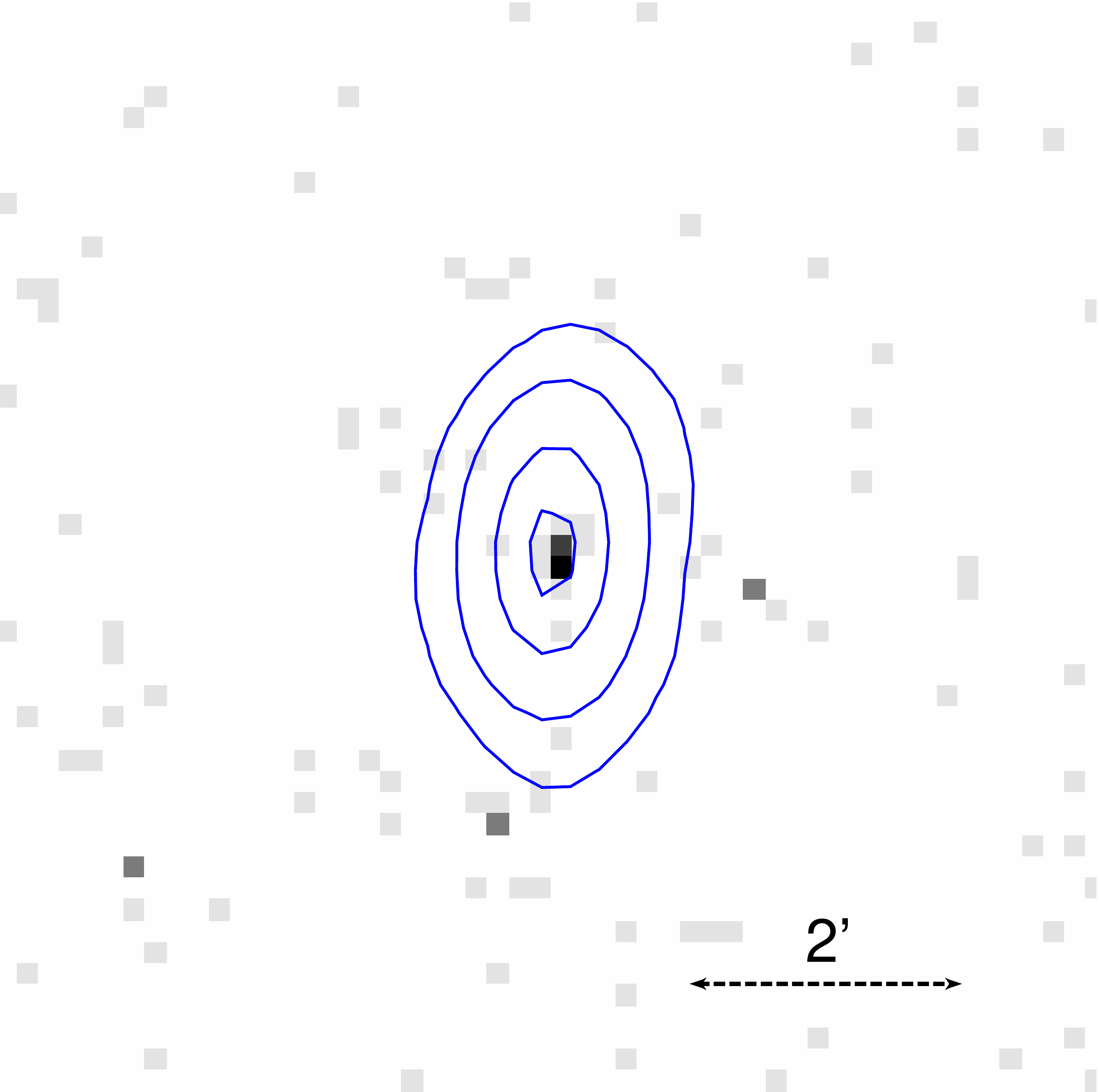}{0.30\textwidth}{(MRC~B0453$-$301)}
}
\gridline{
\fig{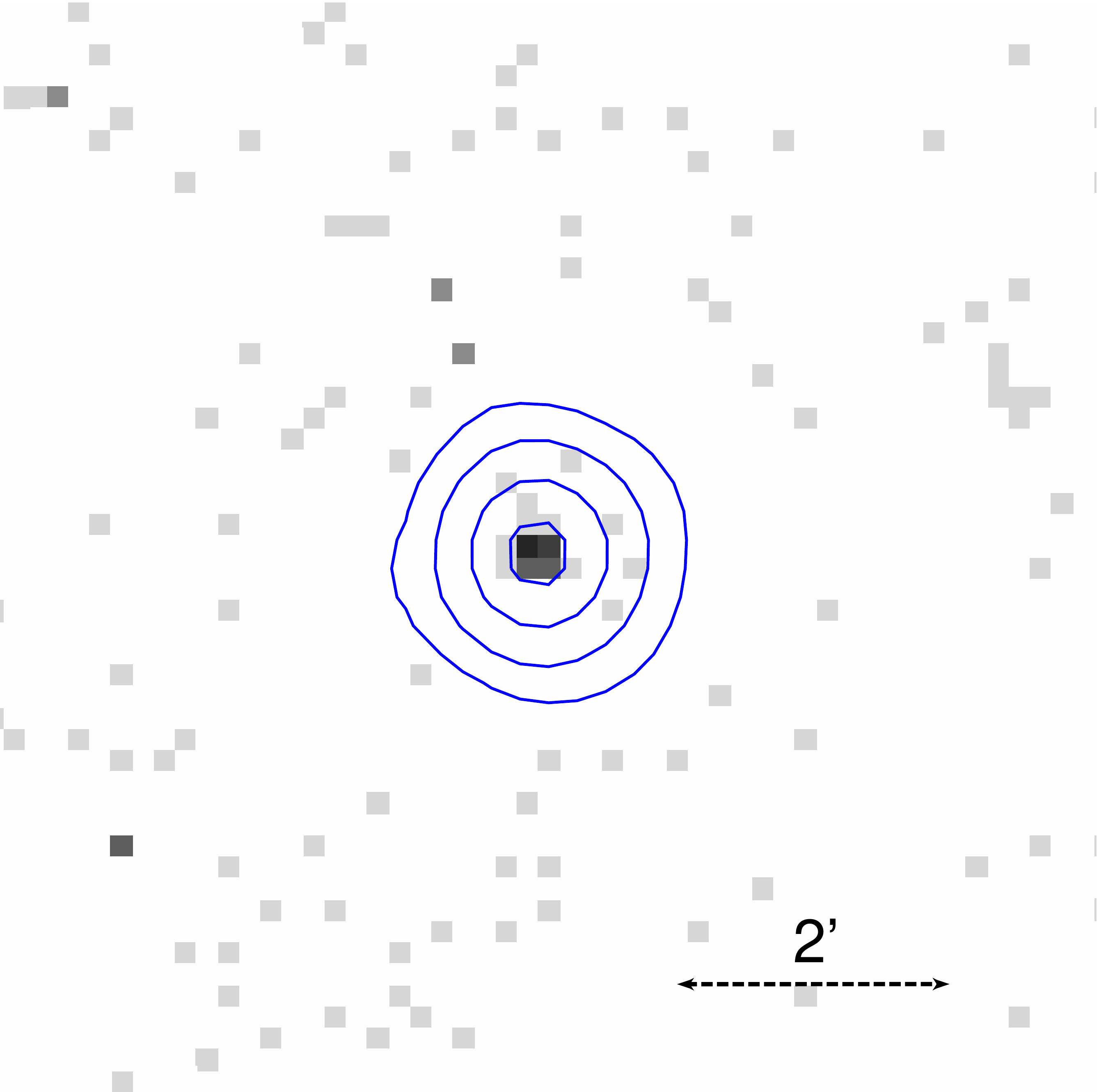}{0.30\textwidth}{(MRC~B0534$-$497)}
\fig{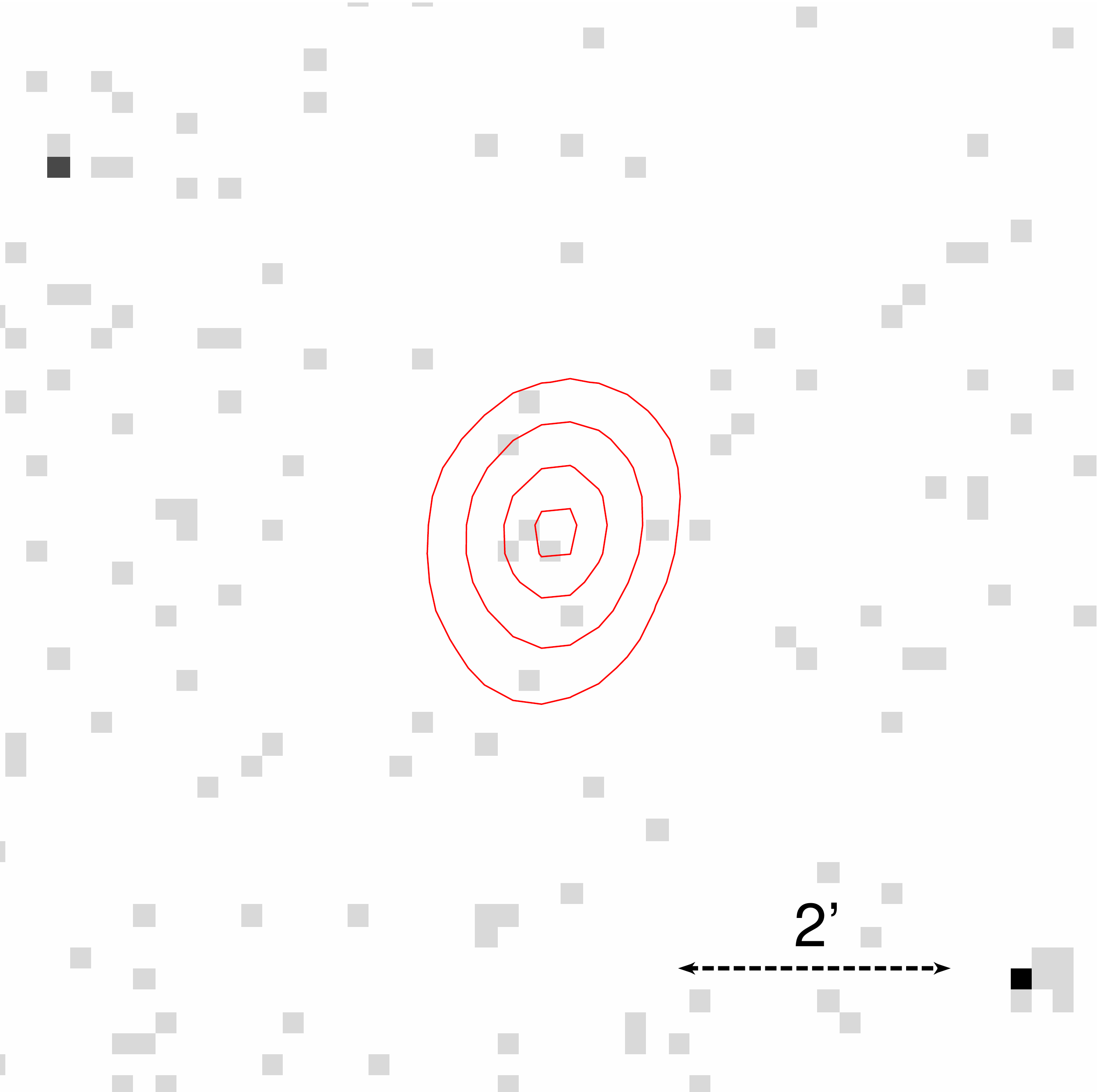}{0.30\textwidth}{(MRC~B0546$-$445)}
\fig{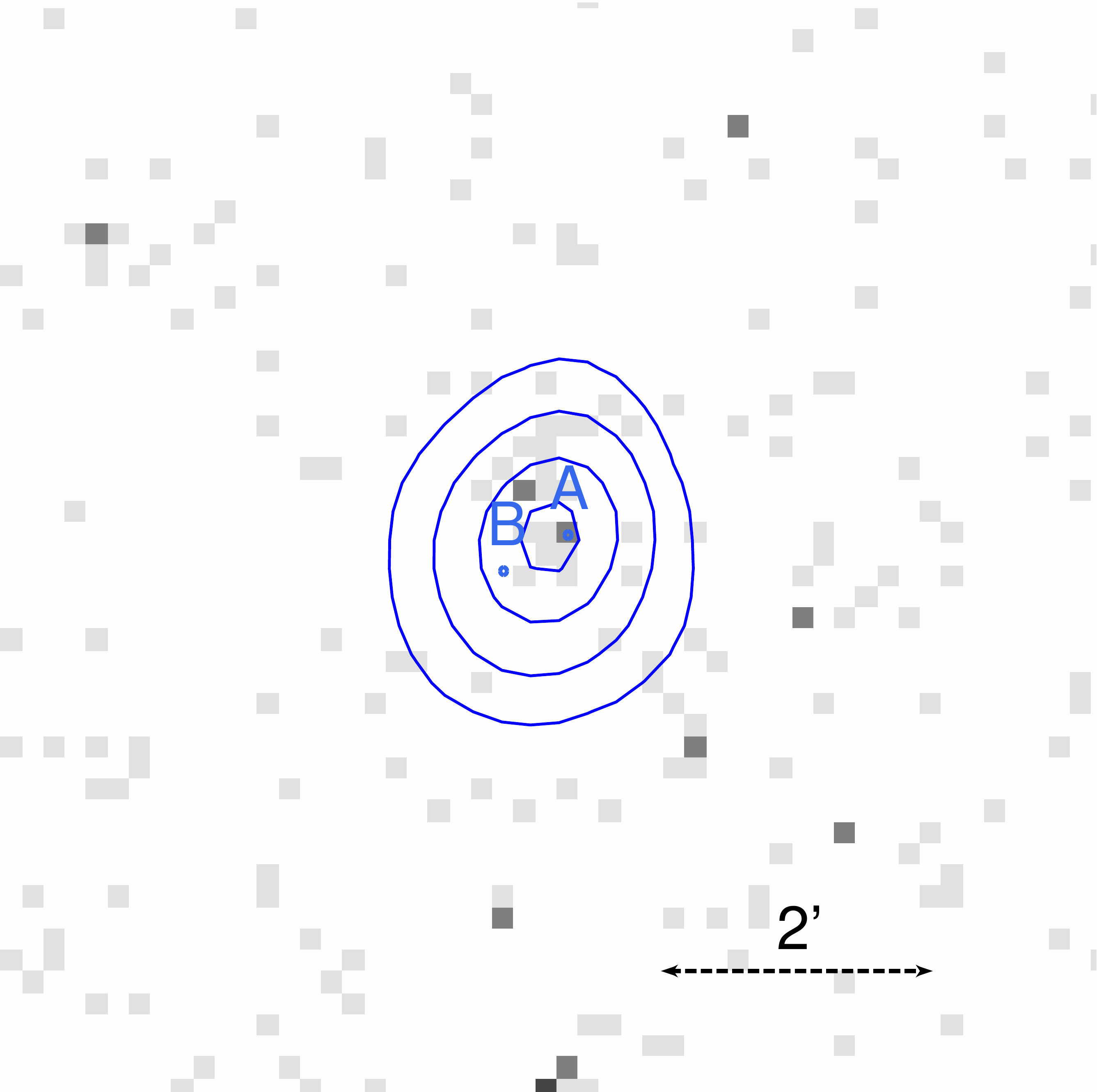}{0.30\textwidth}{(MRC~B0547$-$408)}
}
\gridline{
\fig{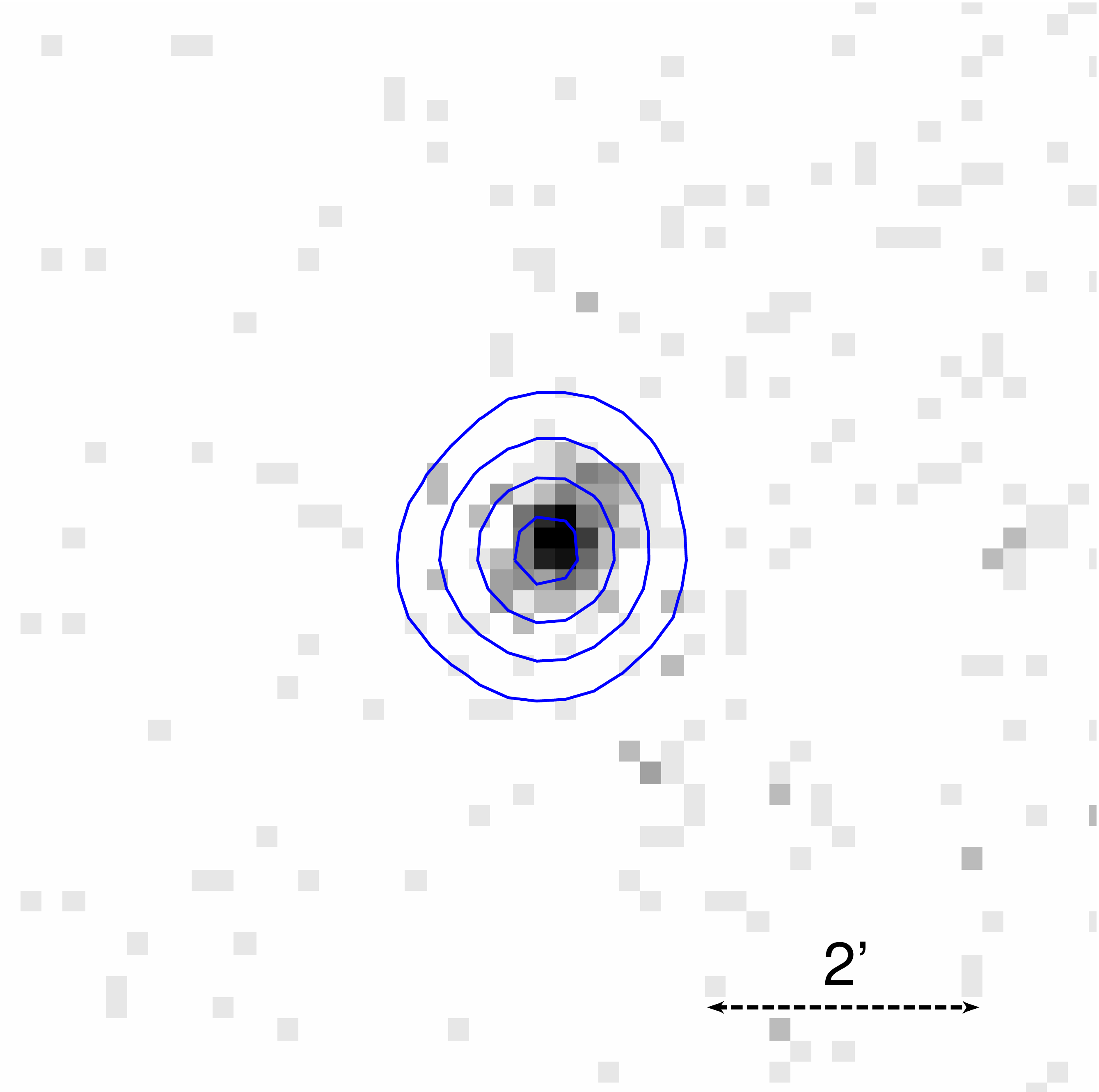}{0.30\textwidth}{(MRC~B0743$-$673)}
\fig{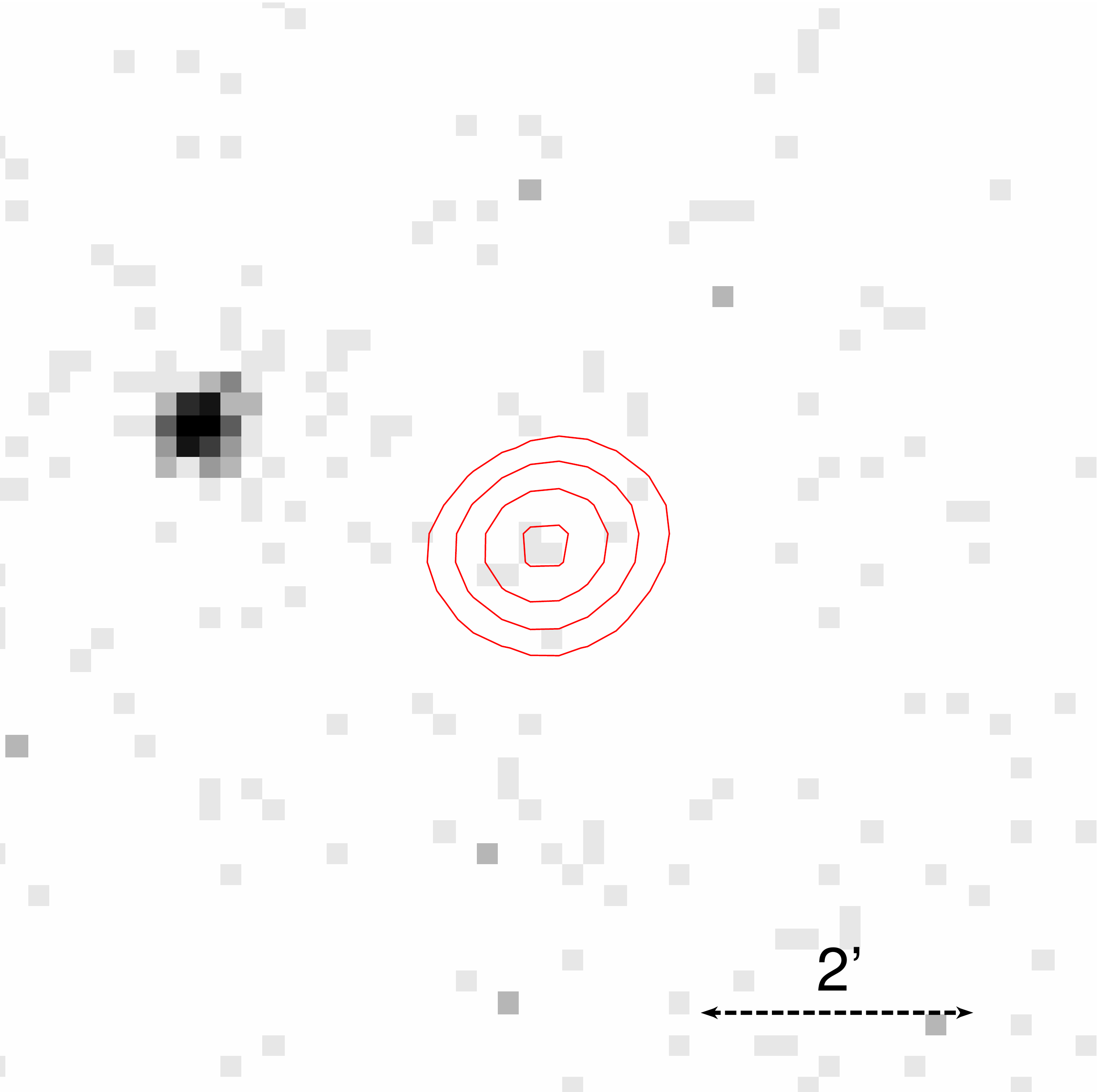}{0.30\textwidth}{(MRC~B0842$-$835)}
\fig{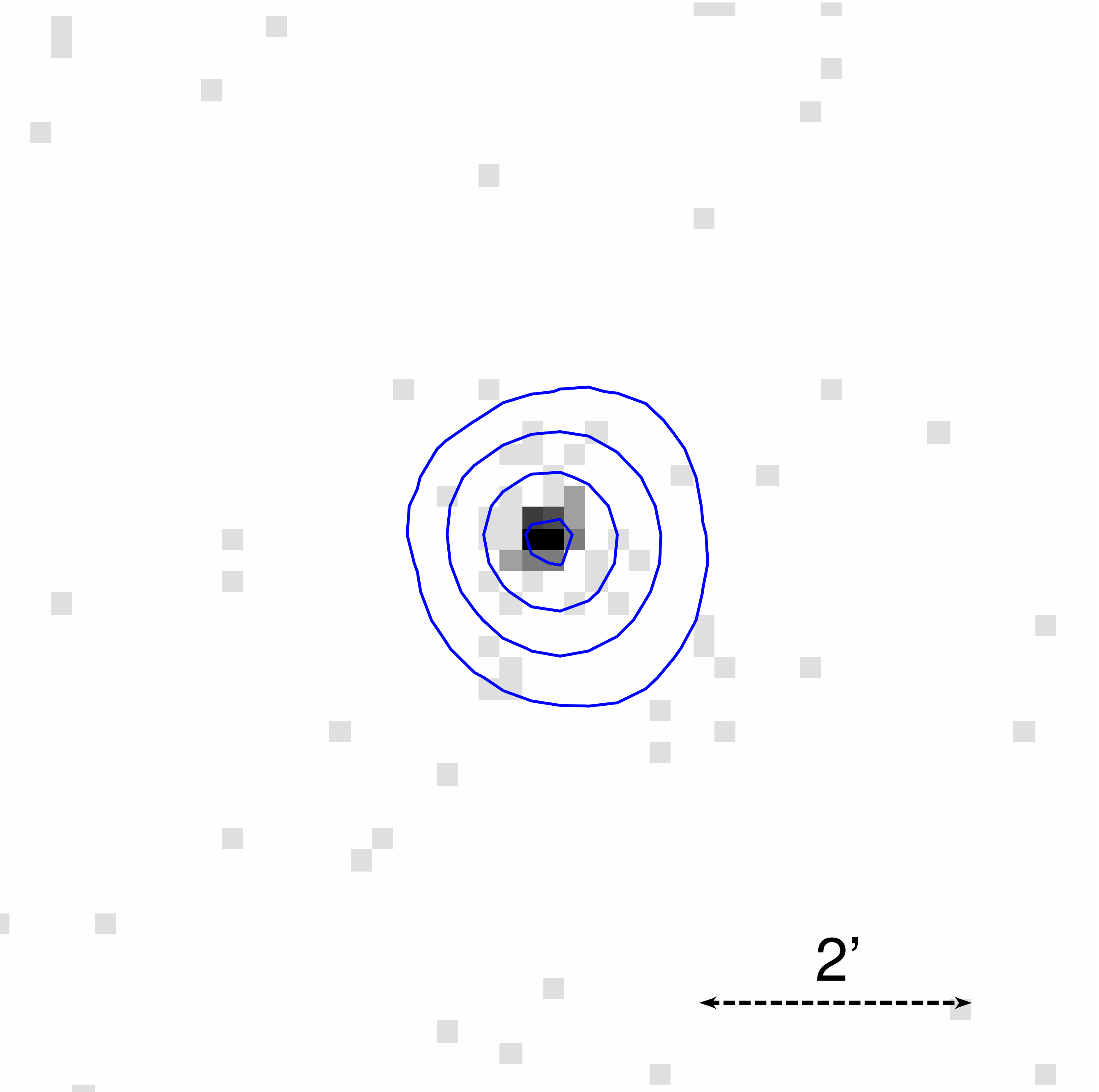}{0.30\textwidth}{(MRC~B0842$-$754)}
}
\caption{{\it (Continued.)}}
\end{figure*}

\setcounter{figure}{0}
\begin{figure*}
\gridline{
\fig{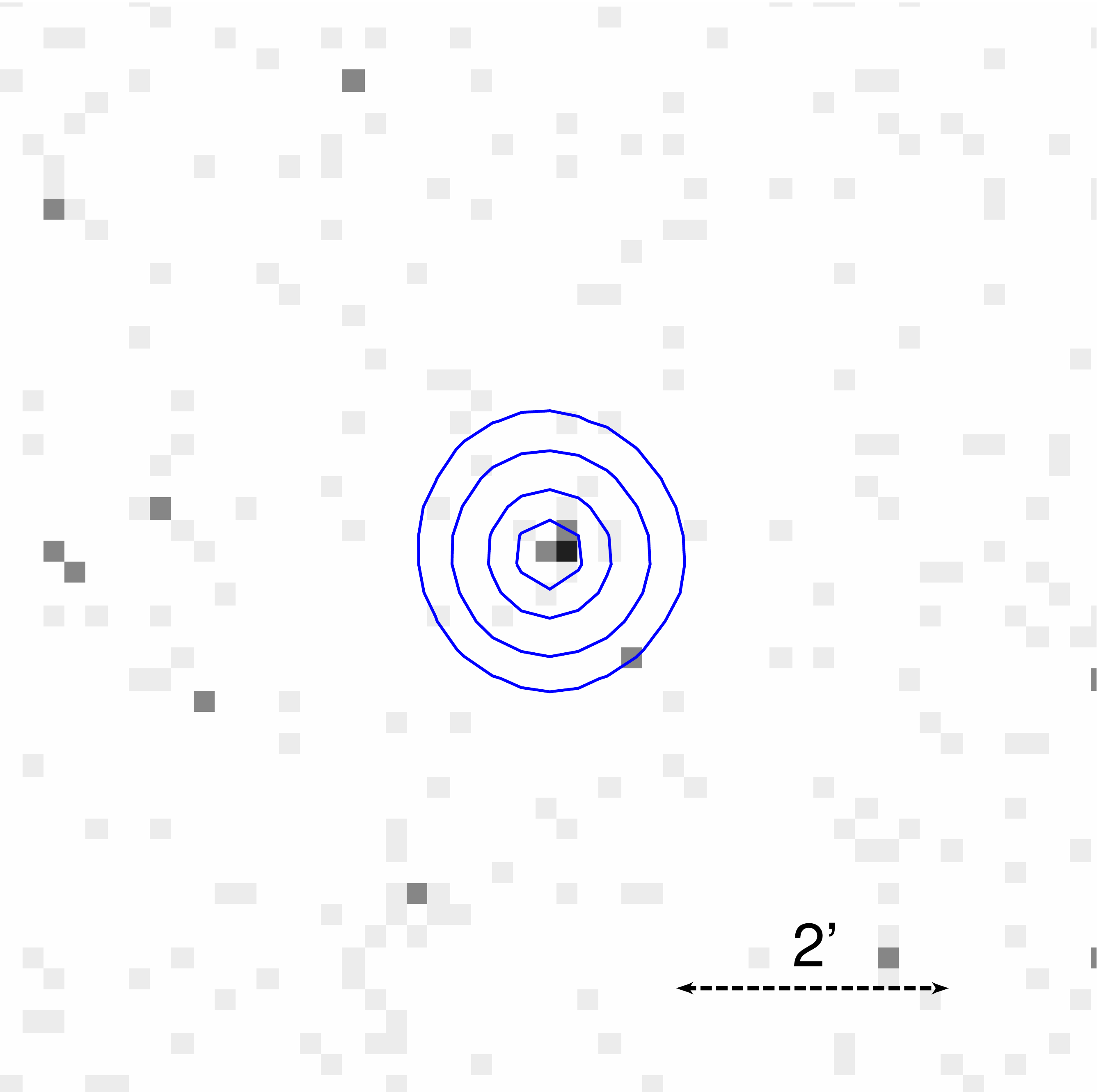}{0.30\textwidth}{(MRC~B0906$-$682)}
\fig{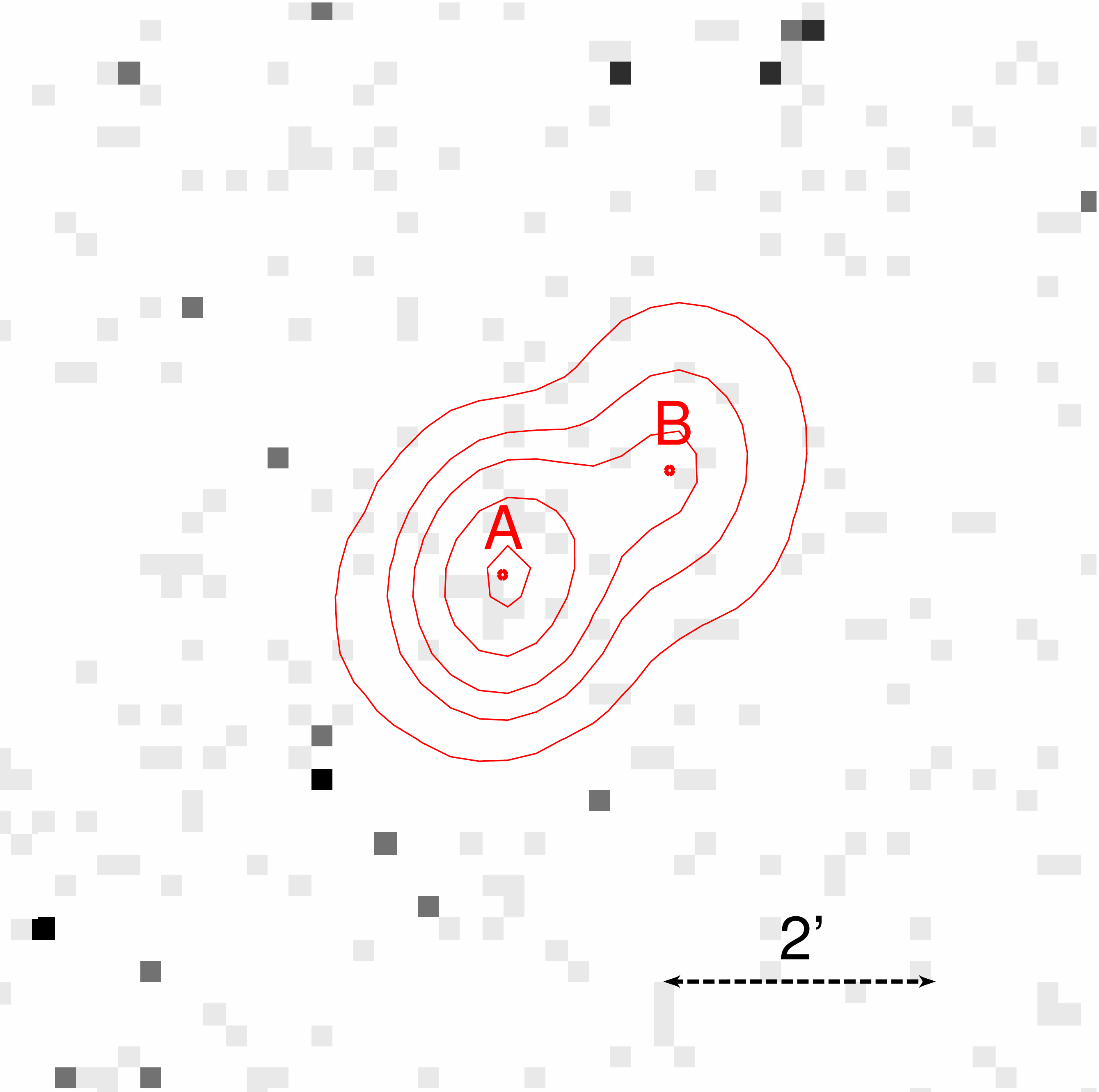}{0.30\textwidth}{(MRC~B1017$-$421)}
\fig{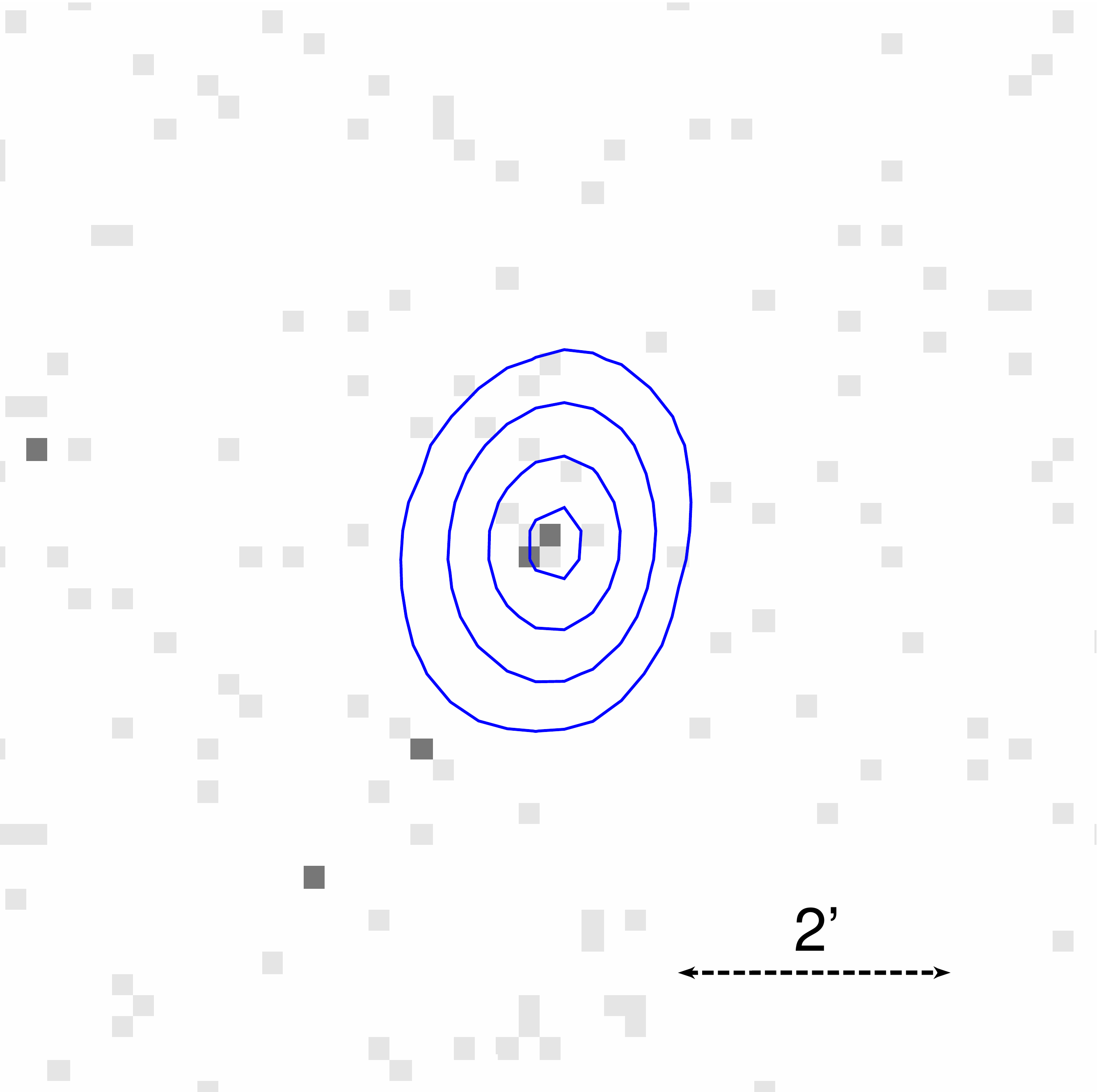}{0.30\textwidth}{(MRC~B1030$-$340)}
}
\gridline{
\fig{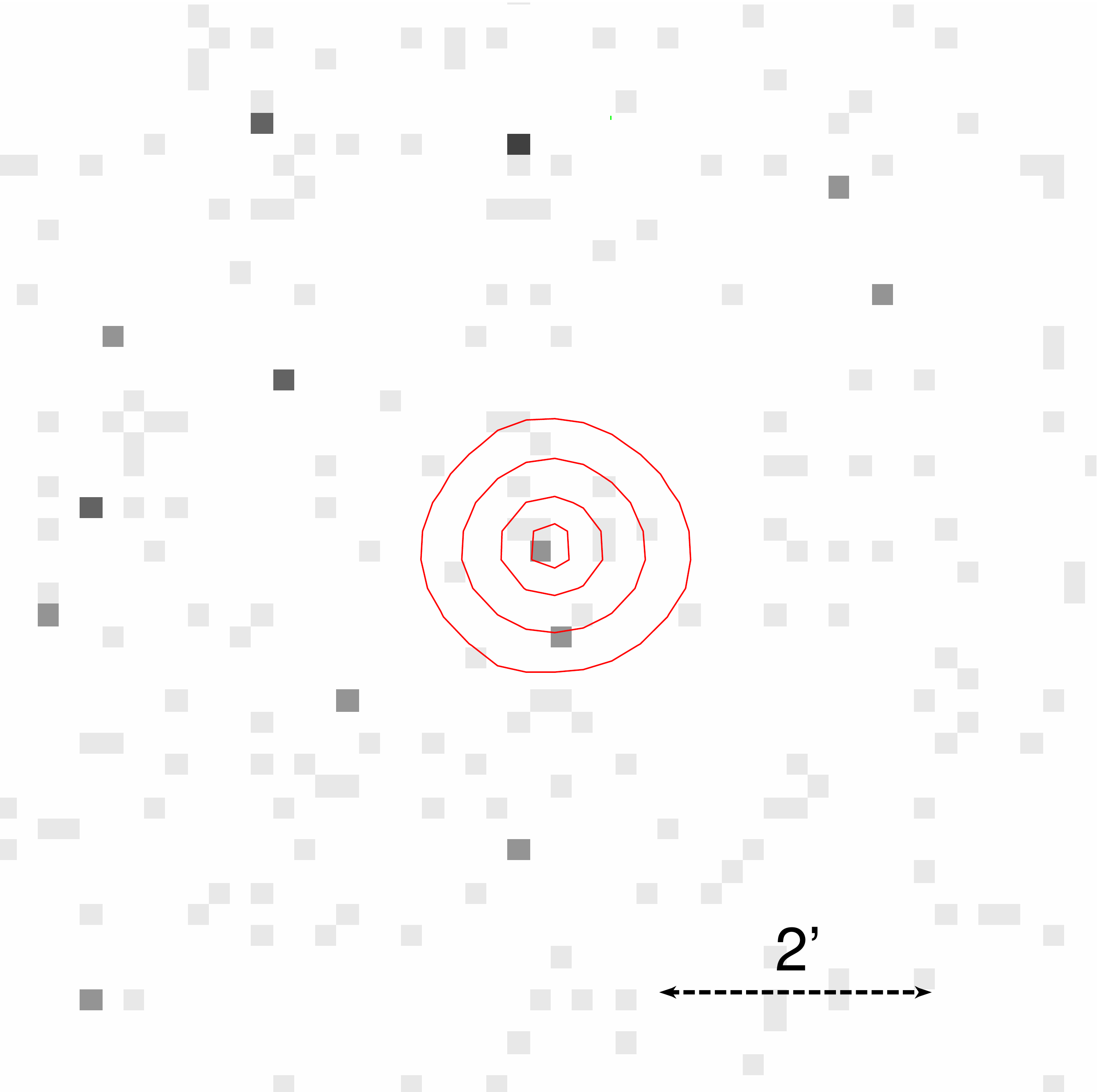}{0.30\textwidth}{(MRC~B1036$-$697)}
\fig{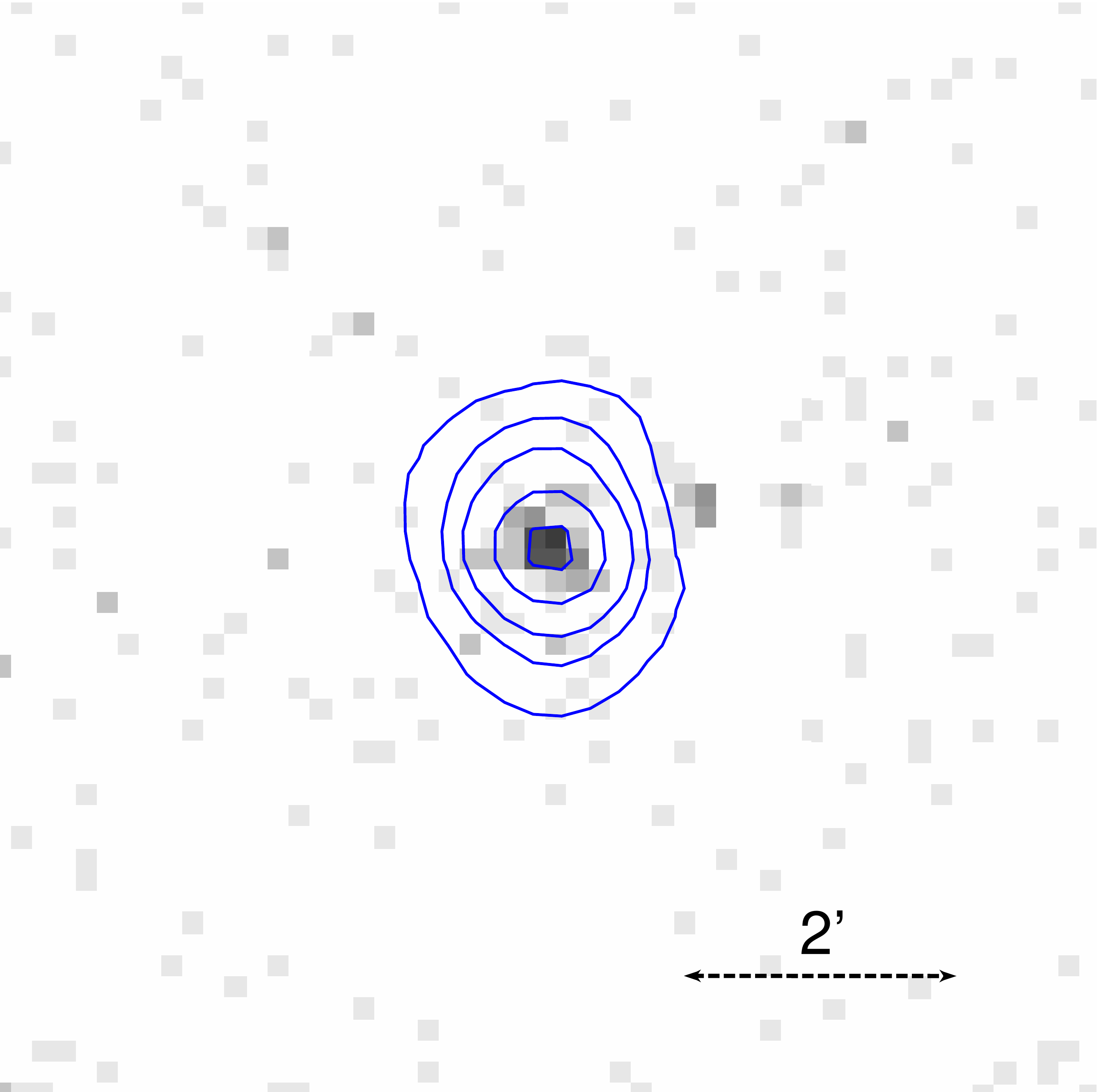}{0.30\textwidth}{(MRC~B1143$-$483)}
\fig{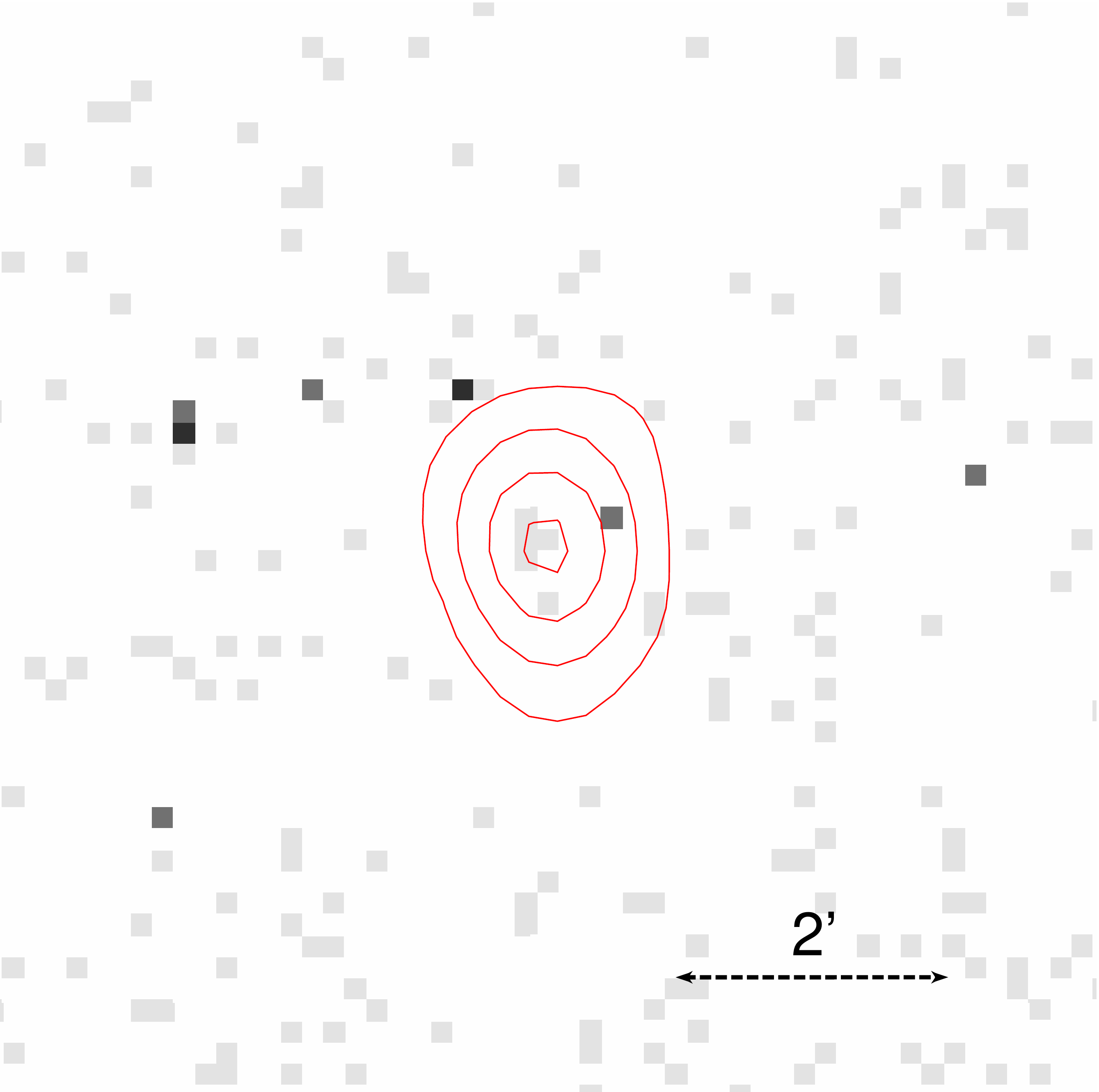}{0.30\textwidth}{(MRC~B1247$-$401)}
}
\gridline{
\fig{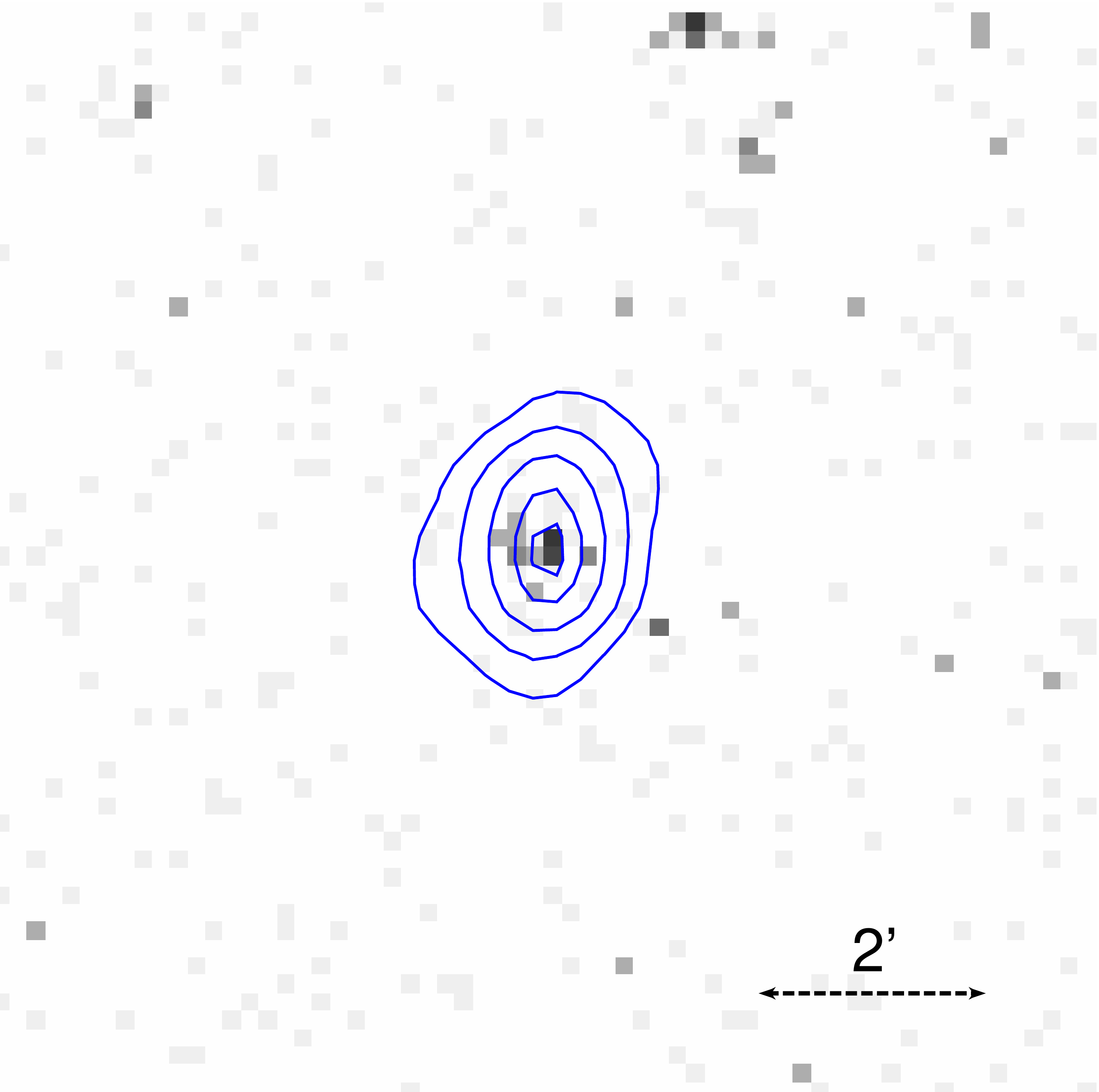}{0.30\textwidth}{(MRC~B1346$-$391)}
\fig{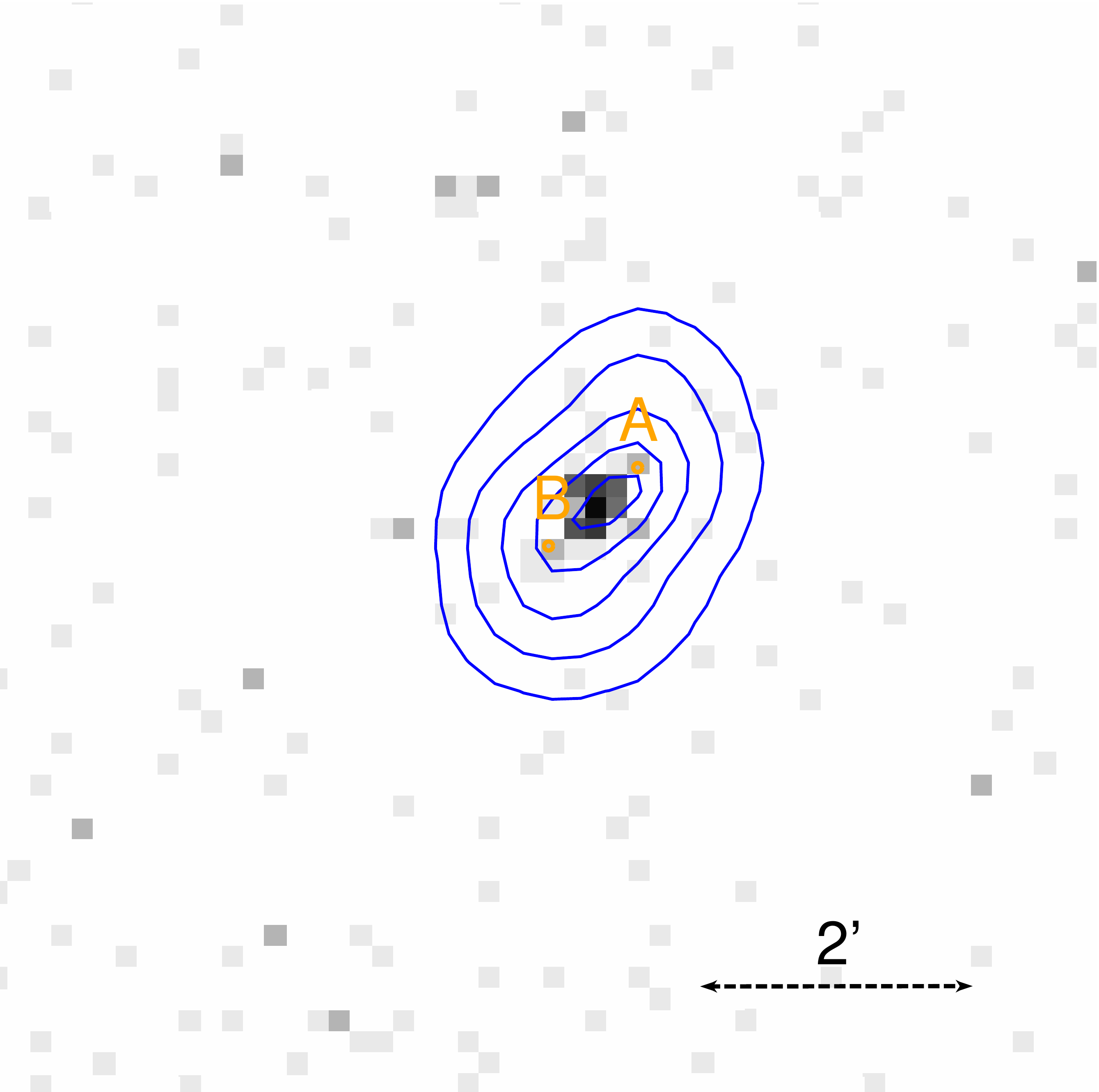}{0.30\textwidth}{(MRC~B1358$-$493)}
\fig{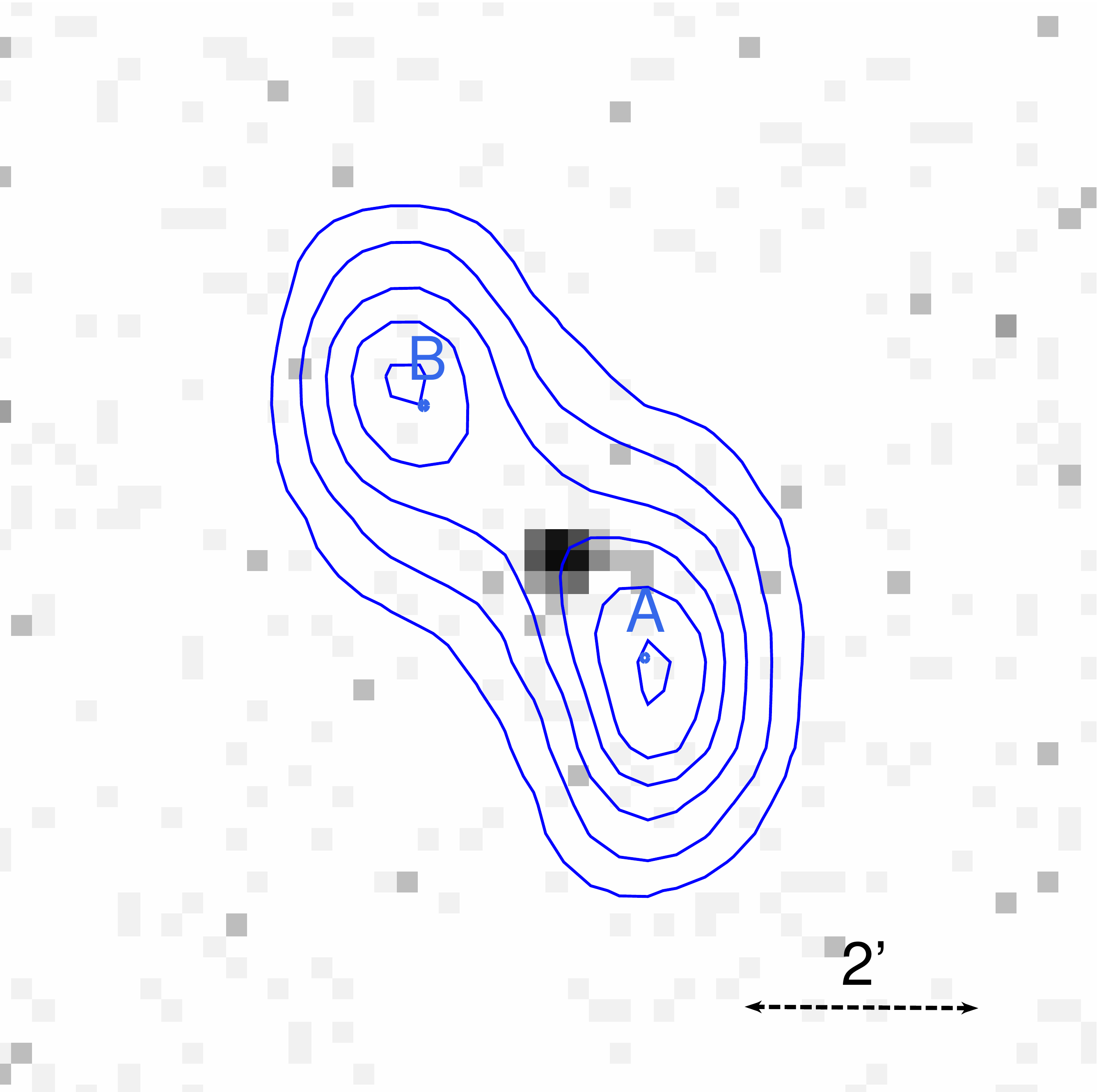}{0.30\textwidth}{(MRC~B1413$-$364)}
}
\caption{{\it (Continued.)}}
\end{figure*}

\setcounter{figure}{0}
\begin{figure*}
\gridline{
\fig{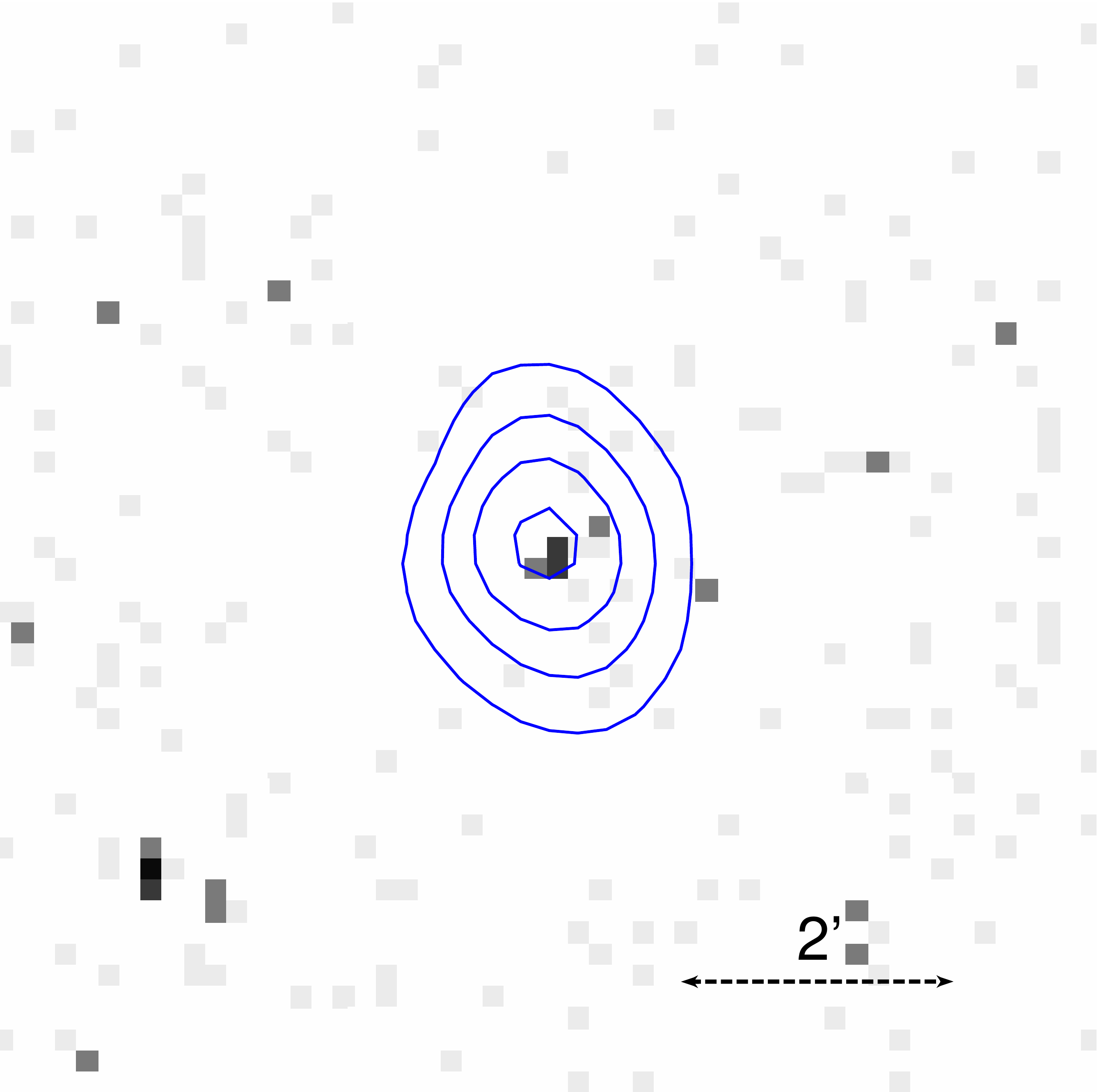}{0.30\textwidth}{(MRC~B1445$-$468)}
\fig{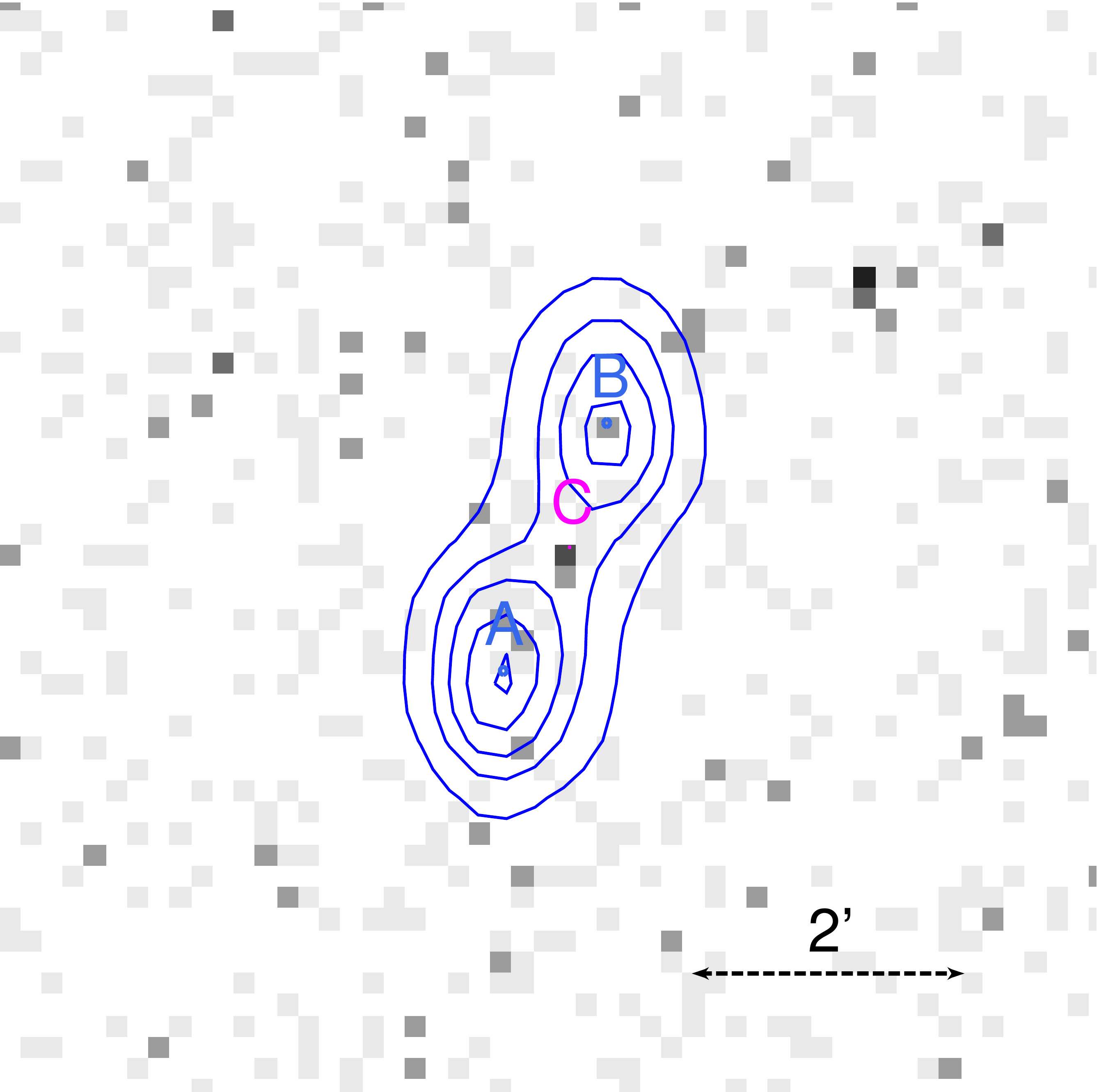}{0.30\textwidth}{(MRC~B1451$-$364)}
\fig{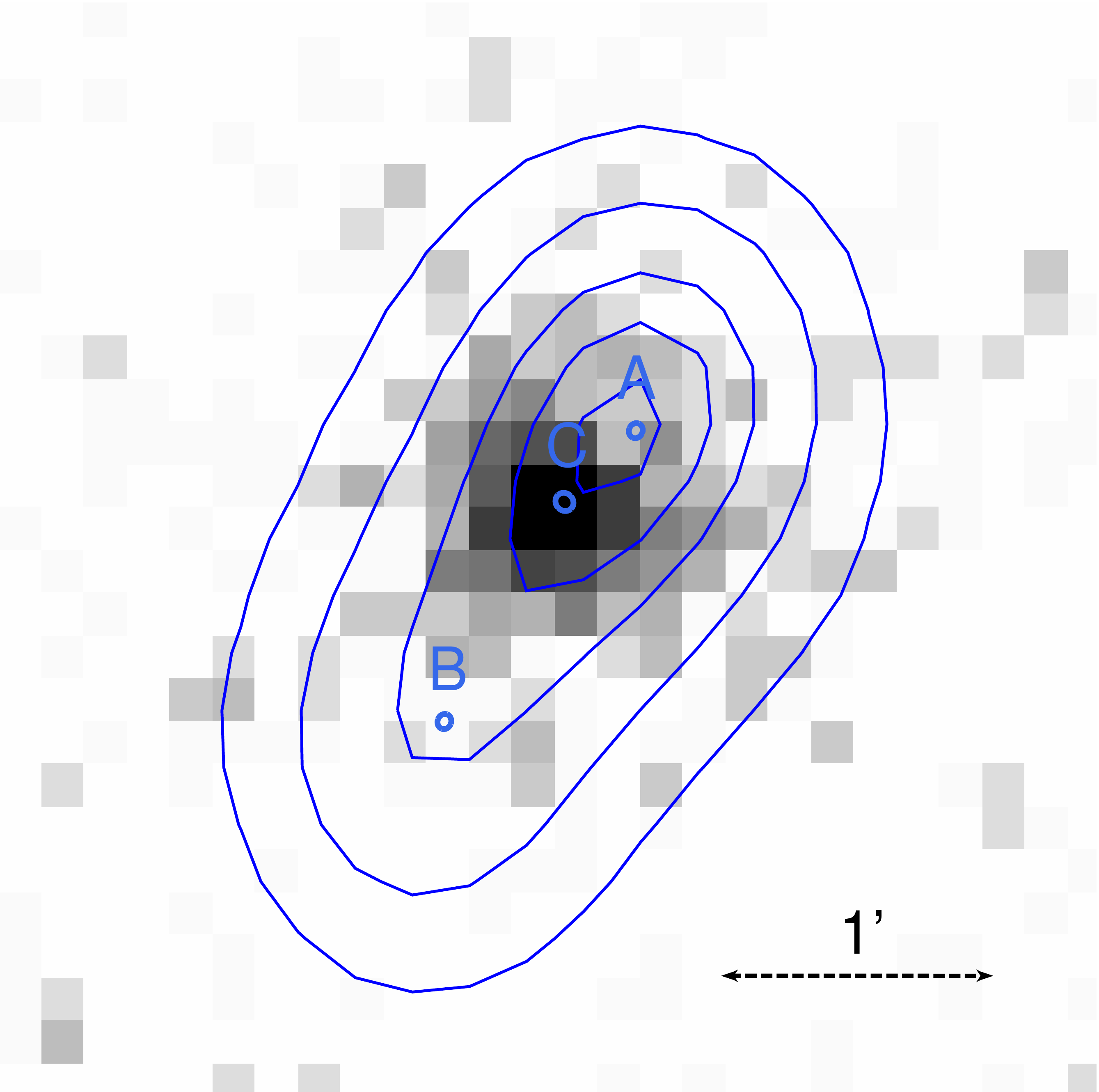}{0.30\textwidth}{(MRC~B1737$-$609)}
}
\gridline{\fig{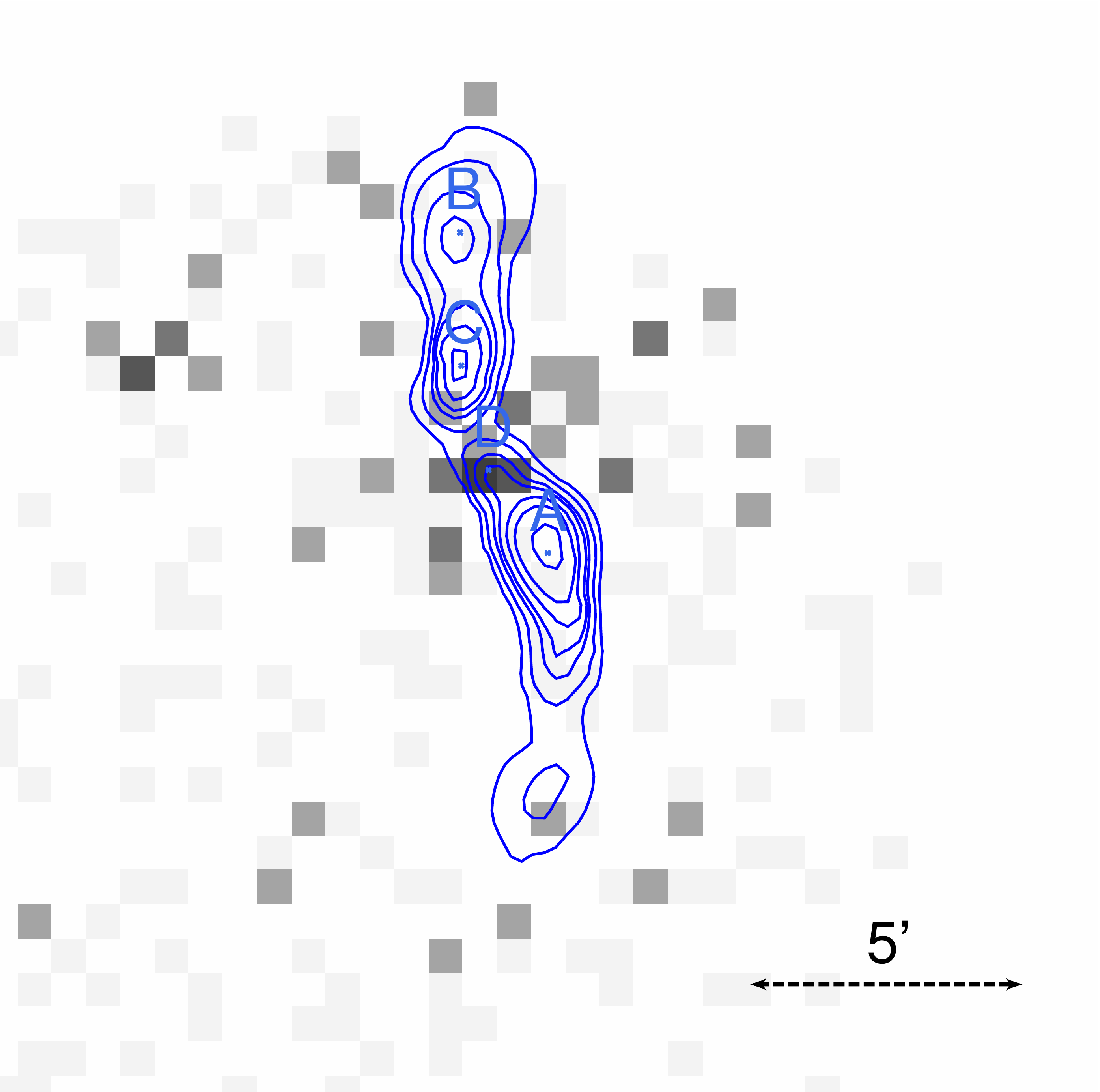}{0.52\textwidth}{(PKS~B2148$-$555)}}
\caption{{\it (Continued.)}}
\end{figure*}

\section{\sw-XRT Data Reduction and Analysis} 
\label{sec:4}

The \sw~observations of the ensemble discussed here include 24 sources observed (mainly between 2015 November and 2016 September) as part of an initial \sw~proposal.
Among these, five sources were observed again between 2019 and 2021 by chance (MRC~B1017$-$421 and MRC~B1036$-$697) or from subsequent \sw~Target of Opportunities (ToO) requests (MRC~B0547$-$408, MRC~B0906$-$682, and MRC~1451$-$364).
In 2021 October, the 24 sources were then supplemented with archival observations for seven additional sources: MRC~B0157$-$311, MRC~B0407$-$658, MRC~B0743$-$673, MRC~B0842$-$754, MRC~B1413$-$364, MRC~B1737$-$609, and PKS~B2148$-$555.
The full observational details are given in Table~\ref{tab:02} (for detections) and Table~\ref{tab:03} (for upper limits).

The X-ray data reduction and procedures adopted in the present analysis are analogous to those already described in \cite{2013ApJS..206...17M,2016MNRAS.460.3829M} and references therein.
Below, we briefly summarize these and include additional details, suited for the present analysis.

X-ray data from the \sw-XRT were retrieved from the \sw~archive and processed with the XRTDAS software package (v3.6.0), developed at the Space Science Data Center (SSDC) of the Italian Space Agency (ASI) and distributed by the NASA High Energy Astrophysics Archive Research Center (HEASARC) within the HEASoft package (v6.28), including a collection of {\sc ftools} to manipulate and analyze FITS files.
All the XRT observations were carried out in the most sensitive photon counting (PC) readout mode. 
Event files were calibrated and cleaned by applying standard filtering criteria with the {\sc xrtpipeline} task and using the latest calibration files available in the \sw~CALDB distributed by HEASARC. 
Events in the 0.3--10 keV energy range, with grades 0--12, were used in the analysis, and exposure maps were also created with {\sc xrtpipeline}.
For sources with several visits, event files and exposure maps from each observation were then accumulated with the {\sc xselect} and {\sc ximage} tools, respectively, to build a single event file and a single exposure map for each source.

Fig.~\ref{fig:01} shows the X-ray maps for all the 31 SMS4 sources included in our sample.
To reveal X-ray emission at the position of each SMS4 source, we simultaneously uploaded both the event file and the exposure map within {\sc ximage}, and used the {\sc background} command to compute the average background intensity over the whole detector.
Then, we performed a local source detection using the {\sc sosta} command within {\sc ximage}.
This task counts the number of events within a specified box and corrects these counts for dead time, vignetting, exposure, and point-spread function (PSF).
As a result, it provides the source intensity and its significance, in terms of the probability $P$ that the signal is a statistical fluctuation of the background.
If the probability is higher than $P = 10^{-3}$, a count rate 3$\sigma$ upper limit is automatically calculated.

In our use of {\sc sosta}, we fixed the side of the box to 14~pixels ($\sim33\arcsec$), a choice that allows the core of the PSF to be fully included in the box and that, based on our experience in the analysis of XRT images, maximizes the signal-to-noise ratio for faint sources.
Furthermore, we used the qualifier {\it /background\_level} to fix the background intensity at the value previously obtained with the {\sc background} command.
For each X-ray image, we used the cursor to precisely center the extraction box on the SMS4 coordinates.
We repeated the source detection procedure adopting G4Jy coordinates in place of the SMS4 ones, which could result in the inclusion of a different number of events in the corresponding boxes, although the significance of the X-ray detections was confirmed in all cases.

\begin{table*}
\scriptsize
\caption{\sw-XRT Detections of SMS4 Sources of Our Sample}
\label{tab:02}
\begin{center}
\begin{tabular}{cccccccccc} 
\hline
 (1)        &  (2)                        &    (3)       &            (4)             &       (5)         &     (6)      &    (7)       & (8)  &  (9)     &         (10)      \\   
R.A.        & Decl.                       & Error Radius &          Count Rate        &       $P$         &  First Obs.  & Latest Obs.  & Obs. & Exposure &      SMS4 Name    \\
($^{h~m~s}$) & ($^{\circ}$~\arcmin~\arcsec) &  (arcsec)    & (10$^{-3}$ counts s$^{-1}$) &                   & (yyyy-mm-dd) & (yyyy-mm-dd) &      &  (s)     &                   \\
\hline                                                                                                                                 
00 10 30.24 &        $-$44 22 55.4        &      6.0     &      ~~~1.7 $\pm$ 0.7      & $3.7\cdot10^{-06}$ & 2016-01-22   & 2016-04-01  &   ~7  &   5495   & {\bf B0007$-$446} \\
00 16 02.80 &        $-$63 10 03.6        &      6.1     &      ~~~2.7 $\pm$ 1.1      & $2.4\cdot10^{-07}$ & 2016-01-27   & 2016-04-24  &   ~9  &   3922   & {\bf B0013$-$634} \\
00 52 14.74 &        $-$43 06 29.8        &      5.3     &      ~~~4.9 $\pm$ 1.2      & $2.2\cdot10^{-16}$ & 2015-11-30   & 2016-03-31  &   ~9  &   5667   & {\bf B0049$-$433} \\
02 00 12.15 &        $-$30 53 27.5        &      5.4     &     ~~35.1 $\pm$ 6.7       & ~~~$< 10^{-16}$    & 2015-05-03   &    ...      &   ~1  &   1221   &      B0157$-$311  \\
02 20 08.00 &        $-$70 22 29.6        &      4.2     &     ~~26.7 $\pm$ 2.9       & ~~~$< 10^{-16}$    & 2015-11-28   & 2016-04-29  &   10  &   4561   & {\bf B0219$-$706} \\
04 20 56.41 &        $-$62 23 41.0        &      5.2     &      ~~~2.2 $\pm$ 0.9      & $3.6\cdot10^{-06}$ & 2015-12-04   & 2016-06-13  &   ~9  &   4261   & {\bf B0420$-$625} \\
04 55 14.05 &        $-$30 06 51.7        &      6.0     &      ~~~6.1 $\pm$ 1.7      & ~~~$< 10^{-16}$    & 2015-12-04   & 2015-12-06  &   ~2  &   3294   & {\bf B0453$-$301} \\
05 36 13.83 &        $-$49 44 23.1        &      5.4     &      ~~~5.7 $\pm$ 1.4      & $2.2\cdot10^{-16}$ & 2015-12-09   & 2016-04-17  &   ~7  &   4426   & {\bf B0534$-$497} \\
05 49 23.24 &        $-$40 51 12.1        &      5.6     &      ~~~1.5 $\pm$ 0.6      & $1.2\cdot10^{-06}$ & 2015-11-28   & 2021-07-18  &   ~8  &   7052   & {\bf B0547$-$408} \\
07 43 32.05 &        $-$67 26 25.0        &      3.7     &     ~~55.7 $\pm$ 3.3       & ~~~$< 10^{-16}$    & 2009-02-16   & 2016-04-21  &   ~9  &   7266   &      B0743$-$673  \\
08 41 27.22 &        $-$75 40 26.4        &      4.3     &     ~~72.2 $\pm$ 8.8       & ~~~$< 10^{-16}$    & 2011-12-24   &    ...      &   ~1  &   1355   &      B0842$-$754  \\
09 06 52.57 &        $-$68 29 37.2        &      5.6     &      ~~~1.7 $\pm$ 0.6      & $1.9\cdot10^{-07}$ & 2015-12-21   & 2021-08-13  &   ~8  &   7499   & {\bf B0906$-$682} \\
10 33 13.45 &        $-$34 18 42.9        &      6.5     &      ~~~2.3 $\pm$ 1.0      & $1.8\cdot10^{-06}$ & 2016-01-05   & 2016-05-27  &   10  &   4107   & {\bf B1030$-$340} \\
11 45 30.99 &        $-$48 36 10.2        &      4.1     &     ~~29.0 $\pm$ 2.9       & ~~~$< 10^{-16}$    & 2016-01-23   & 2016-09-22  &   12  &   4964   & {\bf B1143$-$483} \\
13 49 51.21 &        $-$39 22 51.2        &      4.7     &      ~~~7.2 $\pm$ 1.6      & $1.1\cdot10^{-16}$ & 2015-12-17   & 2016-09 20  &   11  &   4306   & {\bf B1346$-$391} \\
14 01 31.89 &        $-$49 32 36.2        &      4.4     &     ~~20.5 $\pm$ 2.5       & ~~~$< 10^{-16}$    & 2015-12-16   & 2016-08-22  &   11  &   4746   & {\bf B1358$-$493} \\
14 16 33.18 &        $-$36 40 51.3        &      4.1     &     ~~20.3 $\pm$ 2.1       & ~~~$< 10^{-16}$    & 2013-12-20   & 2014-01-02  &   ~5  &   6807   &      B1413$-$364  \\
14 48 28.40 &        $-$47 01 41.4        &      6.0     &      ~~~4.7 $\pm$ 1.4      & $1.7\cdot10^{-12}$ & 2015-12-26   & 2016-05-03  &   ~8  &   3753   & {\bf B1445$-$468} \\
14 54 28.57 &        $-$36 40 04.2        &      4.8     &      ~~~1.3 $\pm$ 0.6      & $3.6\cdot10^{-05}$ & 2015-12-29   & 2021-07-25  &   15  &   8820   & {\bf B1451$-$364} \\
17 42 01.41 &        $-$60 55 13.1        &      3.6     &    190.1 $\pm$ 5.5         & ~~~$< 10^{-16}$    & 2011-11-10   & 2011-11-17  &   ~4  &   9203   &      B1737$-$609  \\
\hline
\end{tabular}
\end{center}
\tablecomments{Columns (1) and (2): the R.A. and decl. of the X-ray detection. Column (3):  the positional error radius, at the 90\% confidence level. Column (4): the 0.3--10 keV count rate, with its uncertainty. Column (5): the probability $P$ that the signal is a statistical fluctuation of the background. Columns (6) and (7): the dates of the first and the latest \sw~observation. Column (8): the number of \sw~observations. Column (9): the total XRT exposure time. Column (10): the name of the corresponding SMS4 source, according to the MRC designation; bold characters emphasize sources in our initial \sw~proposal.} 
\end{table*}

\begin{table*}
\scriptsize
\caption{XRT Count Rate 3$\sigma$ Upper Limits at the Position of the 11 SMS4 Sources of Our sample Not Detected by \sw}
\label{tab:03}
\begin{center}
\begin{tabular}{ccccccc}
\hline
        (1)           &     (2)      &     (3)      &  (4) &  (5)     &        (6)        &        (7)                 \\
      SMS4 Name       &  First Obs.  &  Latest Obs. & Obs. & Exposure &         $P$       &    3$\sigma$ Upper Limit   \\
                      & (yyyy-mm-dd) & (yyyy-mm-dd) &      &  (s)     &                   & (10$^{-3}$ counts s$^{-1}$) \\ 
\hline                                                                                                                                 
{\bf MRC B0223$-$712} &  2015-12-03  &  2016-02-29  &   ~8  &  3822    & $9.2\cdot10^{-02}$ &          ~3.9           \\ 
{\bf MRC B0245$-$558} &  2015-12-11  &  2016-03-29  &   ~9  &  4106    & $1.3\cdot10^{-02}$ &          ~4.3           \\ 
{\bf MRC B0315$-$685} &  2015-11-29  &  2016-04-25  &   10  &  3731    & 1.0               &          ~2.4           \\ 
     MRC B0407$-$658  &  2013-01-03  &     ...      &   ~1  &  1996    & $3.5\cdot10^{-02}$ &          ~7.7           \\ 
{\bf MRC B0411$-$561} &  2015-11-28  &  2016-04-21  &   ~4  &  4244    & 1.0               &          ~2.1           \\  
{\bf MRC B0546$-$445} &  2015-12-11  &  2016-02-05  &   ~9  &  3926    & $1.4\cdot10^{-02}$ &          ~4.5           \\ 
{\bf MRC B0842$-$835} &  2015-12-20  &  2016-01-02  &   ~6  &  4357    & $5.4\cdot10^{-03}$ &          ~4.5           \\  
{\bf MRC B1017$-$421} &  2015-12-30  &  2019-06-25  &   13  &  5536    & $5.9\cdot10^{-03}$ &          ~3.9           \\  
{\bf MRC B1036$-$697} &  2016-01-03  &  2021-04-25  &   10  &  6453    & $1.3\cdot10^{-02}$ &          ~3.0           \\ 
{\bf MRC B1247$-$401} &  2015-12-06  &  2016-08-23  &   ~5  &  4250    & $9.5\cdot10^{-03}$ &          ~4.6           \\ 
     PKS B2148$-$555  &  2019-02-27  &     ...      &   ~1  &  1075    & $2.4\cdot10^{-02}$ &          14.9           \\  
\hline
\end{tabular}
\end{center}
\tablecomments{Column (1): the name in SMS4, with bold characters identifying sources in our initial \sw~proposal. Columns (2) and (3): the dates of the first and the latest \sw~observation. Column (4): the number of \sw~observations. Column (5): the total XRT exposure time. Column (6): the probability $P$ that the signal is a statistical fluctuation of the background. Column (7): the 0.3--10 keV count rate 3$\sigma$ upper limit at the position of the radio coordinates.} 
\end{table*}

Requiring $P < 10^{-4}$, which is an order-of-magnitude-tighter constraint with respect to the default in {\sc sosta}, to establish an X-ray detection, we distinguish detections (Table~\ref{tab:02}) from nondetections (Table~\ref{tab:03}). 
Considering all 31 SMS4 sources, our prescriptions led to 20 X-ray detections. 
These correspond to two additional detections, for MRC~B0547$-$408 and MRC~B1451$-$364, in addition to those reported in the $2^{nd}$ \sw~X-ray Point Source Catalog (2SXPS; \citealp{2020ApJS..247...54E}), whose source detection relies on a blind search using a sliding cell algorithm.
To determine the position, and in particular its uncertainty (90\% confidence level), for all the 20 X-ray detections, we used {\sc xrtcentroid}.

\subsection{The Case of PKS~B2148-555} 
\label{sec:4.1}

Our point-like source detection algorithm did not detect an X-ray source associated with PKS~B2148$-$555.
However, Fig.~\ref{fig:01} definitely shows diffuse X-ray emission in the field of view (FOV) of this radio source.
As reported by \cite{2002MNRAS.331..717L}, this FRI radio galaxy is the brightest galaxy in Abell cluster A3816, with a largest angular size of 14\arcmin.

The only useful observation is part of the \sw~Gravitational Wave Galaxy Survey (SWGSG; see \citealp{2019AAS...23321008T}): its exposure is relatively short ($\sim$1.1~ks), with the source serendipitously in the FOV, close to the edge of the XRT detector.
To reveal the thermal X-ray emission surrounding the galaxy, we filtered the event file in the 0.5--2.0 keV band and chose a circular extraction region.
To reliably include this circle within a useful portion of the XRT detector, we fixed its radius at a maximum of 5\arcmin.
Using the {\sc counts} command within {\sc ximage}, we computed a total of 71 counts within this circle.
After subtracting 38 counts, due to the background contribution within the same circle, we obtain a net number of 33 counts.
Considering an exposure of 1034~s, corrected for the vignetting, we thus obtain a count rate of $3.2\times10^{-02}$ counts s$^{-1}$.
We used the NASA's HEASARC tool WebPIMMS (v4.12a) to convert this rate into unabsorbed flux $S_{unabs}$: assuming the Galactic column density, a metal abundance of 0.4 solar, and a plasma temperature {\it kT} $\sim 1.5$~keV, we obtain $S_{unabs} = 7.0\times10^{-13}$ erg cm$^{-2}$ s$^{-1}$.
At a luminosity distance of 166.5~kpc, this implies a luminosity of $2.3\times10^{42}$ erg s$^{-1}$, typical of a poor cluster or group \citep{2015A&A...573A.118L}.
Note that varying the assumed abundance in the 0.2--1 range and the assumed temperature in the 1--5 keV range only changes the unabsorbed flux, and hence the luminosity, by $\sim$~5\%.

\begin{table*} 
\begin{center}
\scriptsize
\caption{Results from the Analysis of the Extent of the X-Ray Emission}
\label{tab:04}
\begin{tabular}{ccccc}
\hline
   (1)          &         (2)          &    (3)   &    (4)   &      (5)      \\  
SMS4 Name       &        Background    &    $C$   &    $A$   &      $ER$     \\
                & (counts pixel$^{-1}$) & (counts) & (counts) &               \\
\hline                                                                                                                                 
MRC~B1737$-$609 &   $1.08\cdot10^{-2}$  &   1055   &    193   & $5.47\pm0.56$ \\
MRC~B0743$-$673 &   $3.71\cdot10^{-3}$  &    230   &     68   & $3.40\pm0.64$ \\
MRC~B1413$-$364 &   $1.08\cdot10^{-3}$  &     80   &     16   & $5.09\pm1.85$ \\
MRC~B1143$-$483 &   $4.11\cdot10^{-3}$  &     78   &     16   & $4.82\pm1.75$ \\
MRC~B0219$-$706 &   $3.11\cdot10^{-3}$  &     72   &     11   & $6.48\pm2.71$ \\
MRC~B0842$-$754 &   $1.17\cdot10^{-3}$  &     57   &     14   & $4.10\pm1.64$ \\
MRC~B1358$-$493 &   $2.91\cdot10^{-3}$  &     48   &      9   & $5.16\pm2.44$ \\
MRC~B0157$-$311 &   $1.19\cdot10^{-3}$  &     24   &      3   & $8.30\pm6.59$ \\
MRC~B1346$-$391 &   $4.48\cdot10^{-3}$  &     16   &      7   & $2.31\pm1.47$ \\
\hline                                              
PKS~B2148$-$555 &   $1.30\cdot10^{-2}$  &      1   &      5   & $0.19\pm0.28$ \\
\hline
\end{tabular}
\end{center}
\tablecomments{Column (1): the name of the SMS4 source. Column (2): the measured number of background counts, per pixel. Column (3): the background-corrected number of counts $C$ within a circle with a radius of 5 pixels. Column (4): the background-corrected number of counts $A$ within an annulus with inner and outer radii of 10 and 20 pixels, respectively. Column (5): the extent ratio ER, with its 1$\sigma$ error.}
\end{table*}

\subsection{Extent of the X-Ray Emission} 
\label{sec:4.2}

For each point source with sufficient statistics, including PKS~B2148$-$555, which is a special case embedded in extended cluster emission, we compared the radial distribution of the detected events with the one expected from a point-like source, in order to evaluate the possible extent of the X-ray emission.
The fraction of total counts within circles of varying radii, centered at the coordinates of the X-ray source, was derived by \cite{2005SPIE.5898..360M}, who modeled the \sw-XRT PSF profile with a King function.
In our analysis, we used the values computed assuming the on-axis PSF model at 1.5~keV, since most of the photons are found near this energy, where the XRT effective area peaks. 
We also note that the dependence of the PSF profile on energy is mild (\citealp{2005SPIE.5898..360M}).

The two regions that we used for our analysis were a circle, with a radius of 5~pixels ($\sim12\arcsec$), and an annulus with inner and outer radii of 10 and 20 pixels, respectively.
The number of counts, $C$ and $A$, that we extracted from these regions, centered at the coordinates of each X-ray centroid (see Table~\ref{tab:02}), were corrected for the background contribution, computed as already reported in Section~\ref{sec:4}.  
The fraction $ER=C/A$ gives an estimate of the deviation of the distribution of events from a point-like source, and therefore of the extent of the X-ray emission. 
For each source, we then compared the $ER$ value, reported in Table~\ref{tab:04}, with the one expected for a point-like source, which for the regions that we adopted is equal to 5.73.

The extended emission of PKS~B2148$-$555 has already been described in Section~\ref{sec:4.1}; the results obtained following the procedure described here are also reported in Table~\ref{tab:04} for comparison.
The other two sources for which we find evidence of extended emission are MRC~B0743$-$673 and MRC~B1346$-$391.

The information that can be found in the literature for MRC~B1346$-$391 is still rather poor, particularly for its emission in bands different from the radio one.
Conversely, MRC~B0743$-$673 is a well-studied source, being a flat-spectrum radio quasar (FSRQ) with a radio core-jet structure (\citealp{2006MNRAS.371..898S}), included in the {\it Roma}-BZCAT catalog of blazars since its first edition (\citealp{2009A&A...495..691M}).

The other sources listed in Table~\ref{tab:04} do not show evidence of extended X-ray emission. 
For all the remaining X-ray detections, the number of counts was too low to provide statistically significant results.

\begin{table*} 
\begin{center}
\scriptsize
\caption{Results from the Analysis of the Hardness Ratio of X-Ray Detections}
\label{tab:05}
\begin{tabular}{cccc}
\hline
       (1)      &   (2)    &    (3)   &      (4)         \\  
    SMS4 Name   &   $S$    &    $H$   &      $HR$        \\
                & (counts) & (counts) &                  \\
\hline                                                                                                                                 
MRC~B1737$-$609 &  1199    &   222    & $-0.69\pm0.02$   \\
MRC~B0743$-$673 &   258    &    49    & $-0.68\pm0.04$   \\
MRC~B1413$-$364 &    42    &    67    &  ~~$0.23\pm0.16$ \\
MRC~B1143$-$483 &    94    &    15    & $-0.72\pm0.05$   \\
MRC~B0219$-$706 &    78    &    14    & $-0.70\pm0.06$   \\
MRC~B0842$-$754 &    56    &    17    & $-0.53\pm0.10$   \\
MRC~B1358$-$493 &    30    &    41    &  ~~$0.15\pm0.19$ \\
MRC~B0157$-$311 &    24    &     5    & $-0.66\pm0.12$   \\
MRC~B1346$-$391 &    27    &     4    & $-0.74\pm0.10$   \\
\hline
\end{tabular}
\end{center}
\tablecomments{Column (1): the name of the SMS4 source. Column (2): the measured number of counts $S$ in the soft (0.3--3 keV) band. Column (3): the measured number of counts $H$ in the hard (3--10 keV) band. Column (4): the hardness ratio $HR$, with its 1$\sigma$ error.}
\end{table*}

\begin{table*} 
\begin{center}
\scriptsize
\caption{Results from the Spectral Analysis}
\label{tab:06}
\begin{tabular}{cc|cccc}
\hline
    (1)     &  (2)   &         (3)       &         (4)            &               (5)                &      (6)     \\   
SMS4 Name   & Counts &   ${n_{H,\,Gal}}$  &     $\Gamma$          &         $F_{~0.3-10~keV}$           & $\chi^2$/dof \\
            &        &     (cm$^{-2}$)   &                       & ($10^{-12}$ erg cm$^{-2}$ s$^{-1}$) &              \\
\hline                                                                                                                    
B1737$-$609 &  1606  & $6.2\cdot10^{20}$ &  ~~$1.80^{+0.04}_{-0.04}$ &         $7.6^{+0.3}_{-0.3}$       &    64.63/69  \\
B0743$-$673 &   366  & $8.9\cdot10^{20}$ &  ~~$1.60^{+0.08}_{-0.08}$ &         $2.6^{+0.3}_{-0.2}$       &    20.87/16  \\
B1413$-$364 &   121  & $4.0\cdot10^{20}$ &   $-0.24^{+0.16}_{-0.18}$ &         $3.1^{+0.4}_{-0.6}$       &    15.09/4   \\
B1143$-$483 &   120  & $1.0\cdot10^{21}$ &  ~~$1.75^{+0.17}_{-0.17}$ &         $1.0^{+0.1}_{-0.2}$       &     2.85/4   \\
B0219$-$706 &   103  & $6.1\cdot10^{20}$ &  ~~$1.55^{+0.15}_{-0.15}$ &         $1.1^{+0.1}_{-0.1}$       &     3.20/3   \\
B0842$-$754 &    81  & $7.4\cdot10^{20}$ &  ~~$1.40^{+0.18}_{-0.19}$ &         $3.5^{+0.8}_{-0.6}$       &     7.30/2   \\
B1358$-$493 &    80  & $1.5\cdot10^{21}$ &   $-0.20^{+0.23}_{-0.24}$ &         $3.4^{+0.5}_{-0.7}$       &     1.93/2   \\
\hline
\end{tabular}
\end{center}
\tablecomments{The analysis was carried out using an absorbed power law, fixing the hydrogen column density to the Galactic value. Column (1): the MRC name of the SMS4 source. Column (2): the number of counts in the spectrum, corrected for the background contribution. Column (3): the Galactic hydrogen column density ${n_{H,\,Gal}}$. Column (4): the corresponding photon index $\Gamma$, with its 1$\sigma$ error. Column (5): the absorbed flux in the 0.3--10 keV band, with its 1$\sigma$ error. Column (6): the $\chi^2$ value, with the number of degrees of freedom.}
\end{table*}

\subsection{Hardness Ratio and Spectral Analysis} 
\label{sec:4.3}

We used the {\sc counts} command within {\sc ximage} to extract the number of counts within a circle with a radius of 10 pixels, centered at the position of our X-ray detections.
We took into account the two contiguous 0.3--3 keV and 3--10 keV energy bands to distinguish soft ($S$) and hard ($H$) X-ray photons.
Then, we computed the hardness ratio by means of the formula $HR = (H - S) / (H + S)$.
As shown in Table~\ref{tab:05}, all sources appear to be basically soft, excluding MRC~B1358$-$493 and MRC~B1413$-$364.
As reported below, MRC~B1413$-$364 is also one of the two sources detected in the BAT hard X-ray catalogs.

The statistics allowed us to perform a spectral analysis with {\sc xspec} for the seven sources with the highest count rates, among those reported in Table~\ref{tab:05}.  
We extracted source events from a circle with a radius of 20 pixels ($\sim47\arcsec$), covering $\approx$ 90\% of the XRT PSF, centered at the X-ray source coordinates, and background events from a circle with a radius of 50 pixels in its proximity, avoiding eventual spurious sources.
We grouped the energy spectrum files, requiring at least 20 events per bin. 
Then, we fitted the obtained spectra with an absorbed power law, fixing the hydrogen column density to the Galactic value (\citealp{2016A&A...594A.116H}). 
The results of this spectral analysis are shown in Table~\ref{tab:06}: for the photon index $\Gamma$ and the flux $F$ in the 0.3--10 keV band, errors are given at a 1$\sigma$ confidence level.

In most cases, statistics do not allow us to go beyond a simple power-law model fit.
Using the power-law model, and leaving the $n_H$ parameter free to vary, we find that the adoption of the $n_H$ Galactic value adequately describes the spectrum of most sources.   
We report an exception to this behavior for MRC~B1413$-$364, with an improvement of the goodness of fit given by $\chi^2 = 6.37$ (3 dof). 
The statistical significance of this result is given by an {\it F}-test, with a probability of $0.14$ that this improvement is due to chance. 
The column density value that we obtain, $n_H = (2.1^{+1.1}_{-0.9})\times10^{22}$ cm$^{-2}$, is in excess with respect to the Galactic value ($4.0\times10^{20}$ cm$^{-2}$), suggesting the presence of intrinsic absorption.
The best-fit value for the photon index is $\Gamma = 1.25^{+0.63}_{-0.57}$, significantly higher than that found fixing $n_H$ to the Galactic value, as reported in Table~\ref{tab:06}.

Only 2 of the 20 sources detected with the XRT, MRC~B1413$-$364 and MRC~B1737$-$609, are detected in the two series of the catalogs \citep{2010ApJS..186..378T,2010A&A...510A..47S} produced with data from the \sw~Burst Alert Telescope (BAT; \citealp{2005SSRv..120..143B}).    
In the hard BAT telescope band (14--195 keV), SWIFT~J1416.5$-$3671 is brighter than SWIFT~J1742.1$-$6054\footnote{Using the data from the 157-Month catalog (\url{https://swift.gsfc.nasa.gov/results/bs157mon/}), SWIFT~J1416.5$-$3671 has a flux of $(14.71^{+3.87}_{-3.18})\cdot10^{-12}$ erg cm$^{-2}$ s$^{-1}$, while SWIFT~J1742.1$-$6054 has a flux of $(5.71^{+2.26}_{-2.10})\cdot10^{-12}$ erg cm$^{-2}$ s$^{-1}$.}, consistent with our HR analysis, reported in the discussion above and in Table~\ref{tab:05}.


\begin{figure*}
\gridline{\fig{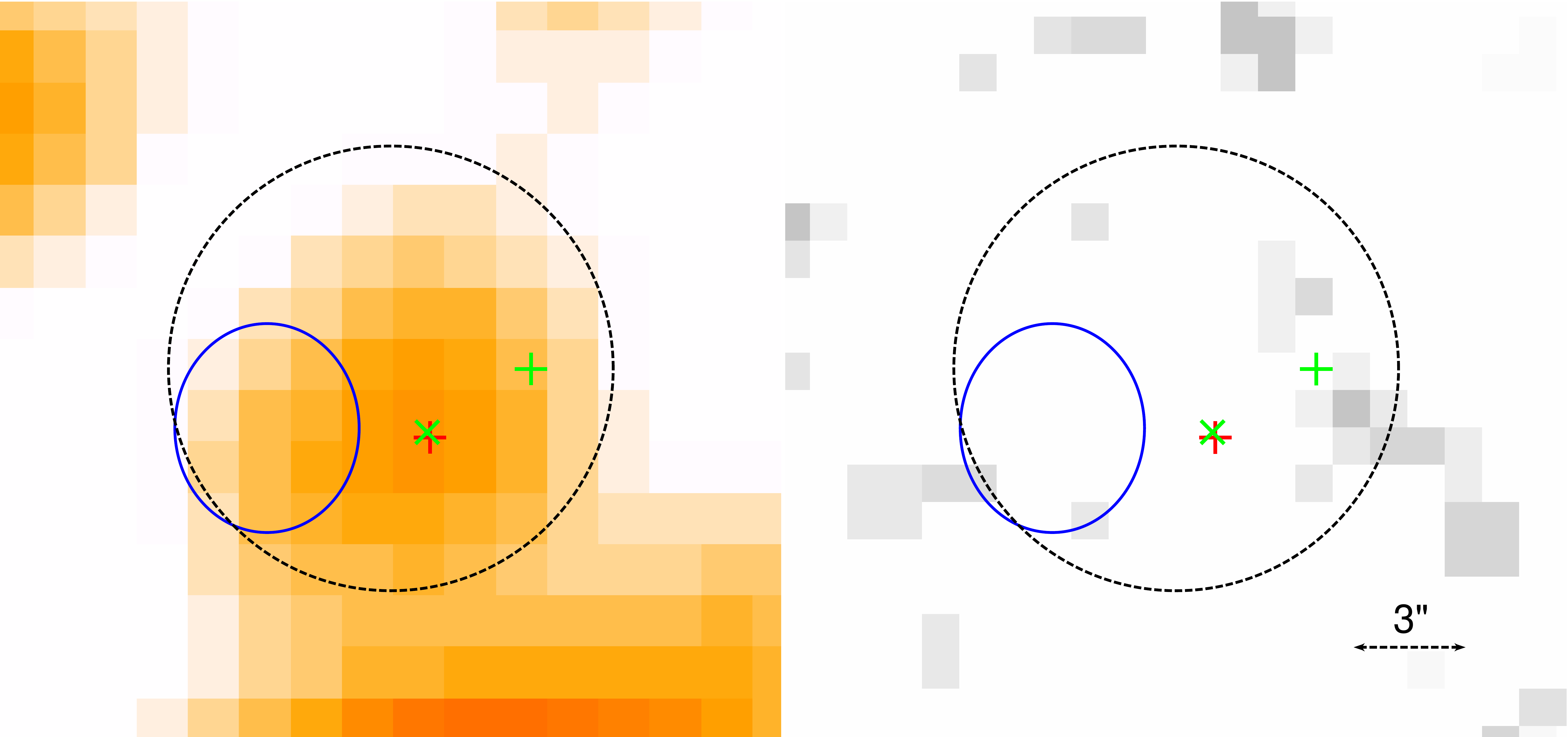}{0.49\textwidth}{(a - MRC~B0007$-$446)} \fig{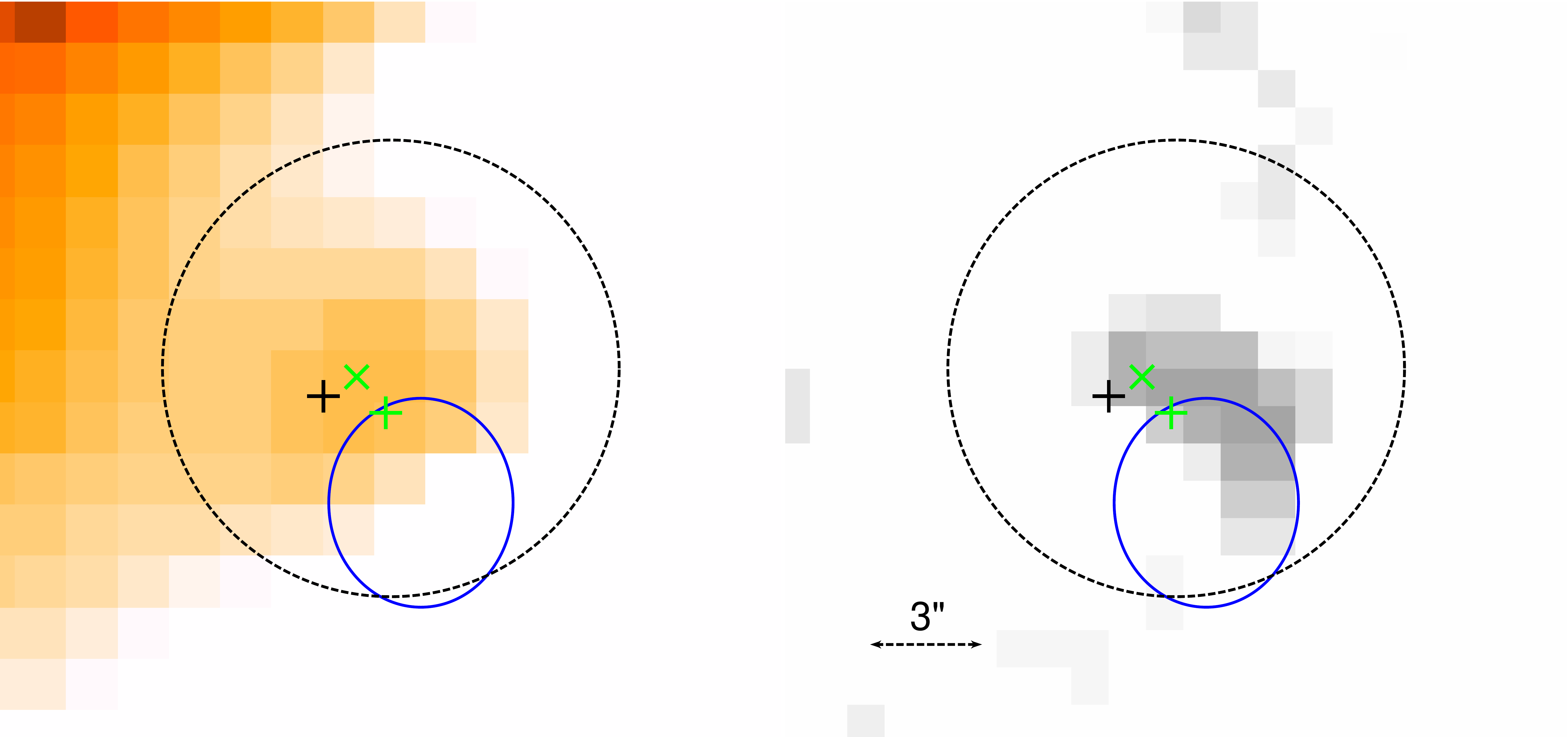}{0.49\textwidth}{(b - MRC~B0013$-$634)}}
\gridline{\fig{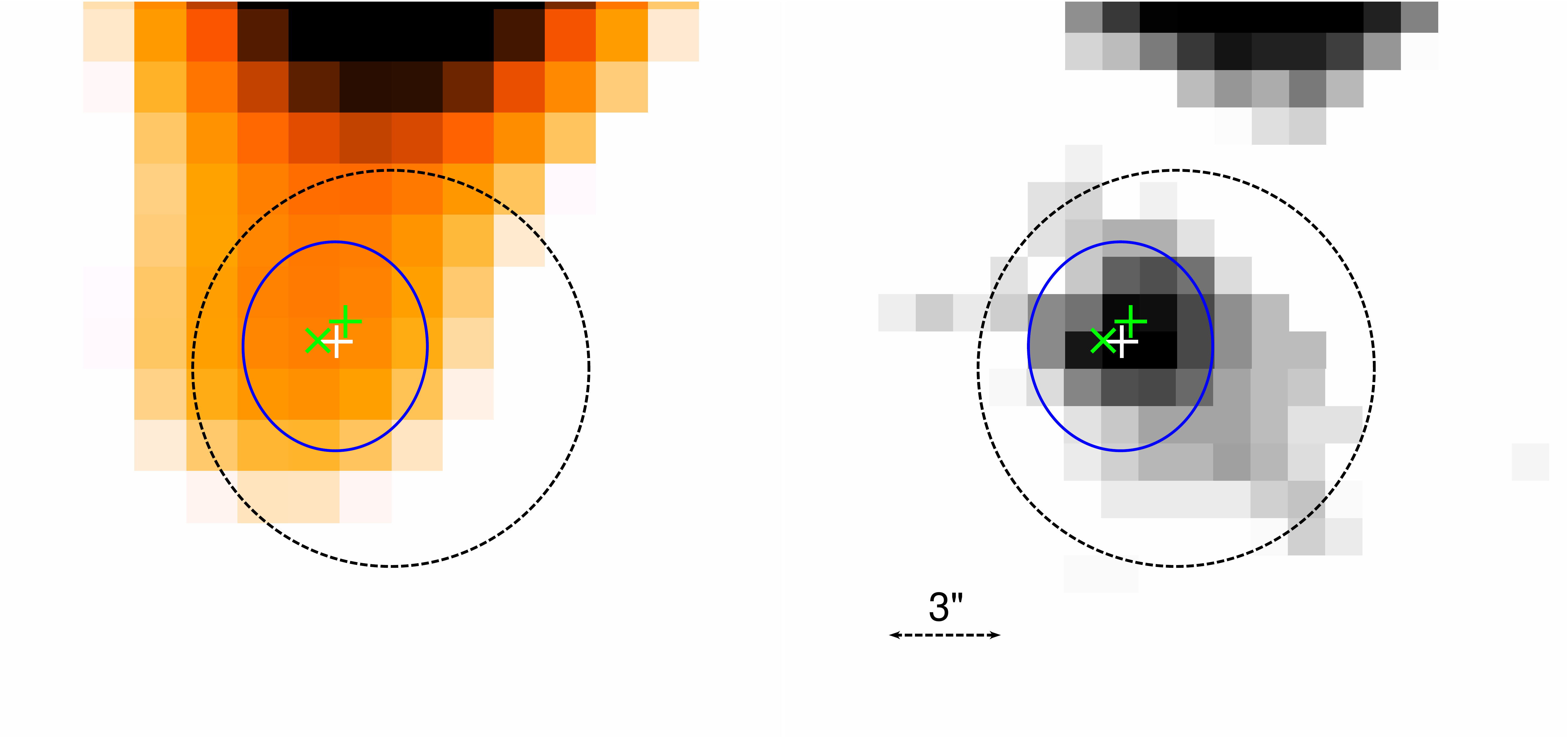}{0.49\textwidth}{(c - MRC~B0049$-$433)} \fig{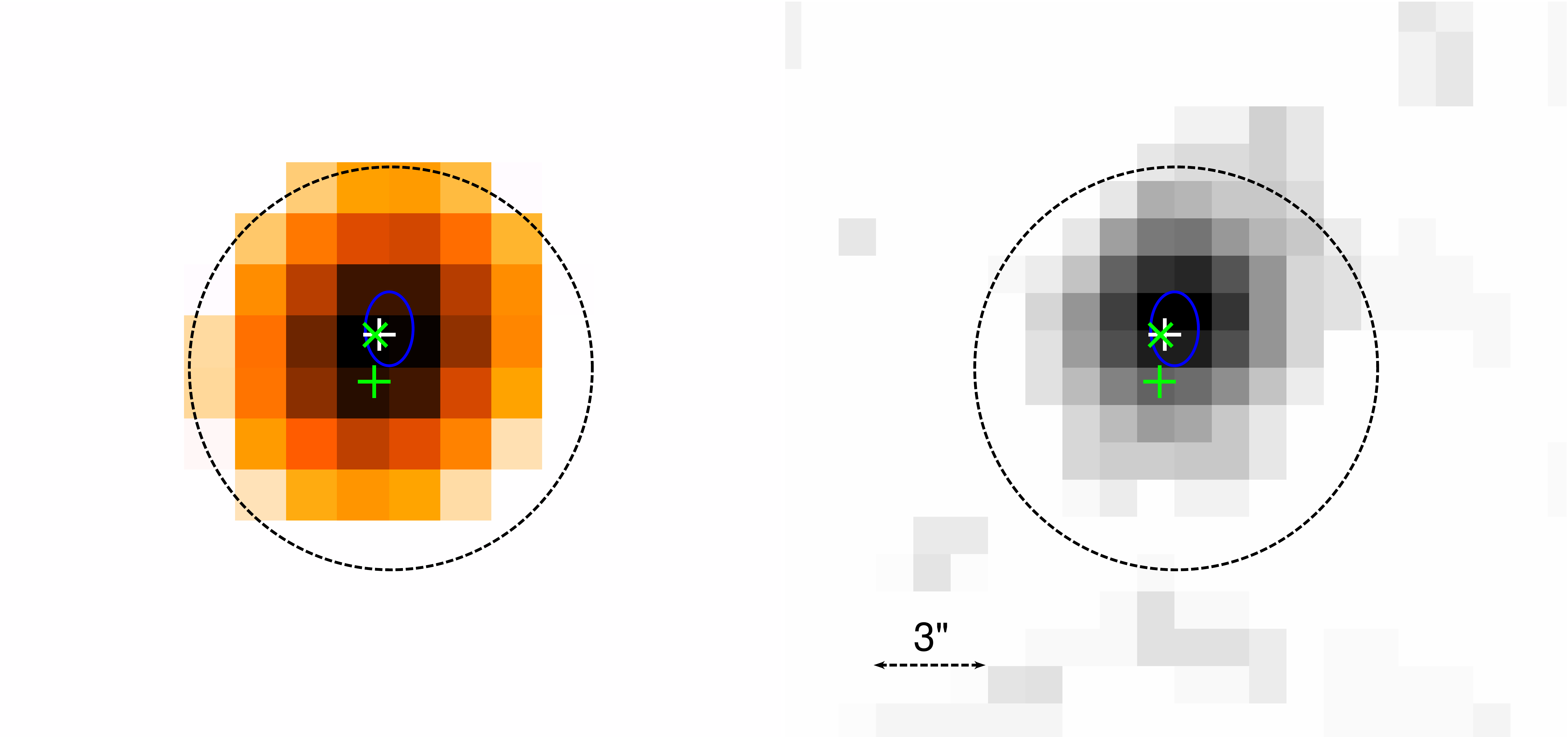}{0.49\textwidth}{(d - MRC~B0157$-$311)}}
\gridline{\fig{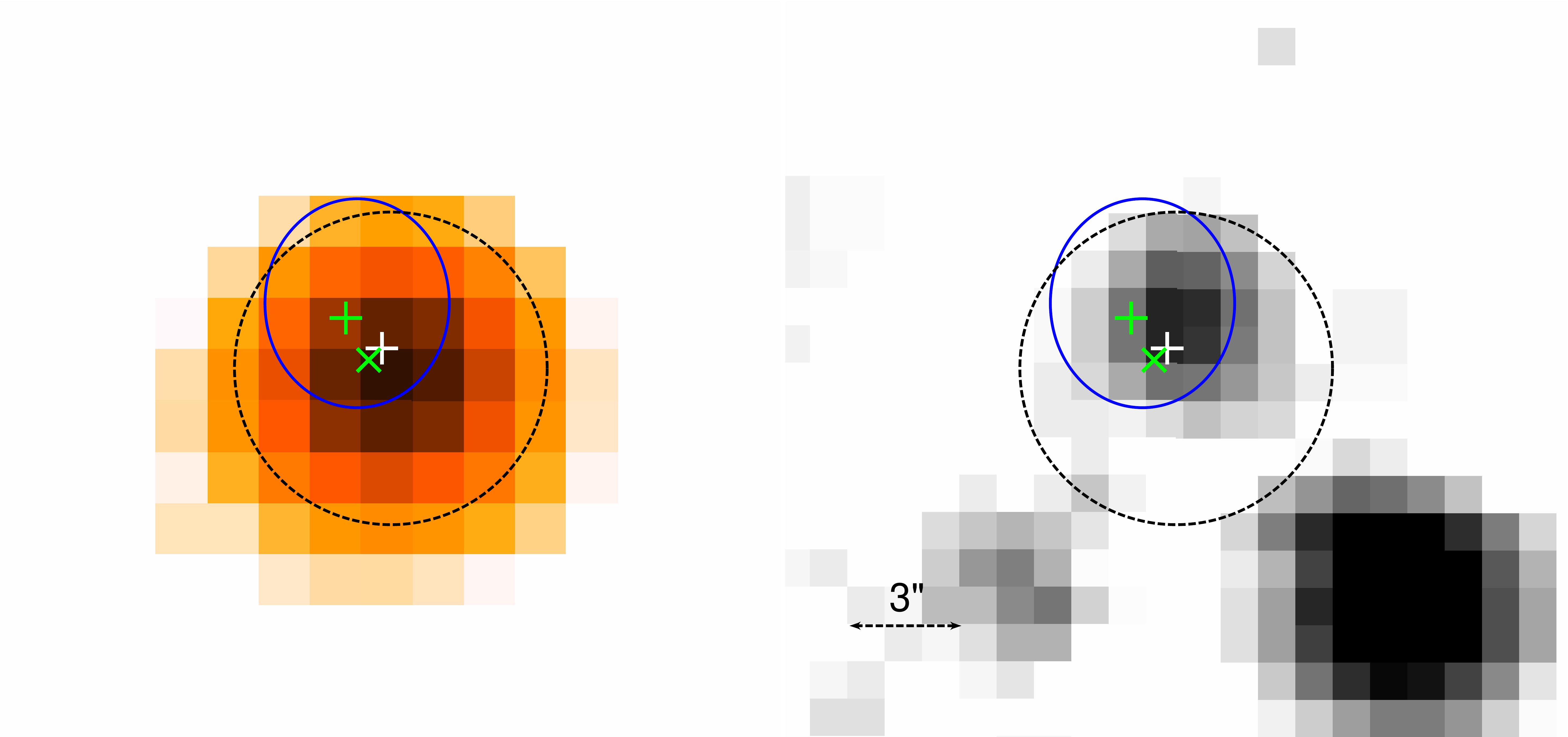}{0.49\textwidth}{(e - MRC~B0219$-$706)} \fig{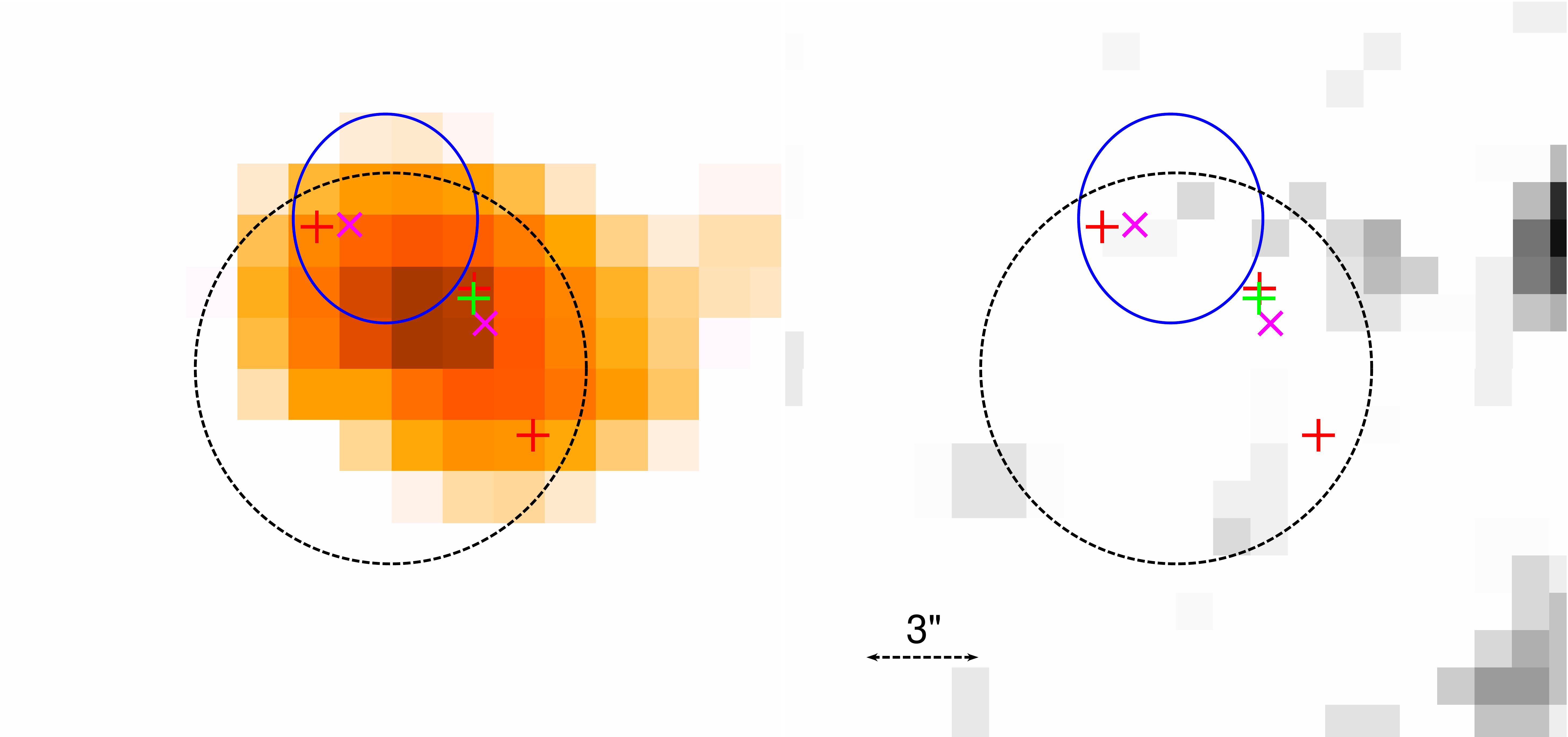}{0.49\textwidth}{(f - MRC~B0420$-$625)}}
\caption{Infrared and optical maps, at the same scale, of the 20 SMS4 sources detected by \sw-XRT in our sample. Infrared maps (left side of each panel) in the W1 filter (3.4 $\mu$m) are from AllWISE, while optical maps (right side; 1~pixel=1\arcsec) in the $r$ filter are from DSS2.  Ellipses mark the positional uncertainty of radio sources: blue is used for G4Jy; magenta is used for SUMSS in panels (o), (p), and (t); and red is used for NVSS in panel (s). Black dashed circles mark the positional uncertainty of the X-ray sources. Positional uncertainties are given at 90\% confidence level for both the radio and the X-ray sources. Crosses (x) and plus signs (+) mark infrared and optical sources, respectively. Green is used for counterparts associated by W20 and by BH06. For other AllWISE and GSC~2.3.2 sources, black or white are equally used to improve the visibility with respect to the map in the background. In the infrared band: in panels (f), (l), (m), and (r), magenta crosses mark sources from CatWISE2020. In the optical band: in panels (a), (f), (h), and (i), red plus signs mark sources from DES~DR2.}
\label{fig:02}
\end{figure*}

\setcounter{figure}{1}
\begin{figure*}
\gridline{\fig{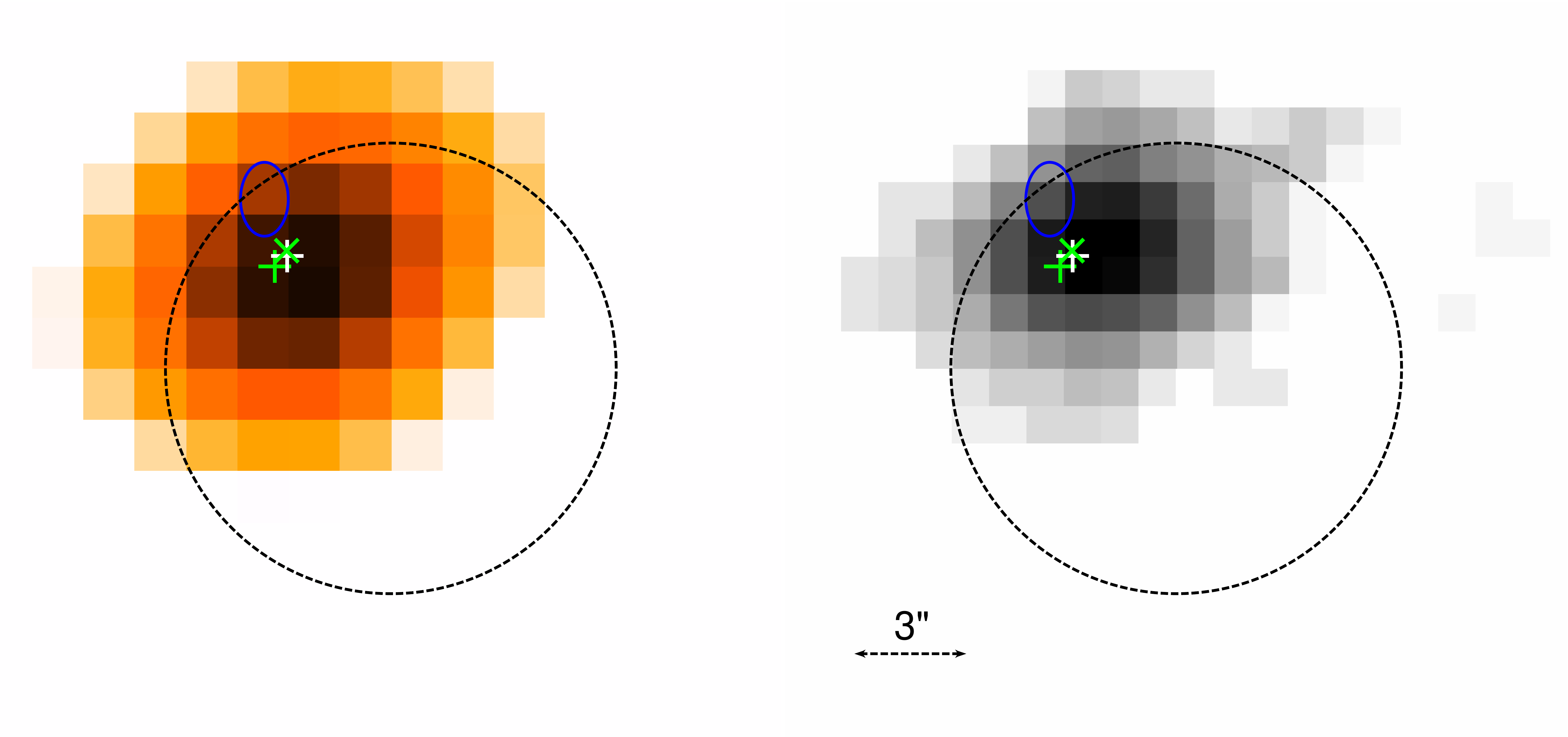}{0.49\textwidth}{(g - MRC~B0453$-$301)} \fig{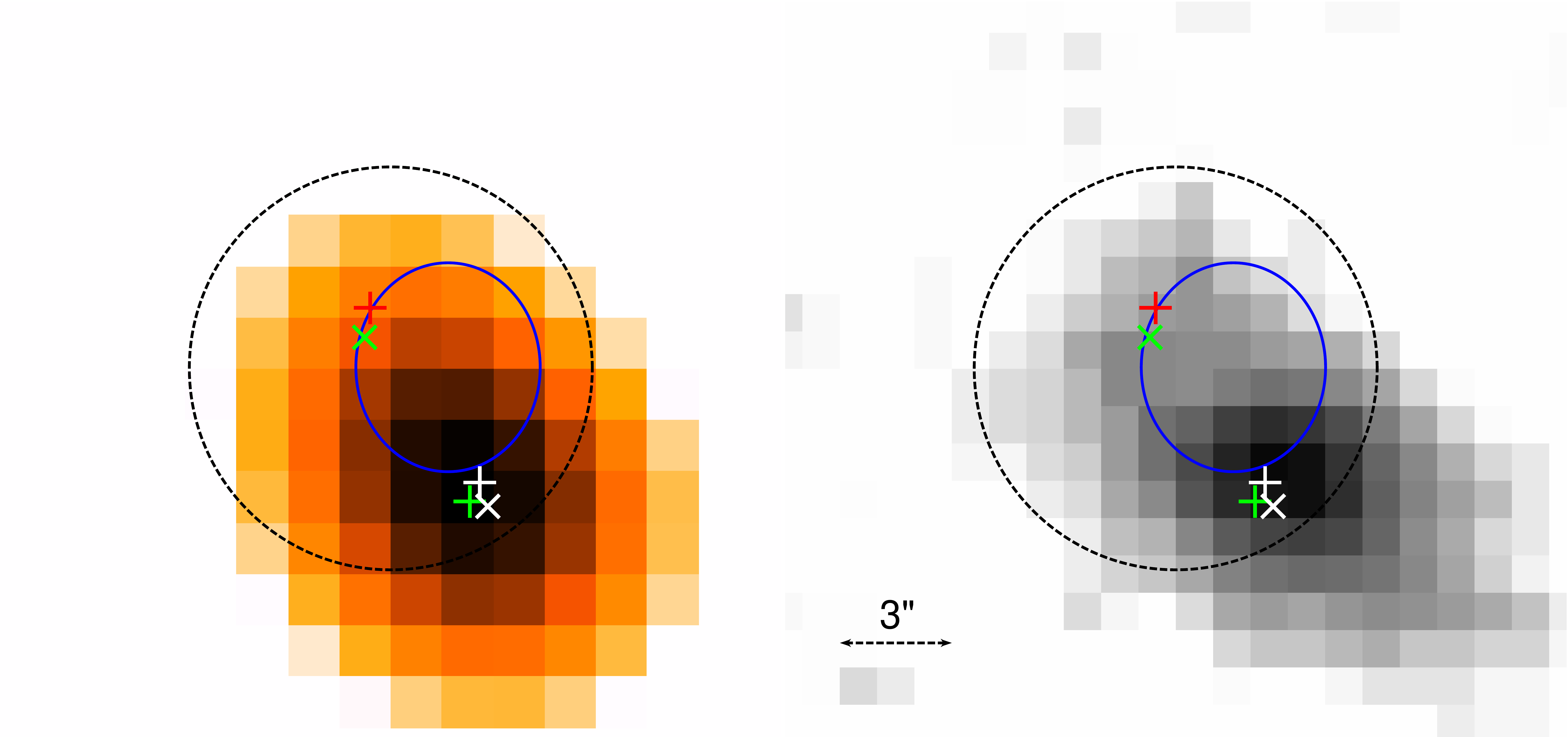}{0.49\textwidth}{(h - MRC~B0534$-$497)}}
\gridline{\fig{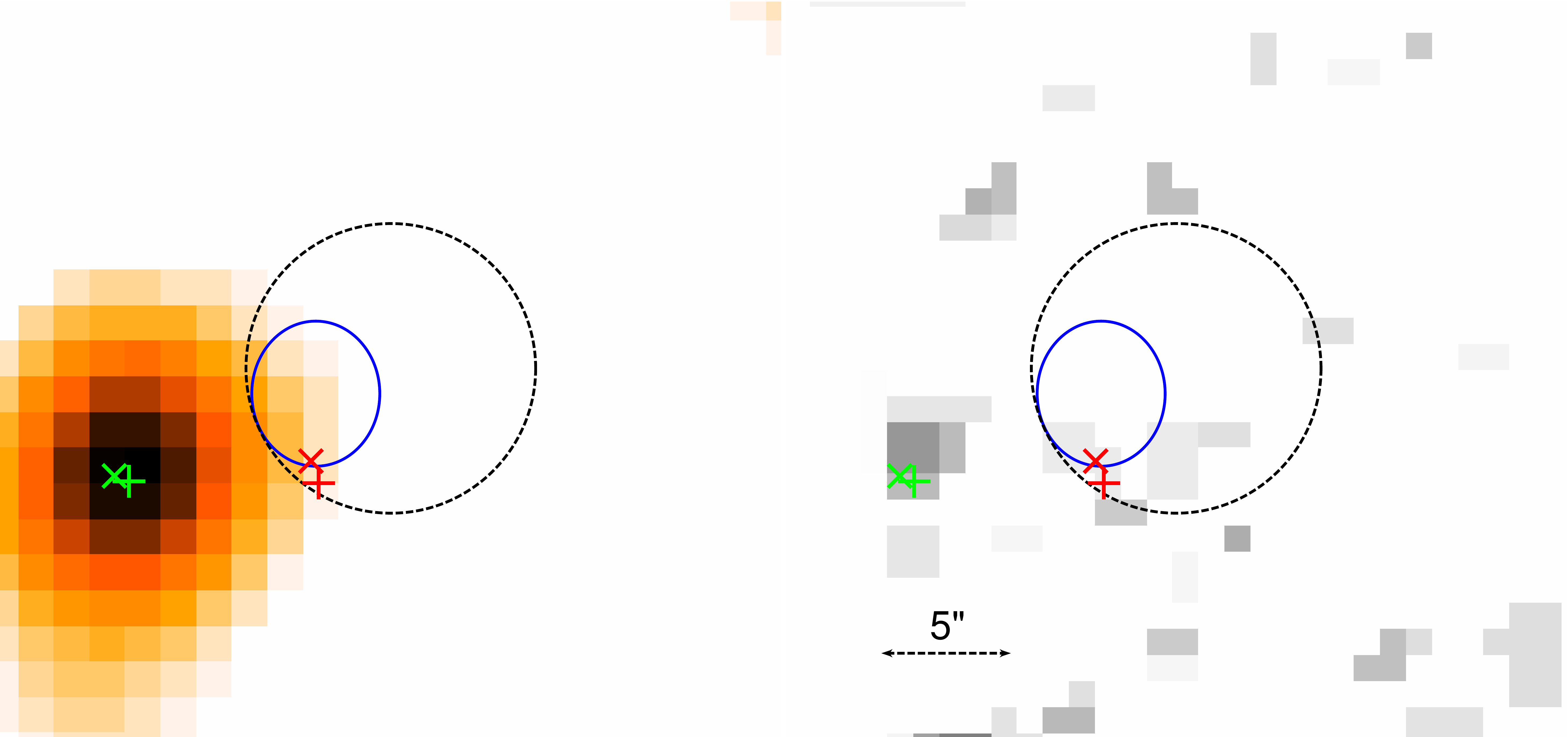}{0.49\textwidth}{(i - MRC~B0547$-$408)} \fig{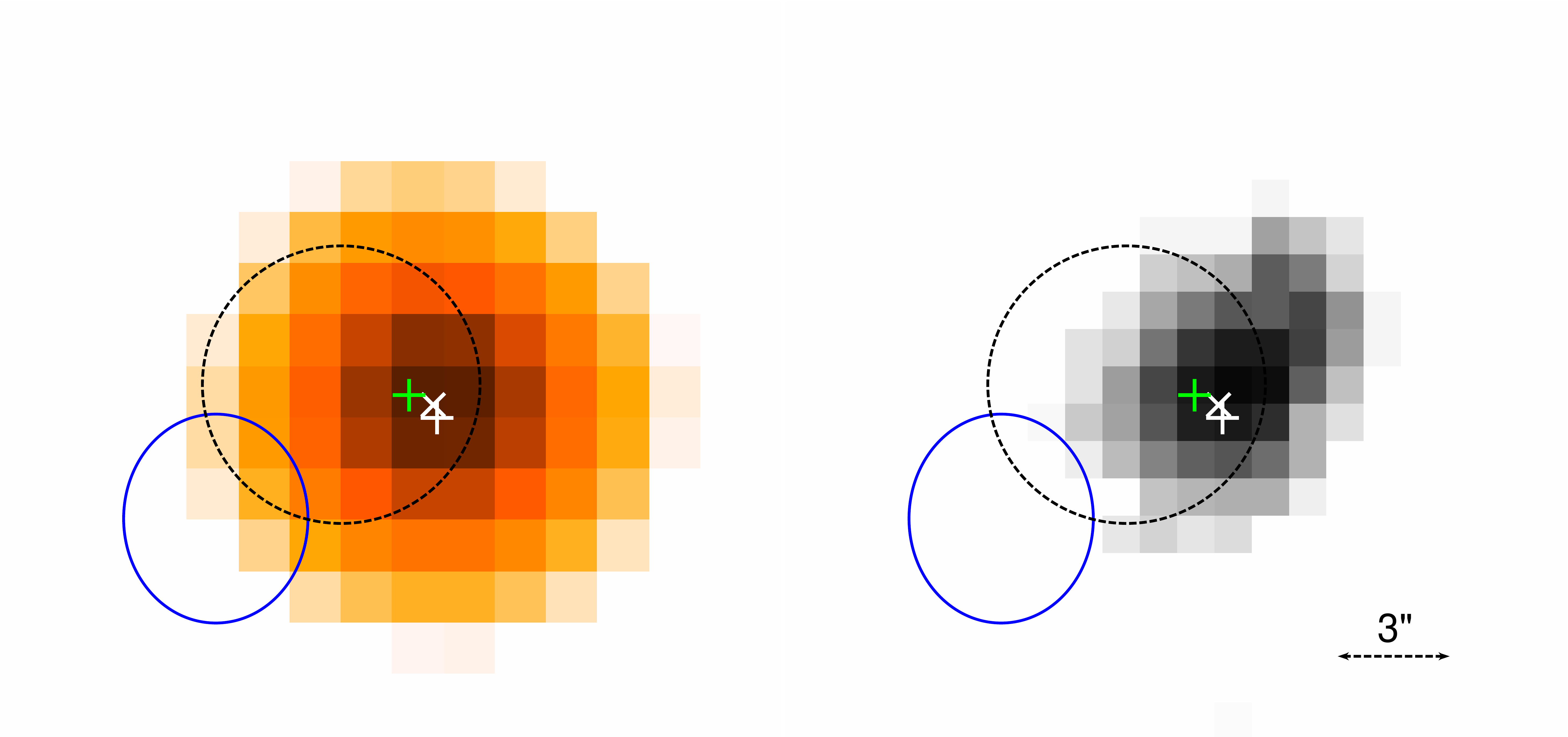}{0.49\textwidth}{(j - MRC~B0743$-$673)}}
\gridline{\fig{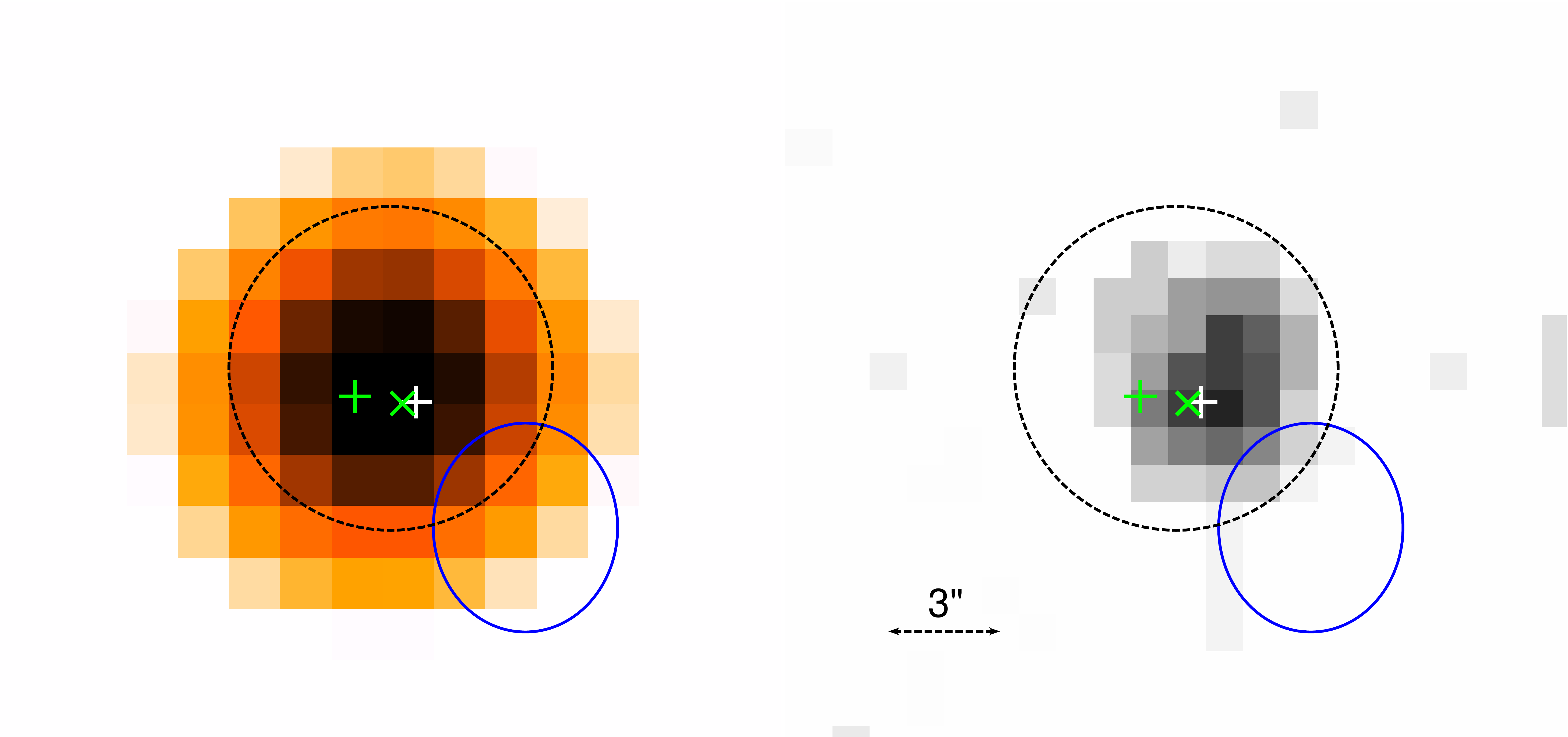}{0.49\textwidth}{(k - MRC~B0842$-$754)} \fig{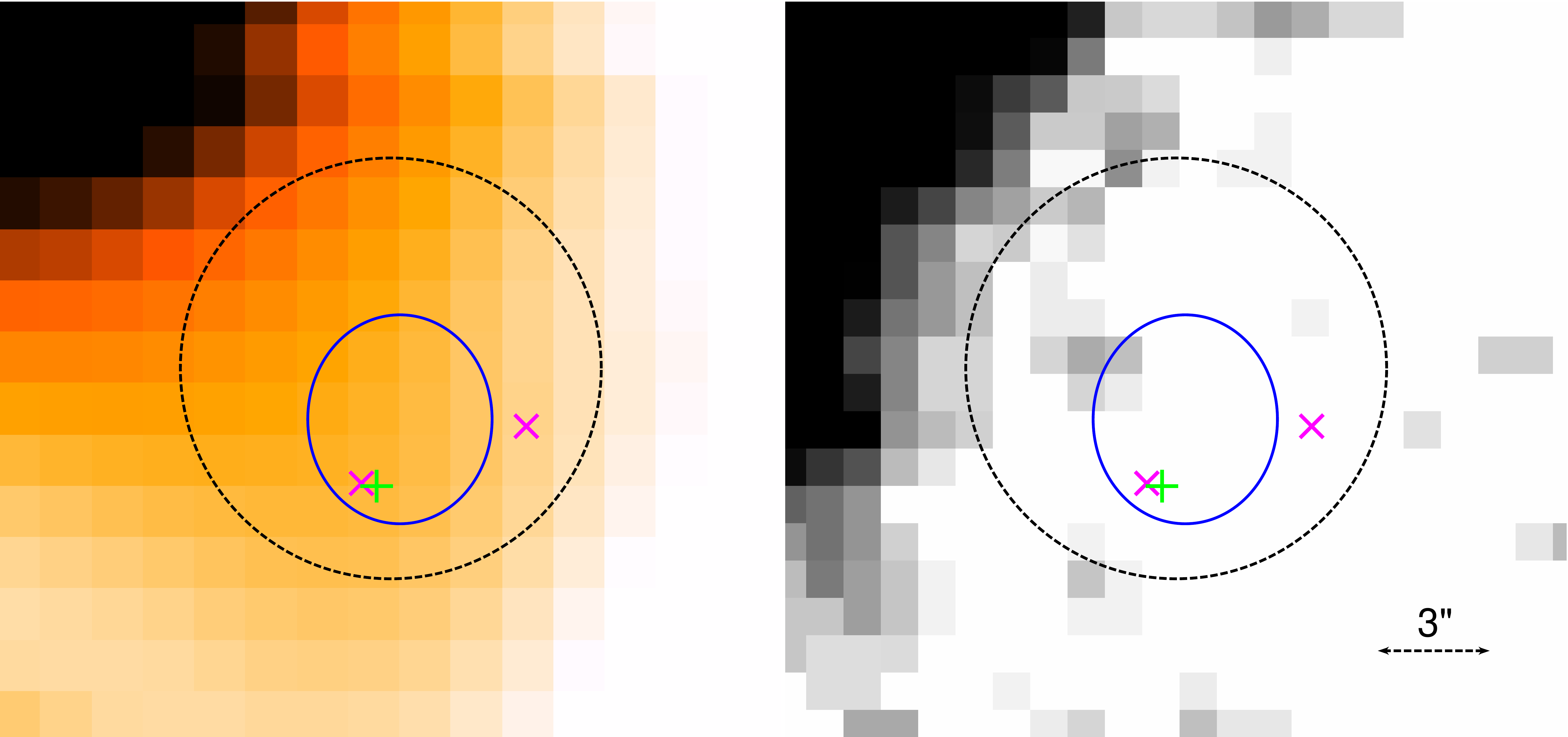}{0.49\textwidth}{(l - MRC~B0906$-$682)}}
\gridline{\fig{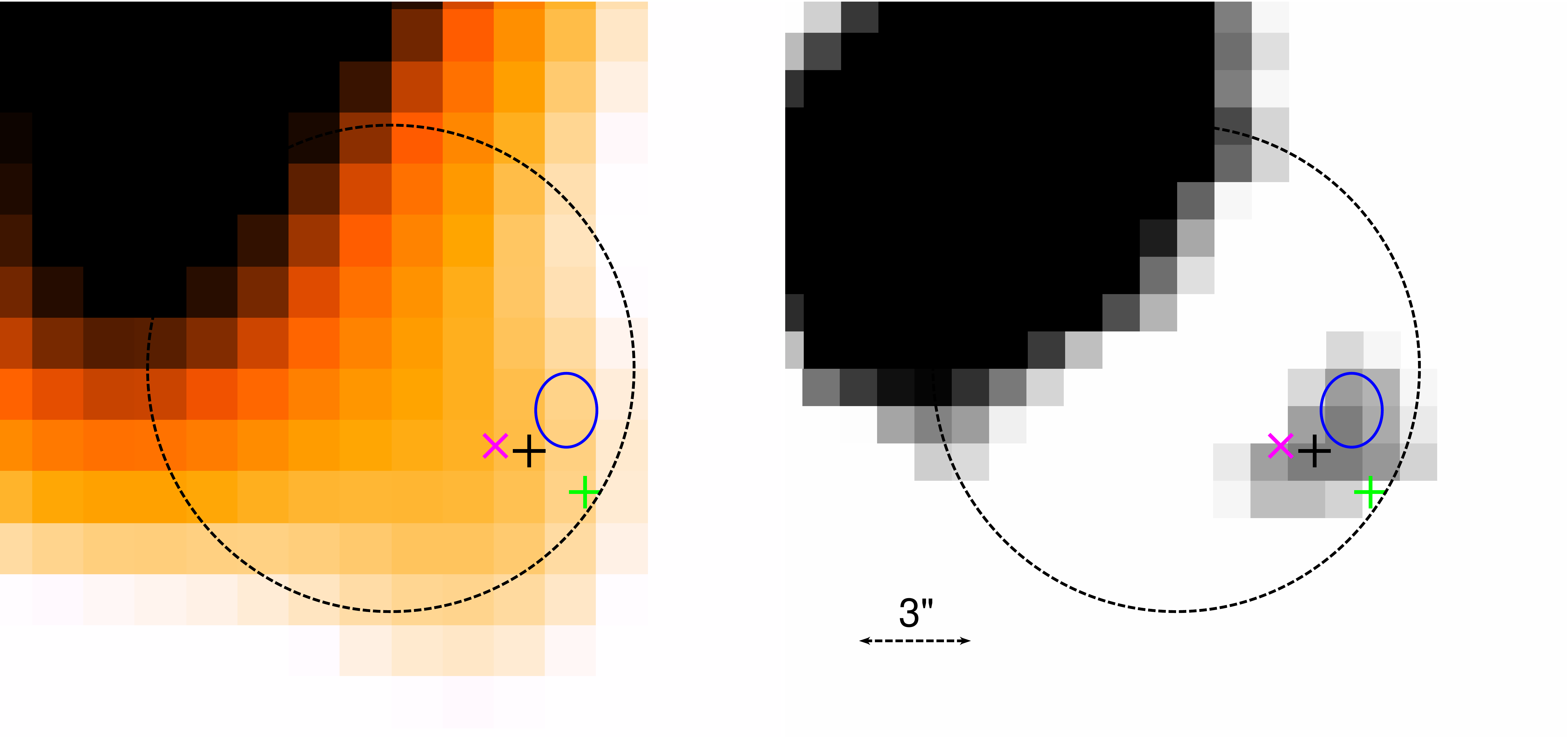}{0.49\textwidth}{(m - MRC~B1030$-$340)} \fig{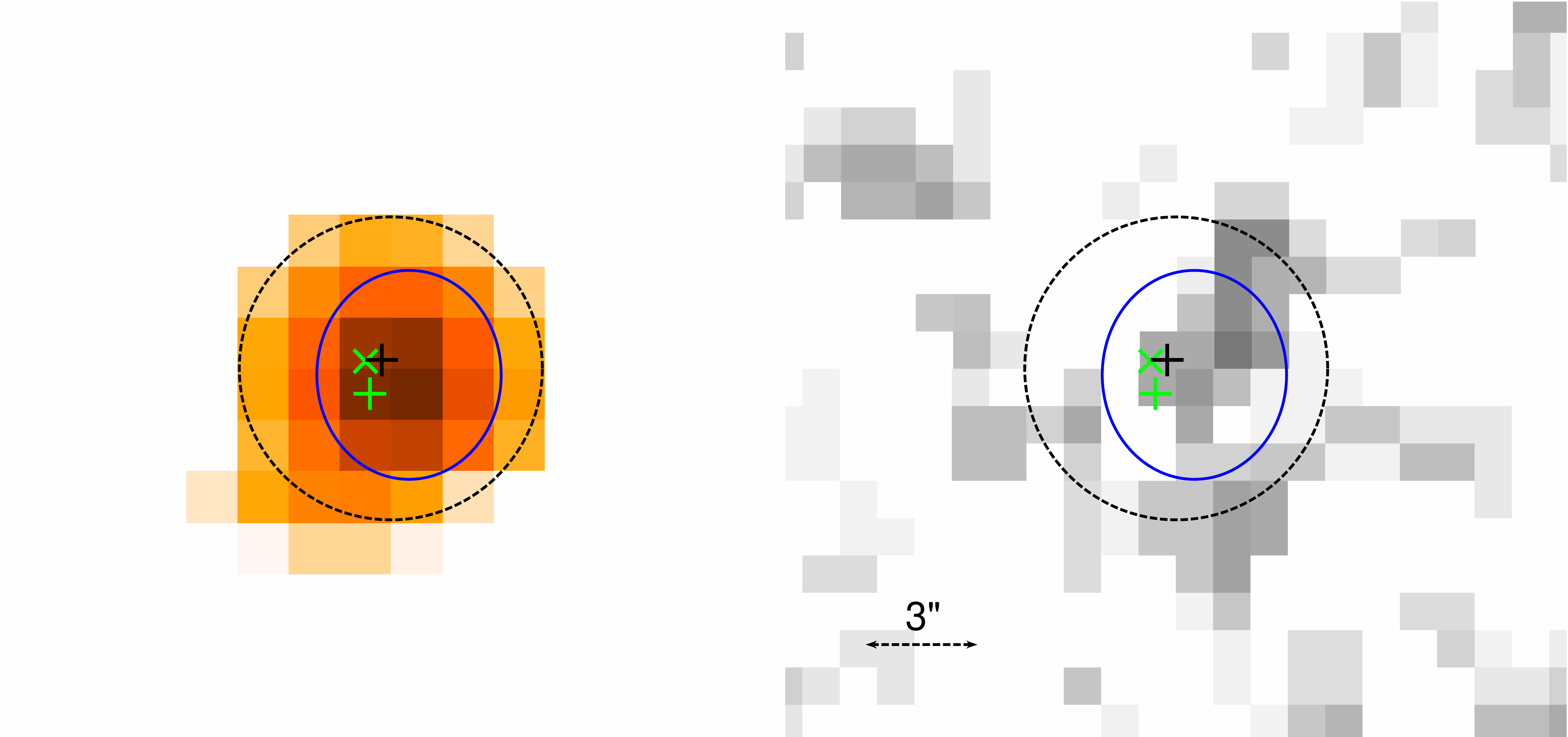}{0.49\textwidth}{(n - MRC~B1143$-$483)}}
\caption{{\it (Continued.)}}
\end{figure*}

\setcounter{figure}{1}
\begin{figure*}
\gridline{\fig{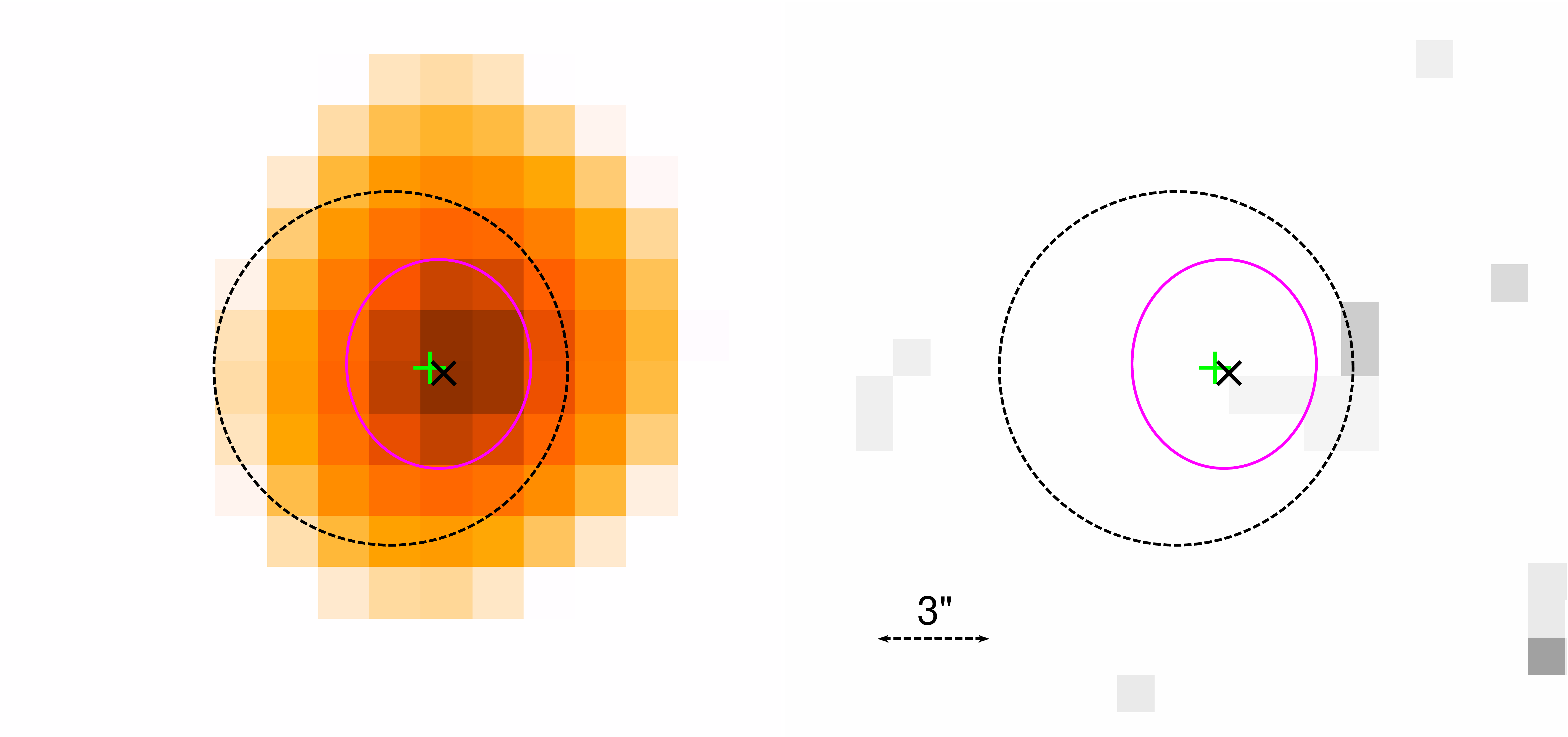}{0.49\textwidth}{(o - MRC~B1346$-$391)} \fig{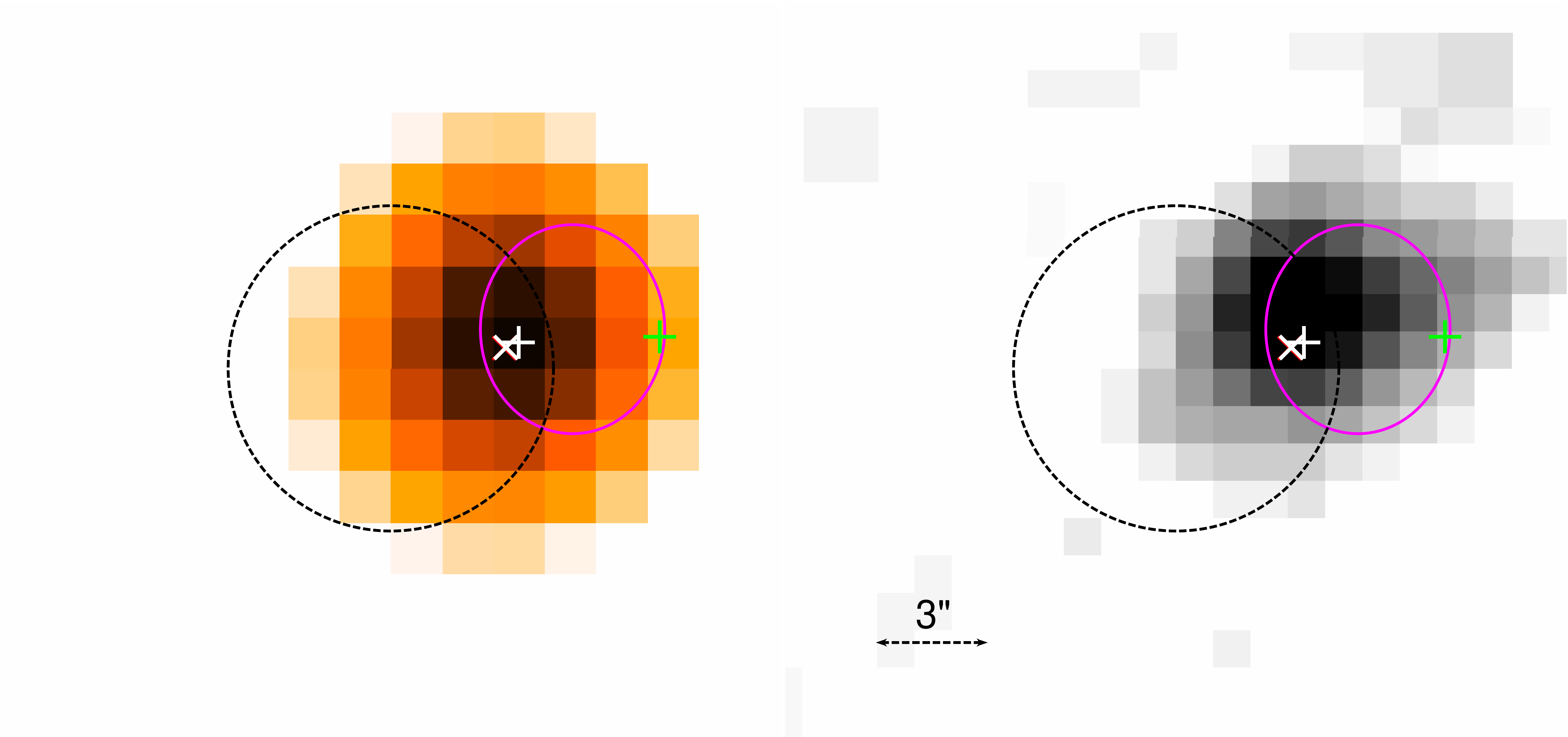}{0.49\textwidth}{(p - MRC~B1358$-$493)}}
\gridline{\fig{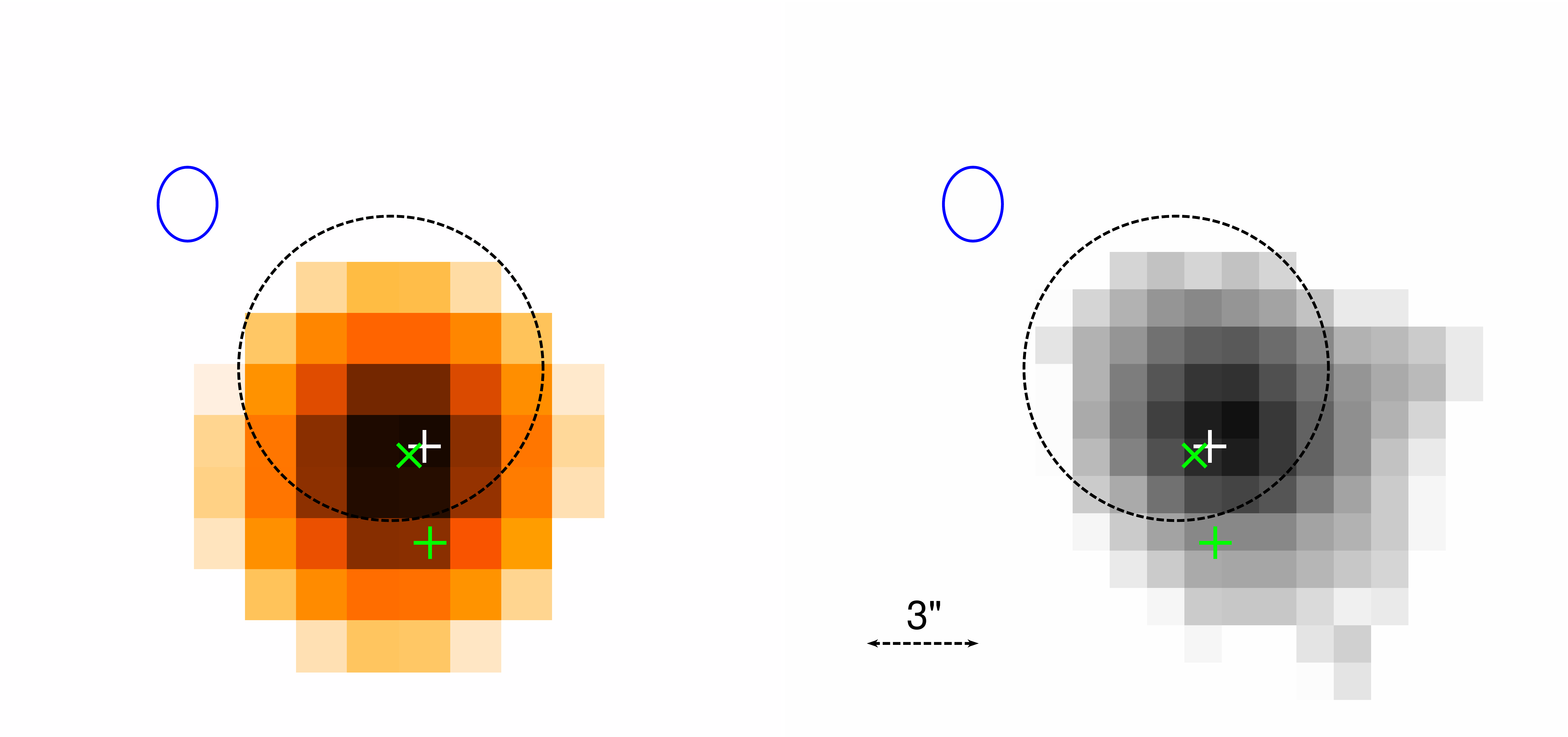}{0.49\textwidth}{(q - MRC~B1413$-$364)} \fig{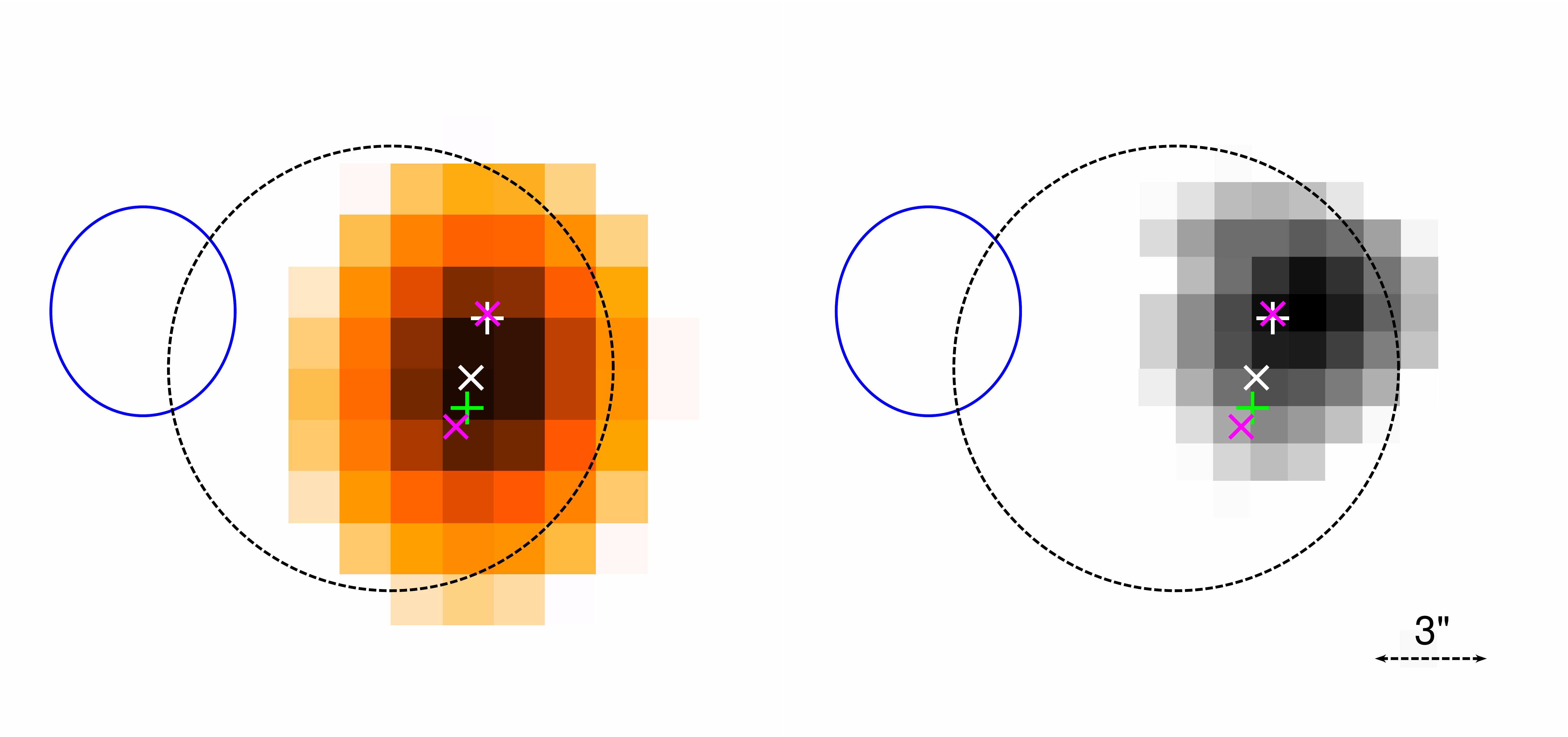}{0.49\textwidth}{(r - MRC~B1445$-$468)}}
\gridline{\fig{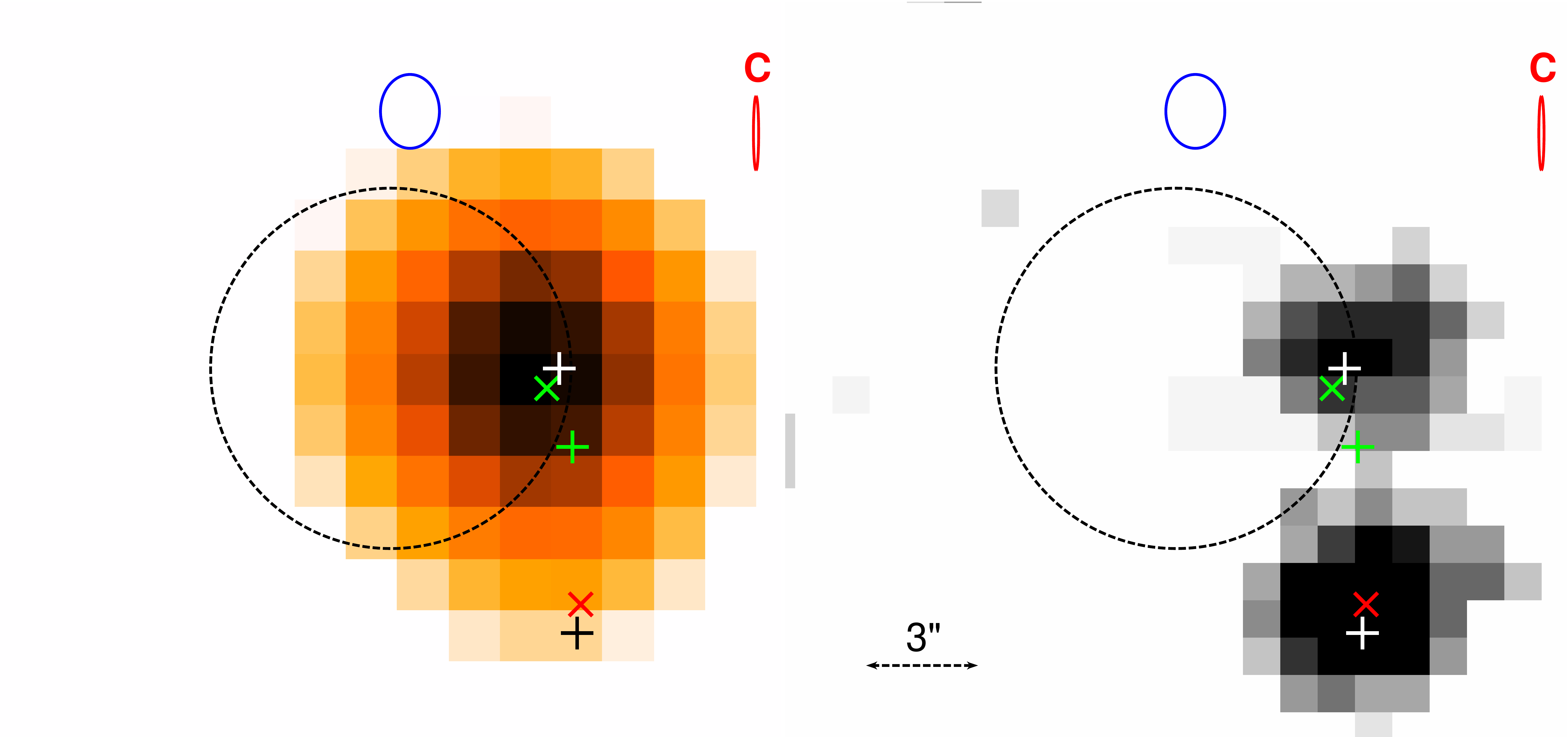}{0.49\textwidth}{(s - MRC~B1451$-$364)} \fig{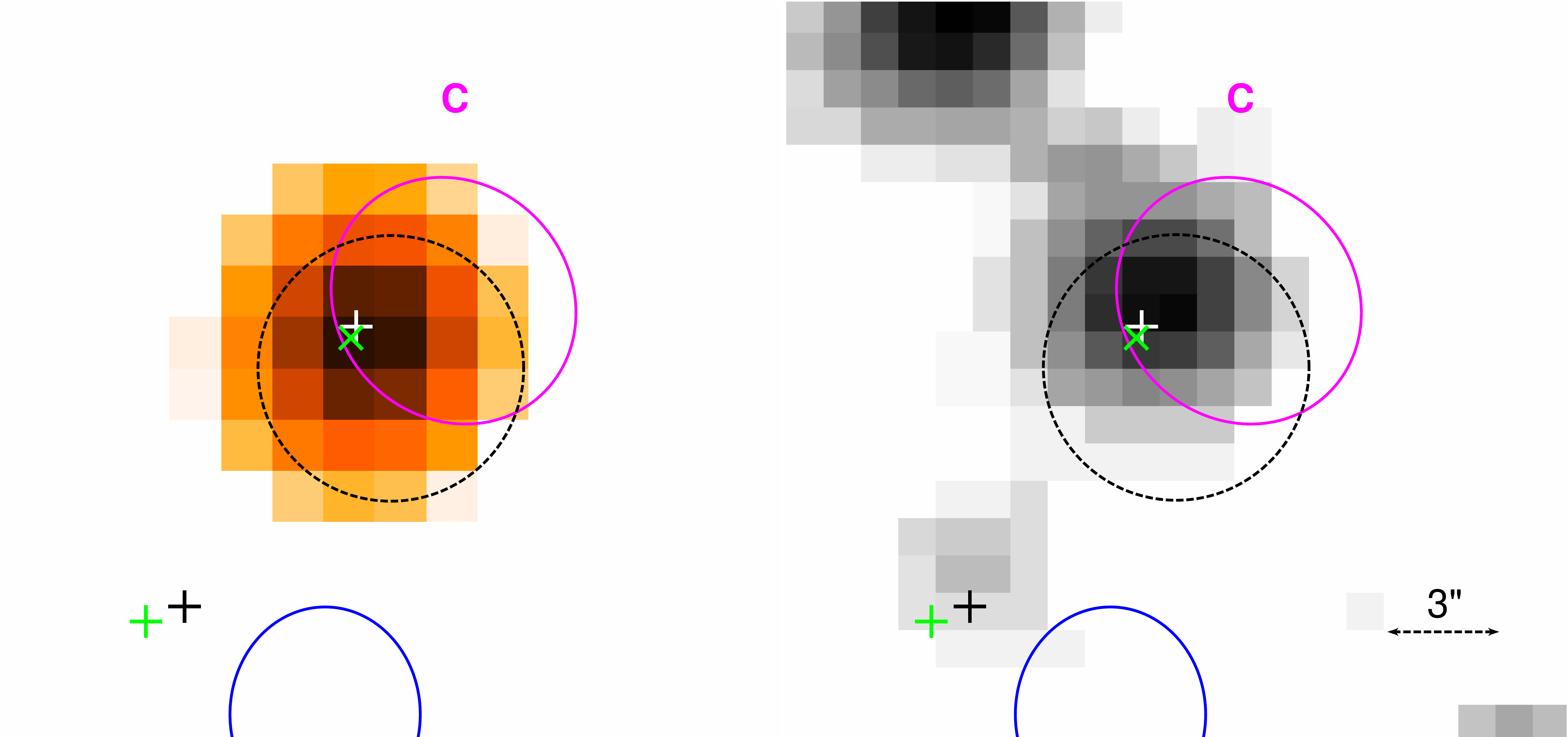}{0.49\textwidth}{(t - MRC~B1737$-$609)}}
\caption{{\it (Continued.)}}
\end{figure*}

\section{Multifrequency Analysis} 
\label{sec:5}

In this section, we first discuss the matching of the SMS4 radio sources with the detected \sw~sources and establish the radio/X-ray association when the radio and the X-ray positional uncertainties overlap. We show that the probability of chance coincidence between a \sw~detection and an SMS4 source is negligible. We then discuss the WISE and optical identifications, mainly utilizing the X-ray position. The association with infrared or optical sources is allowed only if their coordinates lie within the corresponding X-ray positional uncertainty.

All the  sources in our sample were classified by BH06 as quasars and radio galaxies.
In the optical band, emission from the AGN itself, but likely also from the host galaxy, can be detected.
In type II AGNs, the nuclear optical emission might be absorbed and obscured, but the infrared is more likely to reach the observer.
In our efforts to localize the core of the AGN, with our multifrequency analysis, we adopt a simple but conservative criterion, requiring a detection in both the infrared and optical bands.

As a preliminary step to match our X-ray detections with infrared and optical counterparts, we used the SkyView Virtual Observatory to retrieve infrared maps in the W1 filter (3.4~$\mu$m) from the AllWISE Data Release \citep{2014yCat.2328....0C} Images Atlas and optical maps in the $r$~filter (0.62~$\mu$m) from the Space Telescope Science Institute (STScI) $2^{nd}$ Digitized Sky Survey (DSS2).
With these maps, we built the panels shown in Fig.~\ref{fig:02}.

Next, we used TOPCAT to crossmatch our list of X-ray detections with sources from selected radio, infrared, and optical catalogs.


\subsection{Identifying X-Ray Sources with Radio Sources}
\label{sec:5.1}

To reliably associate our 20 X-ray-detected sources, listed in Table~\ref{tab:02}, with corresponding radio sources, we matched the source positions at these two energy ranges.
We used G4Jy (W20) as the main matching radio catalog for the 18 sources that were surveyed by GLEAM, and SUMSS for the remaining two sources (see Table~\ref{tab:01}).
Following W20, the typical rms positional uncertainties for G4Jy sources are $\sigma_{\alpha,S}$ $\approx$ 1\farcs5, $\sigma_{\delta,S}$ $\approx$ 1\farcs7, when their brightness-weighted centroids were computed after a cross-correlation with SUMSS data, while they are $\sigma_{\alpha,N}$ $\approx$ 0\farcs5, $\sigma_{\delta,N}$ $\approx$ 0\farcs6, when the same operation was based on NVSS data.

To match the radio and X-ray positions, we conservatively use a circular confidence region that is sufficiently large to ensure, with high probability, that the radio source lies within the X-ray positional error region.
For this circular region, we use a radius $r_X$ corresponding to the error radius, at 90\% confidence level, for sources at the X-ray limit of sensitivity, to which we add the largest uncertainty $r_r$ for the radio position, at the same confidence level as for the X-rays.
For the X-ray band, we find $r_X=6\farcs5$ from Table~\ref{tab:02}.
For the radio band, we take the larger between $\sigma_{\alpha,S}$ and $\sigma_{\delta,S}$, and multiply it by a factor of 1.645, derived from the Normal distribution, obtaining $r_r=2\farcs8$.
Using a circle with $R=r_r+r_X=9\farcs3$, we find that the combined 90\% confidence positional uncertainty totally includes the corresponding G4Jy or SUMSS positional uncertainty for all \sw-XRT detections.
For MRC~B1737$-$609, we assume the SUMSS C component matches the core, as discussed in Section~\ref{sec:3}, rather than the G4Jy position; in any case, also taking into account the G4Jy centroid itself, the X-ray positional uncertainty covers half of that in the radio band.

After obtaining matches for all our X-ray-detected sources with corresponding radio sources, we computed the probability that these matches might be due to chance.
As a first step, we used WebPIMMS to estimate the flux in the 0.5--2.0 keV band for sources at the limit of sensitivity.
We converted the lowest count rate ($1.3\times10^{-3}$ counts~s$^{-1}$) obtained from our XRT detections, using a power-law model and adopting reasonable spectral assumptions ($n_H = 4\times10^{20}$ cm$^{-2}$, $\Gamma = 1.7$), to obtain $F_{0.5-2~keV} \simeq 2\times10^{-14}$ erg~cm$^{-2}~$s$^{-1}$.
Using the log$N$-log$S$ distribution published by \cite{2004AdSpR..34.2470G}, we derive a density of $\sim30$ sources deg$^{-2}$ brighter than our flux limit. 
Considering that the XRT FOV corresponds to $\sim 5.6\times10^{2}$ arcmin$^2$, $\approx$5 random X-ray sources are expected in each XRT FOV.

In the conservative case of a combined, 90\% confidence positional uncertainty given by a circle with a radius $R=9\farcs3$, five serendipitous X-ray sources cover an area $S_5 \sim 0.38$ arcmin$^2$.
This implies that the chance of a radio/X-ray overlap, in any one field, is at most $P\sim S_5$/FOV$\sim 6.7\times10^{-4}$.
In our complete sample of 31 fields, the expected number of spurious matches is $31 \times P = 0.021.$
Hence, the probability of spurious coincidences between X-ray and radio sources is very small.


\subsection{Crossmatches with Infrared and Optical Catalogs}
\label{sec:5.2}

In the infrared band, the main catalog that we adopted in our crossmatch with the \sw~sources is AllWISE. 
The typical positional uncertainties are in the range 0\farcs03--0\farcs08, with only one source (J001030.14$-$442257.1) with an uncertainty of 0\farcs3. 
In addition, we searched for counterparts in the CatWISE2020 \citep{2021ApJS..253....8M} catalog, which benefits from much longer exposure times and improved source detection algorithms with respect to AllWISE.

In the optical band, the main catalog that we used for source identification is the Second Generation Guide Star Catalog (GSC~2.3.2) by \cite{2008AJ....136..735L}, which has a typical positional uncertainty in the 0\farcs26--0\farcs40 range.
In addition, we searched the Dark Energy Survey (DES) Data Release 2 (DR2; \citealp{2021ApJS..255...20A}).

In our analysis, we took into account the infrared color of our WISE candidates, comparing W1$-$W2 with the threshold (W1$-$W2$~\geq~0.8$~mag) established by \cite{2012ApJ...753...30S} in their simple criterion for selecting AGNs.
Furthermore, we searched for our candidates in previously selected AGN  \citep{2012ApJ...751...52E,2015ApJS..221...12S} and quasar \citep{2019A&A...624A.145S} samples, to strengthen the hypothesis of their extragalactic nature.

As a result of our analysis, we find at least one candidate counterpart, detected in both the infrared and the optical bands, for all our SMS4 sources.
However, we distinguish two classes according to the reliability of the optical and WISE counterparts.
Class~A sources are those whose candidate counterpart is unique since it lies within the XRT positional uncertainty at the 90\% confidence level and shows typical AGN colors.
Class~B sources are those having more than one plausible candidate within, or in the proximity of, the XRT positional uncertainty (90\% c.l.).
All our candidate counterparts are listed in Table~\ref{tab:07}.

To estimate the probability that our counterparts, all with a detection in both the infrared and the optical bands, are by chance aligned with the corresponding X-ray source, we can consider the AllWISE infrared sources alone.  
The W1 magnitudes of our sources lie in the 11.5$-$17.1~mag range (see Table~\ref{tab:07}), with a mean magnitude of W1 = 14.1~mag. 
From the Explanatory Supplement to the AllWISE Data Release Products (\citealp{2013wise.rept....1C}; see Figure~6a at \url{https://wise2.ipac.caltech.edu/docs/release/allwise/expsup/sec2_1.html}), we derived the density of sources with W1 $\le$ 14~mag, which is $\sim1.1\times10^{3}$ sources deg$^{-2}$. 
In a circle with a radius equal to $R$ (see Section~\ref{sec:5.1}), $\sim2.29\times10^{-2}$ sources are then expected. 
For the faintest infrared counterpart (MRC~B0007$-$446) with magnitude W1 = 17.1~mag, the IR source density to this faint limit implies 0.24~sources per $R=9\farcs3$ radius circle.

\begin{table*} 
\begin{center}
\scriptsize
\caption{Our Candidate Counterparts for X-Ray-Detected SMS4 Sources}
\label{tab:07}
\begin{tabular}{ccccccc}
\hline
   (1)                 &  (2)  &             (3)             &  (4)  &   (5)     &            (6)               &    (7)   \\  
SMS4 Name              & Class &      Infrared Source        &  W1   &  W1$-$W2  &        Optical Source        & $\Delta$ \\
                       &       &                             & (mag) &  (mag)    &                              & (arcsec) \\
\hline                                                                                                                                            
  B0007$-$446          &   A   & ~~AW J001030.14$-$442257.1$^a$ & 17.09 &  0.59   &  DES J001030.13$-$442257.4  &    0.34  \\
  B0013$-$634          &   A   & ~~AW J001602.93$-$631003.8$^a$ & 14.65 &  0.49   &  GSC2.3 S19X010969          &    1.03  \\
  B0049$-$433          &   A   & ~~AW J005214.91$-$430629.0$^a$ & 14.29 &  1.02   &  GSC2.3 S2SX008795          &    0.50  \\
  B0157$-$311          &   A   & ~~AW J020012.17$-$305326.6$^a$ & 13.35 &  1.06   &  GSC2.3 S2YS004732          &    0.11  \\
  B0219$-$706          &   A   & ~~AW J022008.11$-$702229.3$^a$ & 14.14 &  1.24   &  GSC2.3 S1G6004776          &    0.48  \\
~~B0420$-$625$^g$      &   B   &   CW J042056.03$-$622339.8     & 16.19 &  0.34   &  BH06 J042056.09$-$622339.1 &    0.76  \\
  B0453$-$301          &   A   & ~~AW J045514.26$-$300648.5$^a$ & 13.94 &  0.47   &  GSC2.3 S239004216          &    0.14  \\
~~B0534$-$497$^g$      &   B   &   AW J053613.56$-$494426.8     & 14.21 &  0.28   &  BH06 J053613.61$-$494426.7 &    0.49  \\
~~B0547$-$408$^b$      &   A   & ~~AW J054924.17$-$405116.2$^a$ & 15.06 &  1.52   &  GSC2.3 S1ZG001633          &    0.58  \\
  B0743$-$673          &   A   &   AW J074331.61$-$672625.5     & 13.16 &  0.93   &  GSC2.3 S4MI002100          &    0.37  \\
  B0842$-$754          &   A   & ~~AW J084127.07$-$754027.7$^a$ & 13.12 &  0.94   &  GSC2.3 S43L002329          &    0.36  \\
  B0906$-$682          &   A   &   CW J090652.71$-$682940.2     & 15.84 &  0.34   &  BH06 J090652.64$-$682940.3 &    0.41  \\
  B1030$-$340          &   A   &   CW J103313.22$-$341844.9     & 15.34 &  0.65   &  GSC2.3 S5M4020229 	        &    0.89  \\
  B1143$-$483          &   A   & ~~AW J114531.05$-$483610.0$^a$ & 13.67 &  1.11   &  GSC2.3 S45R029125 	        &    0.42  \\
  B1346$-$391          &   A   &   AW J134951.08$-$392251.3     & 14.90 &  0.60   &  BH06 134951.12$-$392251.2  &    0.40  \\
  B1358$-$493          &   A   &   AW J140131.57$-$493235.5     & 11.74 &  1.16   &  GSC2.3 S96M055339          &    0.36  \\
~~B1413$-$364$^c$      &   A   & ~~AW J141633.15$-$364053.7$^a$ & 12.29 &  1.03   &  GSC2.3 S9AR000360          &    0.48  \\
~~B1445$-$468$^d$      &   A   &   AW J144828.18$-$470141.6     & 13.78 &  0.68   &  GSC2.3 S9DJ002907          &    1.55  \\
~~B1451$-$364$^e$      &   A   & ~~AW J145428.22$-$364004.7$^a$ & 14.52 &  0.25   &  GSC2.3 S9H2063281          &    0.63  \\
~~B1737$-$609$^f$      &   A   & ~~AW J174201.49$-$605512.2$^a$ & 11.55 &  0.94   &  GSC2.3 S7FE029330          &    0.34  \\
\hline                                                                                       
\end{tabular}
\end{center}
\tablecomments{\scriptsize{Column (1): the SMS4 name, according to the MRC designation. Column (2): the class, reflecting the reliability of the candidate's association (see Section~\ref{sec:5.2}). Column (3): the infrared source identification (AW: AllWISE, \citealp{2014yCat.2328....0C}; CW: CatWISE2020, \citealp{2021ApJS..253....8M}). Column (4): the WISE magnitude in the W1 filter. Column (5): the WISE W1$-$W2 infrared color. Column (6): the optical source identification. Column (7): the angular separation $\Delta$ between the given infrared and optical sources.\\
$^a$ This infrared source was previously associated by W20. \\
$^b$ Bright infrared source, at the boundaries of the XRT 99\% positional uncertainty, with WISE AGN-like colors. \\
$^c$ Extended radio galaxy (see Figure~\ref{fig:01}), with low-frequency G4Jy centroid not reliably matching the core; the BH06 optical counterpart lies 2\farcs6 from the GSC~2.3.2 source. \\
$^d$ The BH06 counterpart lies $\sim2\farcs5$ from the GSC~2.3.2 source; the AllWISE source lies between these optical sources. \\
$^e$ Extended radio galaxy (see Figure~\ref{fig:01}), with low-frequency G4Jy centroid not reliably matching the core; the XRT positional uncertainty (90\% c.l.) includes only one of two close sources. \\
$^f$ Candidate matching the SUMSS C component of an extended radio galaxy, consistent with its radio core.\\
$^g$ As discussed in the text, there are two viable counterparts. The given source IDs are those that include, as an optical counterpart, the one suggested in BH06 (the green plus sign in Fig.~\ref{fig:02}).\\
}
}
\end{table*}


\subsection{Comparison with Earlier Studies}
\label{sec:5.3}
 
While all infrared counterparts associated by W20 are AllWISE sources, optical counterparts were associated by BH06 using either the plates from the UK Schmidt Southern Sky Survey or $R$-band CCD images from a dedicated campaign at the AAT.
We notice that, despite an overall agreement between the coordinates reported in the GSC~2.3.2 catalog and those reported by BH06, they do not strictly coincide.
In four cases (MRC~B1358$-$493, MRC~B1413$-$364, MRC~B1445$-$468, and MRC~B1451$-$364) the angular separations between GSC~2.3.2 and the optical coordinates given by BH06 exceed 2\arcsec, and thus do not allow us to conclude that they are the same astrophysical objects.
Thus, when both the GSC~2.3.2 and the BH06 sources are available and they most likely refer to the same object, we validate the BH06 counterpart, even though we report the GSC~2.3.2 source in Table~\ref{tab:07}.
As discussed below in this section, we treat MRC~B1445$-$468 in the same way; for the remaining three aforementioned sources, we validate the GSC~2.3.2 source as the optical counterpart, rather than the BH06 association.

W20 associated an infrared counterpart to 13 of the 20 SMS4 sources for which \sw~detected X-ray emission, while optical counterparts were associated by BH06 with each of them.
The W20 counterparts are reliably matched by the BH06 counterpart for 8 of the 13 SMS4 sources: MRC~B0013$-$634, MRC~B0049$-$433, MRC~B0157$-$311, MRC~B0219$-$706, MRC~B0453$-$301, MRC~B0547$-$408, MRC~B0842$-$754, and MRC~B1143$-$483.
In the remaining five cases, the match is questionable (MRC~B0007$-$446, MRC~B1413$-$364, and MRC~B1451$-$364) or does not occur (MRC~B0534$-$497 and MRC~B1737$-$609) owing to the large angular separation.

The counterparts of seven among the eight reliable matches described above lie within the corresponding XRT positional uncertainty regions (90\% c.l.).
Conversely, for MRC~B0547$-$408, W20 and BH06 counterparts lie 5\farcs7 from the boundary of this X-ray region. 
Typical AGN-like WISE colors characterize this infrared source.  
Within the same X-ray region, a faint infrared source from CatWISE2020 matches an optical source from DES~DR2. 
We note that the X-ray positional uncertainty, computed using eight counts, fully includes the radio G4Jy centroid, with its positional uncertainty (90\% c.l.).  
However, due to the double radio morphology of MRC~B0547$-$408 (see Section~\ref{sec:3} and Figure~\ref{fig:01}), the position of this G4Jy centroid, lying between the SUMSS lobe components, might not match the actual core position.
For these reasons, the bright WISE candidate, previously selected by W20, and its corresponding optical counterpart selected by BH06, might be the correct match to the radio and X-ray source, even though it lies $11\farcs3$ from the X-ray position.

Focusing on class A sources, there are 12 sources for which the optical and WISE/IR positions are in excellent agreement.
There are an additional six sources where the optical/WISE coordinates differ by more than $2\arcsec$ from the optical candidates given in BH06. 
For four WISE sources from W20, we find an alternative optical candidate to that in BH06, either in GSC~2.3.2 (MRC~B1413$-$364, MRC~B1451$-$364, and MRC~B1737$-$609) or in DES~DR2 (MRC~B0007$-$446). 
Although MRC~B1358$-$493 and MRC~B1445$-$468 have no WISE match listed in W20, we do find WISE counterparts for both sources and matching optical counterparts. 
We discuss each of these six sources:

\begin{itemize}
\item{MRC~B0007$-$446: as shown in Fig.~\ref{fig:02}a, we find an optical source, detected in DES~DR2 rather than in GSC~2.3.2, matching the W20 counterpart; no infrared counterpart is found for the BH06 candidate}. 
\item{MRC~B1358$-$493: missing in GLEAM EGC (see Section~\ref{sec:2}), W20 could not provide an infrared counterpart for this SMS4 source.
However, an AllWISE source with an infrared color (W1$-$W2 = 1.16~mag) fully consistent with AGNs, matched by a GSC~2.3.2 source, lies within the XRT positional uncertainty (90\% c.l.).
The coordinates given by BH06 are slightly different.
The AllWISE source was already included in the W2 sample of \cite{2012ApJ...751...52E}.}
\item{MRC~B1413$-$364: the coordinates of both the W20 infrared and the BH06 optical candidates (Fig.~\ref{fig:02}q) match a source that appears extended in the maps.
According to \cite{1993MNRAS.262..889S}, it is a galaxy at $z=0.075$.    
We find a GSC~2.3.2 source matching the W20 counterpart (W1$-$W2 = 1.03~mag), that was already included both in the W2 sample of \cite{2012ApJ...751...52E} and in the AGN catalog of \cite{2015ApJS..221...12S}. 
For this reason, we give this GSC source as the optical counterpart, rather than the BH06 choice that lies outside the XRT positional uncertainty}.
\item{MRC~B1445$-$468: the angular separation between the BH06 counterpart and the GSC~2.3.2 source is $\sim$2\farcs5: this, in principle, would not guarantee that these objects match each other.
However, a single infrared source, detected in AllWISE, lies between the aforementioned optical sources, at 0\farcs9 and 1\farcs6 from the BH06 and the GSC~2.3.2 source, respectively.
Thus, assuming a reliable association with the AllWISE source, whose infrared color ($W1-W2=0.68$~mag) is close to the 0.8~mag threshold, we have no strong evidence to discard the BH06 counterpart: similar to other cases, we validate this counterpart, reporting at the same time the GSC~2.3.2 source in Table~\ref{tab:07}.}
\item{MRC~B1451$-$364: the difficulty of determining the optical counterpart was first raised by the analysis carried out by \cite{1992ApJS...80..137J}, who quoted four different candidates. 
BH06 also noted more than one candidate.
The conclusions of both these studies were not in agreement, with BH06 supporting a {\it \textquotedblleft diffuse object\textquotedblright} among the four previously reported by \cite{1992ApJS...80..137J}.
The position of this BH06 candidate, marked by a green plus sign in Fig.~\ref{fig:02}s, lies in the middle of two GSC~2.3.2 sources.
The XRT positional uncertainty (90\% c.l.) includes only one of these, also matching the W20 infrared counterpart, AllWISE~J145428.22$-$364004.7, with W1$-$W2 = 0.25~mag.
The other GSC~2.3.2 source also matches an infrared source (CatWISE2020~J145428.15$-$364010.4), with infrared color W1$-$W2 = -0.02~mag, even lower than the AllWISE source.}
\item{MRC~B1737$-$609: the angular separation between counterparts given by W20 and BH06 (Fig.~\ref{fig:02}t) is $\sim$9\farcs5.
The \sw-XRT circular region of uncertainty at 90\% confidence level includes the infrared source selected by W20 (W1$-$W2 = 0.91~mag), also detected in the optical, at variance with the BH06 candidate.
Also in this case, the extragalactic nature of this infrared source is supported by its inclusion in both the \cite{2012ApJ...751...52E} and the \cite{2015ApJS..221...12S} AGN samples.
The same X-ray circle also matches the SUMSS component, named C in both Figure~\ref{fig:01} and Fig.~\ref{fig:02}t, corresponding to the core of the radio galaxy (see Section~\ref{sec:3}).
An optical spectrum, obtained by \cite{2017A&A...602A.124R} for the object corresponding to our AllWISE/GSC~2.3.2 counterpart, revealed broad emission lines over a continuum emission, leading to a redshift $z=0.152$.}
\end{itemize}

In addition to the eight sources with reliable BH06-W20 counterpart matches, the seven remaining SMS4 sources for which we validate the BH06 counterparts are five class~A sources (MRC~B0743$-$673, MRC~B0906$-$682, MRC~B1030$-$340, MRC~B1346$-$391, and MRC~B1445$-$468) and the two class~B sources (MRC~B0420$-$625 and MRC~B0534$-$497).
To these five class~A sources, and also to the already-mentioned MRC~B1358$-$493, we provide an infrared counterpart for the first time, since they were not associated by W20.

Finally, we describe in greater detail the two remaining class~B sources, for which we could not establish firm conclusions about the counterpart.

\begin{itemize}

\item{MRC~B0420$-$625: as shown in Fig.~\ref{fig:02}f, neither an infrared source in the AllWISE Data Release nor an optical source in the GSC~2.3.2 was detected for MRC~B0420$-$625 within the XRT error circle.
However, two infrared sources were detected in CatWISE2020, and both are coincident with optical sources detected in DES~DR2. 
One of these infrared sources, CatWISE2020~J042056.03$-$622339.8 (W1$-$W2 = 0.34~mag), is closer to the center of the source that appears extended in the infrared map, and it matches the optical counterpart found by BH06.
The other source, CatWISE2020~J042056.55$-$622337.1 ($W1-W2=0.29$~mag), lies at the outskirts of the extended source.
Additional information is needed to determine the counterpart.}

\item{MRC~B0534$-$497: both maps in Fig.~\ref{fig:02}h show an extended source, with two AllWISE sources separated by 5\farcs5, on opposite sides of the radio ellipse of G4Jy\#563.
One of the AllWISE sources, AllWISE~J053613.90$-$494422.2 (W1$-$W2 = 0.71~mag): marked by a green cross, corresponds to the counterpart given by W20, and is matched by a DES~DR2 source ($r$=21.1~mag).
The other source is AllWISE~J053613.56$-$494426.8 (W1$-$W2 = 0.28~mag): shown in white, it lies at the center of the extended source in the maps and matches a much brighter optical source ($R$=16.5~mag), detected in GSC~2.3.2; it corresponds to the optical counterpart suggested by BH06.}

\end{itemize}


\section{Summary and Conclusions} 
\label{sec:6}

In 2006, the SMS4 was compiled by extrapolating the flux density, measured at higher frequencies, to 178~MHz, rather than by using actual measurements at that frequency.
Since 2020 the G4Jy \citep{2020PASA...37...18W} has been available, and we crossmatched the SMS4 sample with the G4Jy sample to establish correspondences between sources. 
For the 10 SMS4 sources lacking matches in the G4Jy, data collected from other low radio frequency catalogs were retrieved. 
Based on these data, we establish a fraction of $\sim$35\% of SMS4 sources with flux density at 178~MHz (or comparable frequency) lower than 10.9~Jy, at variance with the extrapolated values.
This result confirms and quantifies the intrinsic selection effect in the SMS4 sample and definitely encourages the use of G4Jy for future comparison with the 3CRR in the Northern Hemisphere.

In 2015, we obtained observations with \sw~for 24 sources classified as radio galaxies by \cite{2006AJ....131..114B}, but not yet observed by either \cha, \sw, or \xmm.
We complemented this list with the seven SMS4 sources observed only by \sw, among the just mentioned missions, and report here detections with the \sw-XRT for 20 of the 31 sources in our sample; 6 of these 20 correspond to less luminous radio sources, with $\overline{S}_{181} \leq 10.9$ Jy.
Furthermore, we highlight the presence of diffuse X-ray emission in the FOV of PKS~B2148$-$555, which is the BCG in the Abell cluster A3816 \citep{2002MNRAS.331..717L}.

The count rate for eight of our X-ray detections is higher than 10$^{-2}$~counts~s$^{-1}$; two of these, MRC~B1413$-$364 and MRC~B1737$-$609, are also found in the \sw-BAT catalogs of hard X-ray sources.
For these eight sources, and also for MRC~B1346$-$391, which is close to this rate level, the statistics allowed us to perform a more detailed analysis, investigating the extent of the X-ray emission, the hardness ratio, and the properties of the spectrum in the 0.3--10~keV band.

In addition to PKS~B2148$-$555, we find evidence of deviation from a point-like emission for MRC~B0743$-$673 and MRC~B1346$-$391.

The $HR$ values of MRC~B1358$-$493 and MRC~B1413$-$364 show that their spectrum in the 0.3--10~keV band is hard; note that MRC~B1413$-$364 is one of the two sources included in the \sw-BAT catalogs.
The results of the spectral analysis show that a fit with a power law, fixing the hydrogen column density $n_H$ to the Galactic value, adequately describes the spectrum for the investigated sources.
However, for MRC~B1413$-$364, an improvement in the fit results after leaving $n_H$ free to vary is found.
We consider this improvement as the sign of intrinsic absorption in the lower energy range of the 0.3--10 keV band.

We matched the 20 X-ray detections with infrared and optical catalogs and required a detection in both the infrared and the optical bands to establish a counterpart at lower frequencies for our X-ray detections.
Based on the analysis of the available information, we rank the 20 SMS4 sources with an XRT detection according to the reliability of the candidate counterpart and establish 18 class~A and two class~B sources. 
Class~A sources have reliable unique candidates; two class~B sources (MRC~B0420$-$625 and MRC~B0534$-$497) each have two possible candidates and thus need further investigation to determine the correct counterpart.

Comparing our results with the counterparts previously proposed by W20 and BH06 for class~A sources, our analysis confirms all 12 infrared counterparts provided by W20 and 13 of the 20 optical counterparts provided by BH06.
Thus, we associate new infrared counterparts for six sources and five optical alternatives to BH06:
\begin{itemize}
\item{For MRC~B0013$-$634, MRC~B0049$-$433, MRC~B0157$-$311, MRC~B0219$-$706, MRC~B0453$-$301, MRC~B0547$-$408, MRC~B0842$-$754, and MRC~B1143$-$483, these infrared and optical counterparts match each other.}
\item{For MRC~B0007$-$446, MRC~B1413$-$364, MRC~B1451$-$364, and MRC~B1737$-$609, our analysis supports the infrared identifications proposed by W20.
For three of these (MRC~B0007$-$446, MRC~B1413$-$364, and MRC~B1451$-$364), we find an optical source, matching the IR counterpart, whose coordinates differ by more than 2\farcs5 from the coordinates of the optical counterparts proposed by BH06, but likely refer to the same object.
For MRC~B1737$-$609, the optical counterpart proposed by BH06 refers to a completely different object from the counterparts that we give, including the infrared source provided by W20 (see Fig.~\ref{fig:02}t).
BH06 relied on the radio location alone and the extended radio structure for MRC~B1737$-$609 yielded an incorrect location for the radio core, while the \sw~X-ray position correctly identified the AGN core.}
\item{For MRC~B1358$-$493, the position given by BH06 lies outside the XRT error circle: our analysis supports a counterpart detected in both AllWISE and GSC~2.3.2, lying in the overlap between the radio and X-ray error regions; in this case, for the first time, we associate an infrared counterpart with this radio source}.
\item{For MRC~B0743$-$673, MRC~B0906$-$682, MRC~B1030$-$340, MRC~B1346$-$391, and MRC~B1445$-$468, not only we do confirm the optical counterparts given by BH06, but we also find for them a matching infrared source in WISE catalogs, providing in this way an infrared counterpart also for these five radio sources for the first time.}
\end{itemize}

In conclusion, 18 class~A SMS4 sources have candidates detected in both the infrared and optical bands, listed in Table~\ref{tab:07}, supporting the multifrequency emission that is expected to characterize quasars and radio galaxies that constitute our sample.
Nine infrared objects, corresponding to our candidate counterparts, had already been included in the sample of $\approx$1.4~million AGNs assembled by \cite{2015ApJS..221...12S}; three additional sources were found in the \cite{2012ApJ...751...52E} and in the \cite{2019A&A...624A.145S} samples, for a total of 12 sources.

The results that we have described fill a gap in our knowledge of the X-ray view of powerful radio sources in the Southern Hemisphere, and identify the most promising sources to be investigated further with narrow-field imaging X-ray instruments. 
Since 2015, when we compiled the list of 45 sources not yet observed by current X-ray observatories with high spatial resolution, excluding the 24 sources for which we obtained \sw~observations, only one source, MRC~B1706$-$606, has been observed by \cha, in 2020.

We provide in the Appendix~\ref{app:1} (Table~\ref{tab:08}) the list of the remaining 56 SMS4 sources that in 2021 September still lacked \cha, \sw, or \xmm~observations.  
After sorting this list by $\overline{S}_{181}$, in 2021 October we successfully proposed \sw~observations for the 18 brightest radio sources.
Their $\overline{S}_{181}$ values are in any case higher than 10.9~Jy and vary in the range between 12.7~Jy (MRC~B2140$-$817) and 34.6~Jy (MRC~B1526$-$423).


\begin{acknowledgments}
The authors are grateful to the referee for constructive comments that helped them to expand their analysis, substantially improving and enriching the paper content.
They acknowledge Francesco Massaro for sowing the seed aiming at the coverage of the whole SMS4 sample in the X-rays with \sw.
They thank Andy Goulding for assistance in understanding the IR emission of AGN and for introducing them to the intricacies of the WISE source catalogs.
They also thank the Swift PI, Brad Cenko, and his deputies for approving the requested ToO observations, and the Science Operations Team for implementing them. 
A.M. thanks Riccardo Campana and Dario Gasparrini for their help and support, when needed.
A.M. acknowledges financial support from the ASI-INAF agreement No.~2014-049-R0 and its No.~2014-049-R1-2016 and No.~2014-049-R2-2017 addenda.
W.F., C.J., and R.K. acknowledge support from the Smithsonian Institution and the Chandra High Resolution Camera Project through NASA contract NAS8-03060.
This research has made use of archival data, software, or online services provided by the ASI Space Science Data Center (SSDC); the High Energy Astrophysics Science Archive Research Center (HEASARC) provided by NASA’s Goddard Space Flight Center; the SIMBAD database operated at CDS, Strasbourg, France; the NASA/IPAC Extragalactic Database (NED) operated by the Jet Propulsion Laboratory, California Institute of Technology, under contract with the National Aeronautics and Space Administration; and the NASA/IPAC Infrared Science Archive, which is funded by the National Aeronautics and Space Administration and operated by the California Institute of Technology. 
\end{acknowledgments}

%

\facilities{\sw~(XRT), SkyView Virtual Observatory (https://skyview.gsfc.nasa.gov/current/cgi/query.pl), IRSA}


\software{HEASoft \citep{2014ascl.soft08004N}, TOPCAT \citep{2005ASPC..347...29T}, SAO Image DS9 \citep{2000ascl.soft03002S,2003ASPC..295..489J}}


\appendix
\section{List of sources not yet observed by narrow-field X-ray instruments}
\label{app:1}

In Table~\ref{tab:08}, we list the SMS4 sources that had not been observed with any narrow-field telescope (\sw, \cha, or \xmm) to allow a precise X-ray location as of 2021 October.
Since the compilation, we have successfully proposed \sw~observations for the 18 brightest sources. 
These sources are marked in bold.

\begin{table*} 
\begin{center}
\tiny
\caption{SMS4 Sources with No Data (2021 September) from Narrow-field X-Ray Telescopes on board \sw, \cha, or \xmm}
\label{tab:08}
\begin{tabular}{ccccccccc}
\hline
        (1)           &     (2)     &             (3)             &   (4)     &   (5)    &      (6)              &   (7)    &      (8)        &          (9)        \\  
    SMS4 Name         & R.A.(J2000) &         Decl.(J2000)        & Redshift  & $S_{178}$ & $\overline{S}_{181}$  &    LAS   & Radio Structure & Optical Counterpart \\
                      & ($^{h~m~s}$) & ($^{\circ}$~\arcmin~\arcsec) &           &  (Jy)    &      (Jy)             & (arcsec) &                 &                     \\
\hline                                                                                                                                           
     MRC B0003$-$833  & 00 06 14.60 & $-$83 06 00.0               & (0.32)    &  13.2    & 10.6                  &      60  &                 &     g               \\
     MRC B0036$-$392  & 00 38 26.89 & $-$38 59 46.7               & 0.592     &  12.7    &  ~\,9.5               &      10  & D2              &     Q               \\
{\bf MRC B0103$-$453} & 01 05 20.88 & $-$45 05 28.2               & (0.71)    &  19.0    & 17.6                  &     140  &                 &     g     d         \\
     MRC B0110$-$692  & 01 11 42.88 & $-$68 59 59.3               & ($>$0.56) &  12.2    & 10.0                  &      56  &                 &     BF              \\
     MRC B0119$-$634  & 01 21 40.36 & $-$63 09 02.0               & 0.837     &  11.4    &  ~\,9.1               &      43  & 2               &     Q               \\
{\bf MRC B0202$-$765} & 02 02 13.53 & $-$76 20 06.8               & 0.38925   &  19.0    & 14.9                  &      20  & 2               &     Q               \\
{\bf MRC B0242$-$514} & 02 43 44.56 & $-$51 12 36.7               & (0.72)    &  19.0    & 13.6                  &      53  & 2               &     g     d         \\
     MRC B0251$-$675  & 02 51 56.30 & $-$67 18 02.6               & 1.002     &  13.5    &  ~\,8.6               &      38  & 2               &     Q               \\
     PKS B0319$-$45   & 03 21 08.10 & $-$45 12 51.0               & 0.0633    &  13.5    & 14.5                  &    1536  &                 &     g               \\
{\bf MRC B0344$-$345} & 03 46 31.10 & $-$34 22 40.0               & 0.0538    &  18.0    & 14.6                  &     264  &                 &     g               \\
{\bf MRC B0427$-$366} & 04 29 40.17 & $-$36 30 55.2               &           &  18.0    & 14.7                  &      14  & 2               &     Q?              \\
     MRC B0456$-$301  & 04 58 26.43 & $-$30 07 22.4               & 0.063     &  13.3    & 11.1                  &     348  &                 &     g               \\
     MRC B0511$-$484  & 05 12 50.79 & $-$48 24 04.3               & 0.30638   &  12.2    & 14.4                  &     132  &                 &     g               \\
     PKS B0511$-$30   & 05 13 34.60 & $-$30 28 24.0               & 0.0583    &  16.0    & 13.2                  &     636  &                 &     g               \\
     MRC B0646$-$398  & 06 48 11.30 & $-$39 57 07.3               &           &  13.9    & 11.0                  &      82  &                 &     Q?              \\
     MRC B0658$-$656  & 06 58 12.84 & $-$65 44 53.3               & ($>$0.14) &  11.0    &  ~\,9.8               &      25  & 2               &     O               \\
     MRC B0719$-$553  & 07 20 14.63 & $-$55 25 16.4               & 0.216     &  11.9    &  ~\,8.9               &      52  &                 &     g               \\
{\bf MRC B1017$-$426} & 10 20 03.77 & $-$42 51 31.5               & 1.28      &  23.0    & 21.3                  &      10  & 2               &     Q               \\
     MRC B1123$-$351  & 11 25 54.46 & $-$35 23 19.7               & 0.032     &  11.3    &  ~\,9.9               &      60  &                 &     g               \\
     MRC B1136$-$320  & 11 39 17.17 & $-$32 22 33.2               & (0.67)    &  14.0    & 12.0                  &      62  &                 &     g     d         \\
     MRC B1143$-$316  & 11 46 20.55 & $-$31 57 14.5               & (1.35)    &  12.7    &  ~\,9.5               &      46  &                 &     g     d         \\
     MRC B1215$-$457  & 12 18 06.23 & $-$46 00 28.6               & 0.529     &  12.8    & 12.4                  &      10  & c               &     Q               \\
     MRC B1421$-$382  & 14 24 16.53 & $-$38 26 49.9               & 0.4068    &  15.0    & 12.5                  &      61  & 2               &     Q               \\
     MRC B1425$-$479  & 14 28 57.20 & $-$48 12 02.0               & (0.11)    &  13.0    & 12.2                  &     270  &                 &     g               \\
{\bf MRC B1526$-$423} & 15 30 14.30 & $-$42 31 53.2               & (0.5)     &  44.0    & 34.6                  &      50  & 2               &     g     d         \\
     MRC B1607$-$841  & 16 19 34.05 & $-$84 18 18.9               & (1.11)    &  11.2    &  ~\,7.1               &      10  & c               &     g     e         \\
     MRC B1716$-$800  & 17 25 25.68 & $-$80 04 45.4               & (0.45)    &  12.7    & 10.8                  &      43  &                 &     g               \\
     MRC B1721$-$836  & 17 33 55.37 & $-$83 42 53.3               & (1.42)    &  13.0    &  ~\,5.5               &      10  & 2               &     g               \\
{\bf MRC B1754$-$597} & 17 59 06.31 & $-$59 46 59.6               & (0.8)     &  20.0    & 25.2                  &      21  & 2               &     g               \\
     MRC B1756$-$663  & 18 01 17.96 & $-$66 23 02.9               & (0.93)    &  11.6    &  ~\,8.5               &      10  & c               &     g               \\
{\bf MRC B1814$-$519} & 18 18 06.96 & $-$51 58 09.6               & (0.48)    &  24.0    & 23.5                  &      10  & 2               &     g               \\
{\bf MRC B1817$-$391} & 18 20 35.31 & $-$39 09 28.4               & (0.91)    &  19.0    & 15.7                  &      16  & 2               &     g     d         \\
{\bf MRC B1817$-$640} & 18 22 16.10 & $-$63 59 19.3               & 0.67      &  28.0    & 21.4                  &      31  & 2               &     g               \\
{\bf MRC B1827$-$360} & 18 30 58.90 & $-$36 02 30.3               & (0.12)    &  33.0    & 30.1                  &      10  & c               &     g               \\
     MRC B1840$-$404  & 18 44 28.35 & $-$40 21 55.8               & (1.77)    &  16.0    & 14.4                  &      46  &                 &     g     d         \\
     MRC B1933$-$587  & 19 37 32.37 & $-$58 38 27.8               & 1.92      &  14.0    & 14.1                  &      10  & T               &     Q               \\
     MRC B1940$-$406  & 19 43 51.87 & $-$40 30 10.2               & (0.18)    &  14.0    & 11.5                  &     126  &                 &     g               \\
{\bf MRC B1953$-$425} & 19 57 15.22 & $-$42 22 20.1               & (0.82)    &  18.0    & 15.9                  &      10  & c               &     g               \\
{\bf MRC B2032$-$350} & 20 35 47.66 & $-$34 54 02.5               & (0.56)    &  27.0    & 23.9                  &      26  & 2               &     g     d         \\
{\bf MRC B2041$-$604} & 20 45 20.72 & $-$60 19 01.3               & 1.464     &  27.0    & 20.8                  &      32  &                 &     g               \\
     MRC B2049$-$368  & 20 52 17.50 & $-$36 40 29.8               & (0.89)    &  11.4    &  ~\,9.5               &      10  & 2               &     g               \\
     MRC B2115$-$305  & 21 18 10.68 & $-$30 19 14.7               & 0.98      &  11.4    & 10.5                  &      12  & 2               &     Q               \\
{\bf MRC B2140$-$434} & 21 43 33.46 & $-$43 12 48.5               & 0.65      &  19.0    & 14.7                  &      56  & 2               &     Q               \\
{\bf MRC B2140$-$817} & 21 47 23.96 & $-$81 32 11.7               & (0.64)    &  26.0    & 12.7                  &      44  & 2               &     g               \\
     MRC B2150$-$520  & 21 54 07.47 & $-$51 50 15.0               & (0.79)    &  18.0    & 15.1                  &      15  & 2               &     g               \\
     MRC B2201$-$555  & 22 05 04.89 & $-$55 17 43.3               & (0.51)    &  13.1    &  ~\,8.9               &      10  & 2               &     g               \\
     MRC B2223$-$528  & 22 27 02.69 & $-$52 33 25.4               & (0.41)    &  17.0    & 13.7                  &      11  & 2               &     g               \\
     MRC B2226$-$411  & 22 29 18.47 & $-$40 51 31.7               & 0.4462    &  11.3    &  ~\,9.3               &      15  & T               &     Q               \\
     MRC B2226$-$386  & 22 29 46.90 & $-$38 23 59.2               & (1.3)     &  14.0    & 11.5                  &      10  & 2               &     g               \\
     MRC B2252$-$530  & 22 55 50.13 & $-$52 45 42.4               & (0.55)    &  12.1    &  ~\,7.9               &      10  & 2               &     g               \\
     MRC B2253$-$522  & 22 56 47.53 & $-$51 58 41.5               & (0.43)    &  15.0    & 10.2                  &      11  & 2               &     g     d         \\
     MRC B2259$-$375  & 23 02 23.86 & $-$37 18 05.8               & (1.14)    &  13.0    & 10.0                  &      10  &                 &     g               \\
     MRC B2323$-$407  & 23 26 34.12 & $-$40 27 17.8               & (0.81)    &  15.0    & 11.9                  &      10  & 2               &     g               \\
{\bf MRC B2331$-$416} & 23 34 26.13 & $-$41 25 25.1               & 0.907     &  30.0    & 23.7                  &      19  & 2               &     g               \\
     MRC B2332$-$668  & 23 35 11.29 & $-$66 37 04.6               & (0.08)    &  12.7    &  ~\,9.7               &      28  & 2               &     O               \\
     MRC B2338$-$585  & 23 41 18.36 & $-$58 16 10.3               & (0.64)    &  15.0    & 11.8                  &      10  & 2               &     g     d         \\ 
\hline
\end{tabular}
\end{center}
\tablecomments{{\scriptsize Column (1): the name in SMS4, according to the MRC or PKS designation, with bold characters to mark those accepted in 2021 October for observation with \sw. Columns (2) and (3): R.A. and decl. of the SMS4 source. Column (4): the redshift, with lower limits and photometric estimates in parentheses. Column (5): the extrapolated flux density $S_{178}$ from SMS4. Column (6): the total integrated flux density $\overline{S}_{181}$ from G4Jy. Column (7): the largest angular size of the radio source at 843~MHz. Column (8): the structural classification of the radio source, following \cite{1974MNRAS.167P..31F}, where 2 = FR2 double; D2 = double, with one component coincident with the optical counterpart; T = core-dominated triple; c = structure not resolved well enough to classify. Column (9): the classification of the associated optical counterpart, where d = extended radio source with more than one candidate; e = compact radio source with large radio-optical offset; g = galaxy; Q = quasar; Q? = quasar candidate; BF = blank field; O = field obscured by star.}}
\end{table*}


\newpage 
\bibliography{aas32335}{}
\bibliographystyle{aasjournal}



\end{document}